\pdfoutput=1
\documentclass[12pt,a4paper]{article}

\usepackage{ifthen}
\newboolean{pdflatex}
\setboolean{pdflatex}{true}

\newboolean{articletitles}
\setboolean{articletitles}{true}

\newboolean{uprightparticles}
\setboolean{uprightparticles}{false}

\def\paperauthors{LHCb collaboration}
\def\paperasciititle{Evidence for the decay B0MuMu and improved measurements of the decay B0sMuMu}
\def\papertitle{Evidence for the decay $B^0 \to \mu^+ \mu^-$ and improved measurements of the decay $B^0_s \to \mu^+ \mu^-$}
\def\paperkeywords{{High Energy Physics}, {LHCb}}
\def\papercopyright{\the\year\ CERN for the benefit of the LHCb collaboration}
\def\paperlicence{CC BY 4.0 licence}
\def\paperlicenceurl{https://creativecommons.org/licenses/by/4.0/}

\newif\ifEnableSectionTOCLinks
\EnableSectionTOCLinksfalse 

\usepackage[top=1in, bottom=1.25in, left=1in, right=1in]{geometry}

\usepackage{microtype}
\usepackage{lineno}  
\usepackage{xspace} 
\usepackage{caption} 

\usepackage{graphicx}  
\usepackage{color}
\usepackage{colortbl}
\graphicspath{{./figs/}} 

\usepackage{amsmath} 
\usepackage{amssymb}
\usepackage{amsfonts}
\usepackage{upgreek} 

\usepackage[normalem]{ulem} 

\newcommand*\patchAmsMathEnvironmentForLineno[1]{%
\expandafter\let\csname old#1\expandafter\endcsname\csname #1\endcsname
\expandafter\let\csname oldend#1\expandafter\endcsname\csname
end#1\endcsname
 \renewenvironment{#1}%
   {\linenomath\csname old#1\endcsname}%
   {\csname oldend#1\endcsname\endlinenomath}%
}
\newcommand*\patchBothAmsMathEnvironmentsForLineno[1]{%
  \patchAmsMathEnvironmentForLineno{#1}%
  \patchAmsMathEnvironmentForLineno{#1*}%
}
\AtBeginDocument{%
\patchBothAmsMathEnvironmentsForLineno{equation}%
\patchBothAmsMathEnvironmentsForLineno{align}%
\patchBothAmsMathEnvironmentsForLineno{flalign}%
\patchBothAmsMathEnvironmentsForLineno{alignat}%
\patchBothAmsMathEnvironmentsForLineno{gather}%
\patchBothAmsMathEnvironmentsForLineno{multline}%
\patchBothAmsMathEnvironmentsForLineno{eqnarray}%
}

\usepackage[pdftex,
            pdfauthor={\paperauthors},
            pdftitle={\paperasciititle},
            pdfkeywords={\paperkeywords}]{hyperref}
\usepackage{hyperxmp}
\hypersetup{
    pdfcopyright={Copyright (C) \papercopyright},
    pdflicenseurl={\paperlicenceurl}
}

\usepackage[colorinlistoftodos,textsize=scriptsize]{todonotes}

\usepackage[bottom,flushmargin,hang,multiple]{footmisc}

\usepackage[all]{hypcap} 

\usepackage{xspace} 
\usepackage{upgreek}

\def\lhcb   {\mbox{LHCb}\xspace}

\def\upgradeone {\mbox{Upgrade~I}\xspace}

\def\MagUp {\mbox{\em Mag\kern -0.05em Up}\xspace}

\ifdefined\ifuprightparticles
\else
\newboolean{uprightparticles}
\setboolean{uprightparticles}{false} 
\fi

\ifthenelse{\boolean{uprightparticles}}%
{

 \def\Pmu         {\ensuremath{\upmu}\xspace}                 
 \def\Pnu         {\ensuremath{\upnu}\xspace}                 
                  
 \def\Ppi         {\ensuremath{\uppi}\xspace}

 \def\Ppsi        {\ensuremath{\uppsi}\xspace}

 \def\PDelta      {\ensuremath{\Delta}\xspace}                 
 \def\PXi         {\ensuremath{\Xi}\xspace}                 
 \def\PLambda     {\ensuremath{\Lambda}\xspace}                 
 \def\PSigma      {\ensuremath{\Sigma}\xspace}                 
 \def\POmega      {\ensuremath{\Omega}\xspace}                 
 \def\PUpsilon    {\ensuremath{\Upsilon}\xspace}
 \let\oldPi\Pi
 \def\PPi         {\ensuremath{\oldPi}\xspace}

 \def\PB      {\ensuremath{\mathrm{B}}\xspace}                 
 \def\PD      {\ensuremath{\mathrm{D}}\xspace}                 
 \def\PJ      {\ensuremath{\mathrm{J}}\xspace}                 
 \def\PK      {\ensuremath{\mathrm{K}}\xspace}                 
 \def\Pb      {\ensuremath{\mathrm{b}}\xspace}                 
 \def\Pc      {\ensuremath{\mathrm{c}}\xspace}

 \def\Ps      {\ensuremath{\mathrm{s}}\xspace}

 \def\thebaroffset{0.0em}
}
{

 \def\Pmu         {\ensuremath{\mu}\xspace}                 
 \def\Pnu         {\ensuremath{\nu}\xspace}                 
                  
 \def\Ppi         {\ensuremath{\pi}\xspace}

 \def\Ppsi        {\ensuremath{\psi}\xspace}                 
                  
 \mathchardef\PDelta="7101
 \mathchardef\PXi="7104
 \mathchardef\PLambda="7103
 \mathchardef\PSigma="7106
 \mathchardef\POmega="710A
 \mathchardef\PUpsilon="7107
 \mathchardef\PPi="7105
 \def\PB      {\ensuremath{B}\xspace}                 
 \def\PD      {\ensuremath{D}\xspace}                 
 \def\PJ      {\ensuremath{J}\xspace}                 
 \def\PK      {\ensuremath{K}\xspace}                 
 \def\Pb      {\ensuremath{b}\xspace}                 
 \def\Pc      {\ensuremath{c}\xspace}

 \def\Ps      {\ensuremath{s}\xspace}

 \def\thebaroffset{0.18em}
}
\newcommand{\offsetoverline}[2][\thebaroffset]{\kern #1\overline{\kern -#1 #2}}%

\makeatletter
\ifcase \@ptsize \relax
  \newcommand{\miniscule}{\@setfontsize\miniscule{4}{5}}
\or
  \newcommand{\miniscule}{\@setfontsize\miniscule{5}{6}}
\or
  \newcommand{\miniscule}{\@setfontsize\miniscule{5}{6}}
\fi
\makeatother

\DeclareRobustCommand{\optbar}[1]{\shortstack{{\miniscule (\rule[.5ex]{1.25em}{.18mm})}
  \\ [-.7ex] $#1$}}

\def\mup        {{\ensuremath{\Pmu^+}}\xspace}
\def\mun        {{\ensuremath{\Pmu^-}}\xspace} 

\def\neub       {{\ensuremath{\overline{\Pnu}}}\xspace}

\def\squark    {{\ensuremath{\Ps}}\xspace}

\def\cquark    {{\ensuremath{\Pc}}\xspace}

\def\bquark    {{\ensuremath{\Pb}}\xspace}

\def\pion   {{\ensuremath{\Ppi}}\xspace}

\def\pip    {{\ensuremath{\pion^+}}\xspace}
\def\pim    {{\ensuremath{\pion^-}}\xspace}

\def\kaon    {{\ensuremath{\PK}}\xspace}

\def\KorKbar {\kern \thebaroffset\optbar{\kern -\thebaroffset \PK}{}\xspace}

\def\Kp      {{\ensuremath{\kaon^+}}\xspace}
\def\Km      {{\ensuremath{\kaon^-}}\xspace}

\def\D       {{\ensuremath{\PD}}\xspace}

\def\DorDbar {\kern \thebaroffset\optbar{\kern -\thebaroffset \PD}\xspace}

\def\Dp      {{\ensuremath{\D^+}}\xspace}
\def\Dm      {{\ensuremath{\D^-}}\xspace}

\def\DpDm    {\ensuremath{\Dp {\kern -0.16em \Dm}}\xspace}

\def\B       {{\ensuremath{\PB}}\xspace}
\def\Bbar    {{\ensuremath{\offsetoverline{\PB}}}\xspace}

\def\BorBbar {\kern \thebaroffset\optbar{\kern -\thebaroffset \PB}\xspace}
\def\Bz      {{\ensuremath{\B^0}}\xspace}
\def\Bzb     {{\ensuremath{\Bbar{}^0}}\xspace}
\def\Bd      {{\ensuremath{\B^0}}\xspace}

\def\BdorBdbar {\kern \thebaroffset\optbar{\kern -\thebaroffset \Bd}\xspace}
\def\Bu      {{\ensuremath{\B^+}}\xspace}

\def\Bp      {{\ensuremath{\Bu}}\xspace}

\def\Bs      {{\ensuremath{\B^0_\squark}}\xspace}
\def\Bsb     {{\ensuremath{\Bbar{}^0_\squark}}\xspace}
\def\BsorBsbar {\kern \thebaroffset\optbar{\kern -\thebaroffset \Bs}\xspace}

\def\Bds     {{\ensuremath{\B_{(\squark)}^0}}\xspace}

\def\jpsi     {{\ensuremath{{\PJ\mskip -3mu/\mskip -2mu\Ppsi}}}\xspace}

\def\Y#1S{\ensuremath{\PUpsilon{(#1S)}}\xspace}

\def\Lz          {{\ensuremath{\PLambda}}\xspace}

\def\LorLbar     {\kern \thebaroffset\optbar{\kern -\thebaroffset \PLambda}\xspace}

\def\Lc          {{\ensuremath{\Lz^+_\cquark}}\xspace}

\def\Lb           {{\ensuremath{\Lz^0_\bquark}}\xspace}

\newcommand{\decay}[2]{\ensuremath{\mathinner{#1\!\to #2}}\xspace}

\def\to                 {\ensuremath{\rightarrow}\xspace}

\def\BdTopipi     {\decay{\Bd}{\pip\pim}}
\def\BdToKpi      {\decay{\Bd}{\Kp\pim}}
\def\BsToKK       {\decay{\Bs}{\Kp\Km}}

\def\AT#1     {\ensuremath{A_{\mathrm{T}}^{#1}}\xspace}           

\def\Bsmm     {\decay{\Bs}{\mup\mun}}
\def\Bdmm     {\decay{\Bd}{\mup\mun}}

\def\C#1      {\ensuremath{\mathcal{C}_{#1}}\xspace}                       
\def\Cp#1     {\ensuremath{\mathcal{C}_{#1}^{'}}\xspace}                    
\def\Ceff#1   {\ensuremath{\mathcal{C}_{#1}^{\mathrm{(eff)}}}\xspace}        
\def\Cpeff#1  {\ensuremath{\mathcal{C}_{#1}^{'\mathrm{(eff)}}}\xspace}       
\def\Ope#1    {\ensuremath{\mathcal{O}_{#1}}\xspace}                       
\def\Opep#1   {\ensuremath{\mathcal{O}_{#1}^{'}}\xspace}                    

\newcommand{\aunit}[1]{\ensuremath{\text{\,#1}}}       

\newcommand{\tev}{\aunit{Te\kern -0.1em V}\xspace}
\newcommand{\gev}{\aunit{Ge\kern -0.1em V}\xspace}
\newcommand{\mev}{\aunit{Me\kern -0.1em V}\xspace}
\newcommand{\kev}{\aunit{ke\kern -0.1em V}\xspace}
\newcommand{\ev}{\aunit{e\kern -0.1em V}\xspace}
 
\newcommand{\mevc}{\ensuremath{\aunit{Me\kern -0.1em V\!/}c}\xspace}
\newcommand{\gevc}{\ensuremath{\aunit{Ge\kern -0.1em V\!/}c}\xspace}
\newcommand{\mevcc}{\ensuremath{\aunit{Me\kern -0.1em V\!/}c^2}\xspace}
\newcommand{\gevcc}{\ensuremath{\aunit{Ge\kern -0.1em V\!/}c^2}\xspace}

\def\cm   {\aunit{cm}\xspace}

\def\sec  {\ensuremath{\aunit{s}}\xspace}

\def\ps   {\ensuremath{\aunit{ps}}\xspace}

\newcommand{\chisq}{\ensuremath{\chi^2}\xspace}

\newcommand{\chisqip}{\ensuremath{\chi^2_{\text{IP}}}\xspace}

\def\gsim{{~\raise.15em\hbox{$>$}\kern-.85em
          \lower.35em\hbox{$\sim$}~}\xspace}
\def\lsim{{~\raise.15em\hbox{$<$}\kern-.85em
          \lower.35em\hbox{$\sim$}~}\xspace}

\def\sPlot{\mbox{\em sPlot}\xspace}

\def\sqs   {\ensuremath{\protect\sqrt{s}}\xspace}

\def\pt         {\ensuremath{p_{\mathrm{T}}}\xspace}

\def\evtgen     {\mbox{\textsc{EvtGen}}\xspace}

\def\geant      {\mbox{\textsc{Geant4}}\xspace}

\def\photos     {\mbox{\textsc{Photos}}\xspace}

\def\pythia     {\mbox{\textsc{Pythia}}\xspace}

\def\tell1  {TELL1\xspace}
\def\ukl1   {UKL1\xspace}

\newcommand{\ie}{\mbox{\itshape i.e.}\xspace}

\newcommand{\lhcborcid}[1]{\href{https://orcid.org/#1}{\hspace*{0.1em}\raisebox{-0.45ex}{\includegraphics[width=1em]{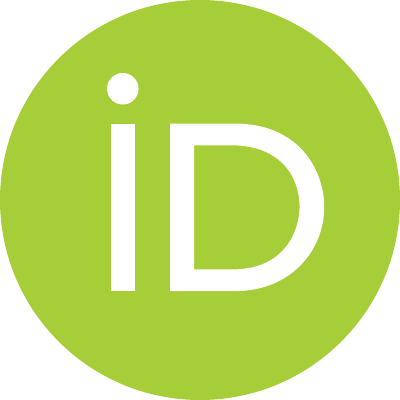}}}}

\hypersetup{
  colorlinks   = true, 
  urlcolor     = blue, 
  linkcolor    = blue, 
  citecolor    = red   
}

\ifEnableSectionTOCLinks
    \usepackage[explicit]{titlesec} 
    
    \let\oldcontentsline\contentsline
    \renewcommand

    \titleformat{\section}{\normalfont\Large\bf}{\hyperlink{tocsection.\thesection}{{\thesection} \parbox[t]{\dimexpr\textwidth-1pc}{#1}}}{1pc}{}

    \titleformat{\subsection}{\normalfont\bf}{\hyperlink{tocsubsection.\thesubsection}{{\thesubsection} \parbox[t]{\dimexpr\textwidth-1pc}{#1}}}{1pc}{}

    \titleformat{name=\section,numberless}[display]{}{}{0pt}{\normalfont\Huge\bfseries #1}
\fi

\usepackage{cite} 
\usepackage{LHCb/mciteplus}

\usepackage{placeins}
\newcommand{\BRof}[1]{\ensuremath{{\cal B}(#1)}\xspace}

\newcommand{\Bsmumu}{\decay{\Bs}{\mup\mun}}
\newcommand{\Bdmumu}{\decay{\Bd}{\mup\mun}}
\def\Rmumu{\ensuremath{\mathcal{R}_{\mu\mu}}\xspace}
\newcommand{\Bmumu}{\decay{B^0_{(s)}}{\mup\mun}}
\newcommand{\BuToJpsiK}{\decay{\Bp}{\jpsi\Kp}}

\newcommand{\Jpsimumu}{\decay{\jpsi}\mup\mun}
\newcommand{\PhiKK}{\decay{\phi} \Kp \Km}

\newcommand{\BsToJpsiphi}{\decay{\Bs}{\jpsi\phi}}

\def\ADG{\ensuremath{\mathcal{A}^{\mu\mu}_{\Delta\Gamma_s}}\xspace}
\newcommand{\Bhh}{\decay{\Bds}{h^+ h^{\prime -} }}

\begin{document}

\renewcommand{\thefootnote}{\fnsymbol{footnote}}
\setcounter{footnote}{1}


\begin{titlepage}
\pagenumbering{roman}

\vspace*{-1.5cm}
\centerline{\large EUROPEAN ORGANIZATION FOR NUCLEAR RESEARCH (CERN)}
\vspace*{1.5cm}
\noindent
\begin{tabular*}{\linewidth}{lc@{\extracolsep{\fill}}r@{\extracolsep{0pt}}}
\ifthenelse{\boolean{pdflatex}}
{\vspace*{-1.5cm}\mbox{\!\!\!\includegraphics[width=.14\textwidth]{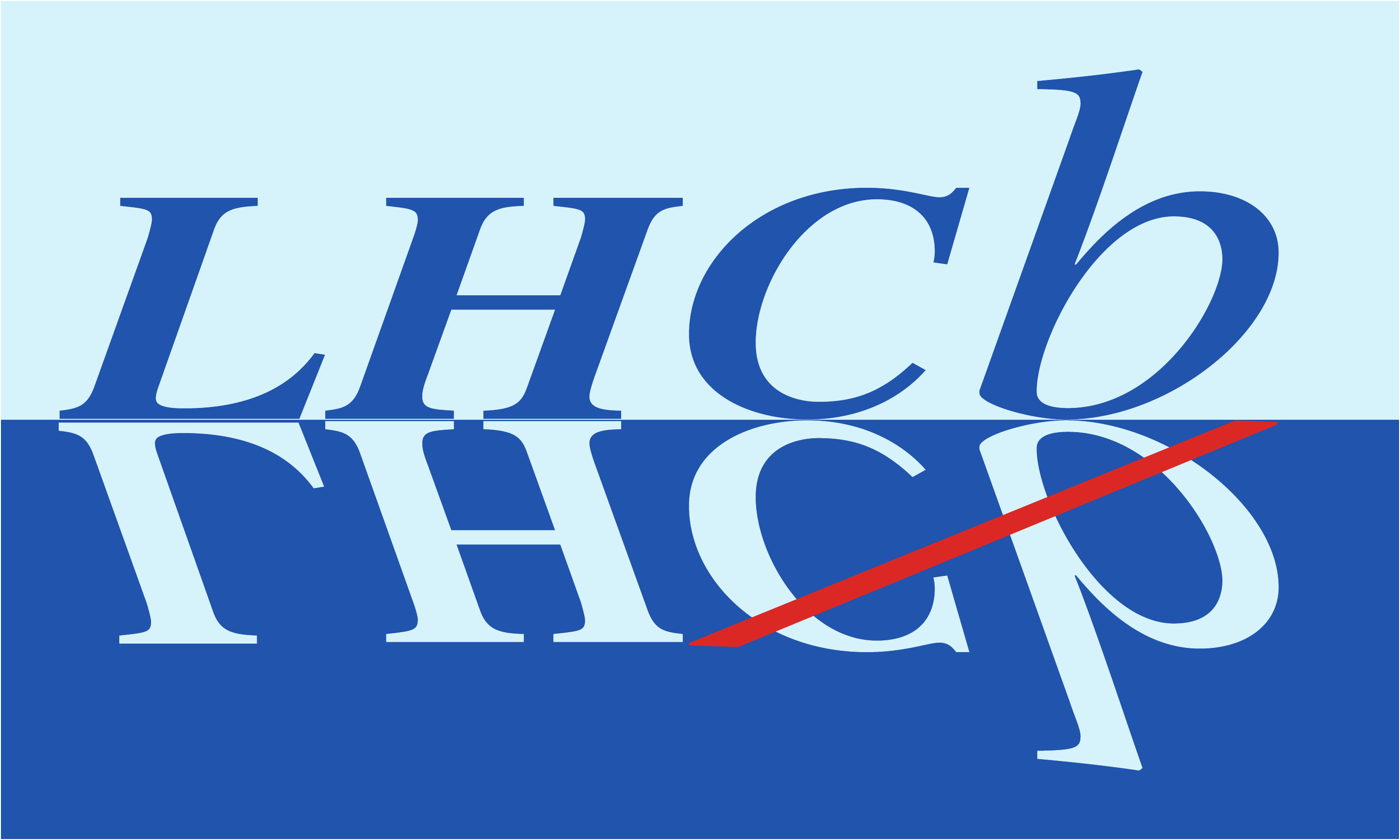}} & &}%
{\vspace*{-1.2cm}\mbox{\!\!\!\includegraphics[width=.12\textwidth]{LHCb/Figures/lhcb-logo.eps}} & &}%
\\
 & & CERN-EP-2026-287 \\  
 & & LHCb-PAPER-2026-053 \\  
     & & November 2, 2026 \\ 
 & & \\
\end{tabular*}

\vspace*{4.0cm}

{\normalfont\bfseries\boldmath\huge
\begin{center}
  \papertitle 
\end{center}
}

\vspace*{2.0cm}

\begin{center}
\paperauthors\footnote{Authors are listed at the end of this paper.}
\end{center}

\vspace{\fill}

\begin{abstract}
\noindent
The first single-experiment evidence for the rare decay $B^0 \to \mu^+ \mu^-$ and the world's most precise measurements of the $B^0_s \to \mu^+ \mu^-$ branching fraction and effective lifetime are reported. The results are obtained using Run~3 proton-proton collision data corresponding to an integrated luminosity of 18 fb$^{-1}$ recorded with the LHCb Upgrade I detector at a centre-of-mass energy of $13.6$ TeV between 2024 and 2025. Their combination with the Run~1 and Run~2 LHCb results further improves the precision. All measurements are found to be in agreement with the Standard Model expectations.

\end{abstract}

\vspace*{2.0cm}

\begin{center}
  Submitted to Physical Review X
\end{center}

\vspace{\fill}

{\footnotesize 
\centerline{\copyright~\papercopyright. \href{\paperlicenceurl}{\paperlicence}.}}
\vspace*{2mm}

\end{titlepage}


\newpage
\setcounter{page}{2}
\mbox{~}

\renewcommand{\thefootnote}{\arabic{footnote}}
\setcounter{footnote}{0}

\cleardoublepage


\pagestyle{plain} 
\setcounter{page}{1}
\pagenumbering{arabic}

\section*{Introduction}
The Standard Model of particle physics (SM) describes the known elementary particles and their strong, weak and electromagnetic interactions within a single quantum field theory. It has been tested with remarkable precision over half a century, most recently at the Large Hadron Collider (LHC) at CERN, where the discovery of the Higgs boson~\cite{ATLAS:2012yve,CMS:2012qbp} completed its particle content. Nevertheless, the SM has no candidate for dark matter, which is about five times more abundant than ordinary matter~\cite{Planck:2018vyg}, and it cannot account for the observed asymmetry between matter and antimatter in the universe~\cite{Sakharov:1967dj}.

Extensions of the SM that address these shortcomings generally introduce new particles. These may be searched for at colliders if sufficient energy is available, but particles too heavy to be produced can still reveal their presence through precise measurements of the properties and interactions of known particles. By judicious choice of observables, it is possible to gain sensitivity to unobserved particles through their quantum mechanical interference with SM processes, even when their rest mass may be orders of magnitude higher than the energy scale of the collider. This strategy has a long historical pedigree: the suppression of strangeness-changing neutral-current decays led to the prediction of the charm quark~\cite{Glashow:1970gm} before its direct discovery~\cite{E598:1974sol,SLAC-SP-017:1974ind}, while the rate of $\Bz$--$\Bzb$ mixing~\cite{ARGUS:1987xtv} and precision electroweak measurements~\cite{Langacker:1991an} provided early indirect evidence for a heavy top quark, prior to its direct observation in 1995~\cite{CDF:1995wbb,D0:1995jca}.

Processes in which a beauty ($b$) quark couples to either a strange ($s$) or down ($d$) quark and emits a lepton-antilepton pair ($\ell^+\ell^-$) are particularly sensitive probes of New Physics (NP). Since the only gauge field in the SM that can change flavour directly is the charged-current $W^{\pm}$ boson, these flavour-changing neutral current (FCNC) transitions proceed via higher-order interactions involving the electroweak and top quark fields, which causes them to be highly suppressed. However, this suppression makes these processes particularly sensitive to contributions from high-scale NP, which can alter their branching fractions, effective lifetimes, charge-parity asymmetries and angular observables. Measurements of these processes can probe energy scales up to 100\tev~\cite{Altmannshofer:2012az}, significantly above the direct reach of the LHC.

Over the past decade, several persistent deviations from SM predictions have been reported in the differential branching fractions~\cite{LHCb-PAPER-2014-006,LHCb-PAPER-2021-014,LHCb-PAPER-2025-041,CMS:2024syx} and angular observables~\cite{LHCb-PAPER-2025-041,LHCb-PAPER-2020-041,LHCb-PAPER-2021-022,CMS:2024atz,ATLAS:2018gqc,Belle:2016fev} of $b\to s \ell^+ \ell^-$ decays, where $\ell=e,\mu$, hinting at effects from NP. However, the interpretation of these results is ambiguous due to the difficulty in accounting theoretically for contributions from quarks and gluons, particularly in respect of hadronic form factors and charm loops~\cite{Horgan:2013hoa,Bharucha:2015bzk,Gubernari:2018wyi,Parrott:2022zte,Khodjamirian:2010vf,Lyon:2014hpa,Ciuchini:2015qxb,Gubernari:2020eft,Gubernari:2022hxn,LHCb-PAPER-2024-011,LHCb-PAPER-2023-033}. 

\begin{figure}[t]
    \centering
    \includegraphics[width=0.49\linewidth]{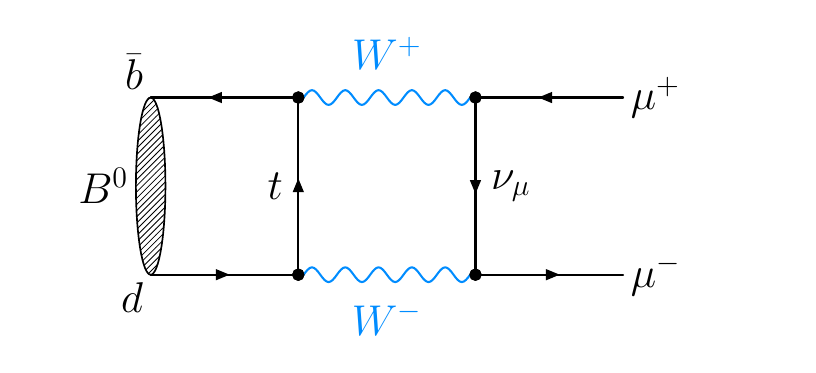}
    \includegraphics[width=0.49\linewidth]{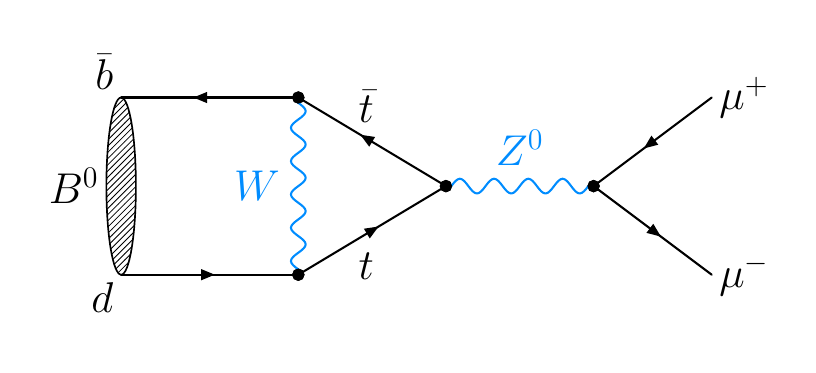}
    \includegraphics[width=0.49\linewidth]{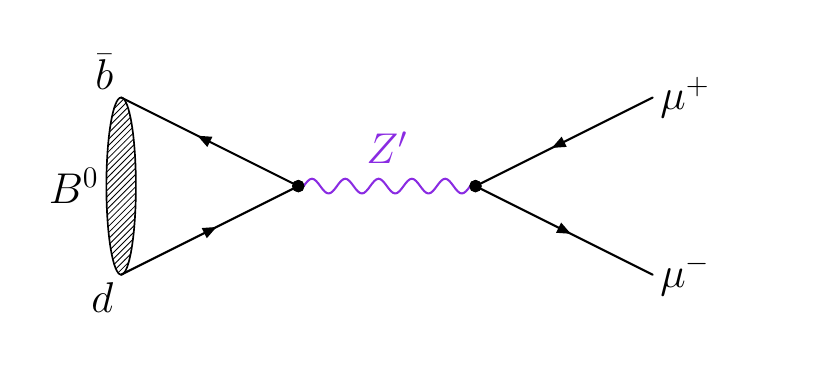}
    \includegraphics[width=0.49\linewidth]{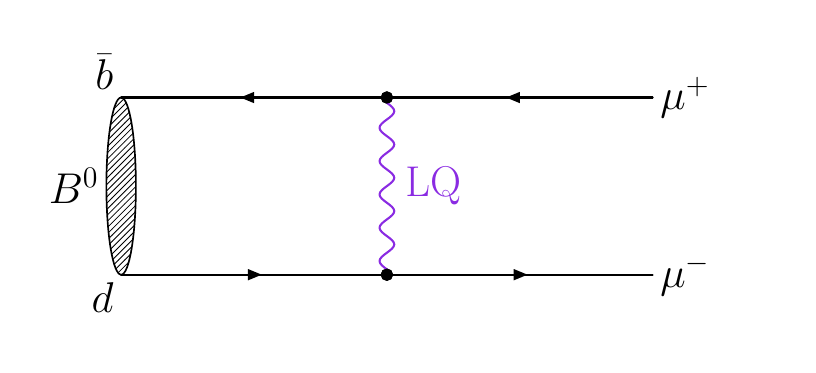}
    \caption{Example Feynman diagrams for the $B^0 \to \mu^+\mu^-$ decay:  (top left) SM box, (top right) SM triangle, (bottom left) NP $Z^{'}$ boson and (bottom right) NP vector leptoquark. The corresponding diagrams for $B^0_s \to \mu^+\mu^-$ are obtained by replacing the $d$ quark with an $s$ quark.}
    \label{fig:feynman}
\end{figure}

The decays \Bsmumu and \Bdmumu are examples of FCNC transitions, which involve the annihilation of a beauty antiquark with another down-type quark: $\bar{b}s\to \mu^+ \mu^-$ and $\bar{b}d\to \mu^+ \mu^-$ respectively.\footnote{Inclusion of charge-conjugated processes is implied throughout this article, unless otherwise specified.} Example SM and NP Feynman diagrams are shown in Fig.~\ref{fig:feynman}. These decays are further helicity suppressed compared with $b\to s \ell^+ \ell^-$ decays due to the initial $B_{(s)}^0$ meson having total angular momentum zero and the $W^{\pm}$ and $Z$ bosons having reduced couplings to right-helicity fermions and left-helicity antifermions. Consequently, these decays are very rare, with predicted branching fractions of $\BRof \Bsmumu = (3.66 \pm 0.14)\times 10^{-9}$ and $\BRof \Bdmumu = (1.03 \pm 0.05)\times 10^{-10}$ in the SM~\cite{Beneke:2019slt}. The purely leptonic final state means that the theoretical uncertainties associated with their branching fractions are small compared with those of $b\to s \ell^+ \ell^-$ decays. Residual hadronic effects are controlled via precise lattice QCD calculations~\cite{FLAG2024}.

The decays \Bmumu are particularly sensitive to contributions from additional Higgs scalar bosons, $Z^\prime$ vector bosons or leptoquarks, and provide complementary information on potential NP contributions to the deviations seen in $b\to s \ell^+ \ell^-$ decays. Precise measurements of their properties are thus highly motivated~\cite{Buras:2024per}. The suppression of the \Bdmm decay rate, relative to that of the \Bsmm decay, is predominantly due to the hierarchy of CKM matrix elements. The ratio of branching fractions is a theoretically clean observable and in the SM is given by 
\begin{equation}
\Rmumu
\equiv
\frac{\BRof \Bdmumu}{\BRof \Bsmumu} = \frac{\tau_{\Bd}\Gamma_{\rm H}}{\hbar}\left( \frac{f_\Bd}{f_\Bs}\right)^2 \left\vert \frac{V_{td}}{V_{ts}}\right\vert^2 \frac{\sqrt{M_\Bd^2-4m_\mu^2}}{\sqrt{M_\Bs^2-4m_\mu^2}},
\end{equation}
where $\tau_\Bd$ denotes the \Bd lifetime, $\Gamma_{\rm H}$ the decay width of the heavy \Bs mass eigenstate, $M_{\Bd}$ and $M_{\Bs}$ the \Bds meson masses, $f_{\Bd}$ and $f_{\Bs}$ their decay constants, and $m_\mu$ the muon mass. The SM prediction $\Rmumu^{\rm{SM}}=0.0280\pm0.0016$~\cite{Beneke:2019slt} also holds in all NP theories satisfying the Minimal Flavour Violation (MFV) hypothesis~\cite{DAmbrosio:2002vsn}. The measurement of \Rmumu can therefore reveal new flavour structures beyond the CKM paradigm.

In addition to the branching fractions and their ratio, the effective lifetime of the \Bsmm decay can be measured. It is defined from the decay-time distribution $\langle\Gamma\rangle$, where the angular brackets denote the sum over \Bs and \Bsb decays, as
\begin{equation}
 \tau_{\mu\mu} \equiv \frac{\int^\infty_0 t \left<\Gamma \left( \Bsmm \right) \right> dt }{\int^\infty_0 \left<\Gamma \left( \Bsmm \right) \right> dt }
 =  \frac{\tau_{\Bs}}{1 - y_s^2} \left[ \frac{1+ 2 \ADG y_s + y_s^2}{1 + \ADG y_s} \right],
\end{equation}
where $t$ is the decay time, $\tau_{\Bs}=2\hbar/(\Gamma_\text{L}+\Gamma_\text{H})$ and $y_s = \tau_{\Bs}(\Gamma_\text{L}-\Gamma_\text{H})/(2\hbar)$ with $\Gamma_\text{L}$ and $\Gamma_\text{H}$ denoting the decay widths of the light and heavy \Bs mass eigenstates, respectively. The effective lifetime provides complementary sensitivity to NP~\cite{DeBruyn:2012wk} via the quantity \ADG, which encodes the relative contributions of the \Bs light and heavy mass eigenstates to the decay. In the SM, the decay proceeds purely from the longer-lived heavy-mass eigenstate, such that $\ADG=1$ up to $\mathcal{O}(\alpha_{\text{em}}^2)$ corrections~\cite{Beneke:2019slt}. However, NP contributions may change its value in the range $[-1, +1]$, resulting in a smaller effective lifetime even in cases where \BRof \Bsmm agrees with the SM.

The search for the decay \Bdmumu began with the CLEO experiment in 1984, which set the first upper limit of $\BRof \Bdmm <  2.0 \times 10^{-4}$ at 90\% confidence~\cite{CLEO:1984nby}. The \Bsmumu decay was first observed through a joint analysis of LHCb and CMS Run~1 data~\cite{LHCb-PAPER-2014-049}, followed by the first single-experiment observation by LHCb in 2017~\cite{LHCb-PAPER-2017-001}. The most recent measurements of $\BRof \Bsmumu$ have been performed by the ATLAS~\cite{ATLAS:2018cur}, CMS~\cite{CMS:2022mgd} and LHCb~\cite{LHCb-PAPER-2021-007,LHCb-PAPER-2021-008} experiments using Run~1 and 2 data. The first evidence for the \Bdmumu decay reported by the joint LHCb and CMS analysis~\cite{LHCb-PAPER-2014-049} was not confirmed in subsequent analyses, and the current experimental limit is $\BRof \Bdmm <  1.6 \times 10^{-10}$ at 90\% confidence~\cite{LHCb-CONF-2020-002}. The effective lifetime of \Bsmumu decays has also been measured by LHCb~\cite{LHCb-PAPER-2021-007,LHCb-PAPER-2021-008}, CMS~\cite{CMS:2022mgd} and ATLAS~\cite{ATLAS:2023trk}.

This article presents measurements of the \Bsmm branching fraction and effective lifetime and a search for the \Bdmm decay, using proton-proton ($pp$) collision data recorded by the LHCb experiment in 2024 and 2025 during Run~3 of the LHC at a centre-of-mass energy $\sqs = 13.6 \tev$, corresponding to an integrated luminosity of 18~fb$^{-1}$. In addition, results are reported combining the Run~3 data with the previously analysed Run~1 and 2 data, which were recorded at $\sqs=7 \tev$ (2011), $8 \tev$ (2012) and $13 \tev$ (2015--2018), with a total integrated luminosity of 9~fb$^{-1}$. The results are the most precise to date by a single experiment and supersede those presented in Refs.~\cite{LHCb-PAPER-2021-007, LHCb-PAPER-2021-008}.

\section*{Detector and simulation}

The LHCb \upgradeone detector~\cite{LHCb-DP-2022-002} is a single-arm forward spectrometer covering the pseudorapidity range $2<\eta<5$, designed for the study of particles containing $b$ or $c$ quarks. The detector includes a high precision tracking system consisting of a silicon-pixel vertex detector (VELO) surrounding the interaction region, a large-area silicon-strip detector located upstream of a dipole magnet, and three stations of scintillating-fibre detectors located downstream of the magnet~\cite{LHCb-TDR-015}. The identification of particles~\cite{LHCb-TDR-014} is performed with two ring-imaging Cherenkov (RICH) detectors, an electromagnetic calorimeter, a hadronic calorimeter and a muon system. The \upgradeone detector represents a major enhancement of the LHCb experiment, allowing operation at an instantaneous luminosity of $2\times10^{33}\,\cm^{-2}\sec^{-1}$, five times that of the previous detector. To enable this, the tracking and RICH subsystems have been replaced with higher granularity detectors, while the existing muon system has been upgraded with finer readout granularity and improved shielding. The \upgradeone detector maintains or improves the excellent track reconstruction and particle identification (PID) performance of the original \lhcb detector in the higher pile-up environment of Run~3~\cite{LHCb-DP-2026-006}. Furthermore, the initial hardware trigger stage has been replaced with an all-software trigger~\cite{LHCb-TDR-016}, leading to substantial improvements in efficiencies for hadronic and electron final states. While the \Bmumu yield per unit luminosity is similar to that obtained in Runs~1 and~2, the analysis presented in this article benefits from a substantial increase in integrated luminosity in Run~3 and from improvements in the muon identification~\cite{LHCb-DP-2020-002,LHCb-DP-2025-007}, particularly in rejecting misidentified hadronic backgrounds.

Simulated samples are used to model the detector response. In the simulation, \Bds mesons are generated using \pythia~\cite{Sjostrand:2007gs} with a specific LHCb configuration; decays of unstable particles are described by \evtgen~\cite{Lange:2001uf} and final-state radiation is generated using \photos~\cite{davidson2015photos}.
The interaction of the generated particles with the detector, and its response, are implemented using the \geant toolkit~\cite{Agostinelli:2002hh,Allison:2006ve} as described in Ref.~\cite{LHCb-PROC-2011-006}. 

\section*{Analysis method}
The analysis follows a similar strategy to that adopted with the previous version of the LHCb experiment for Run~1 and 2 data~\cite{LHCb-PAPER-2021-007, LHCb-PAPER-2021-008}, but introduces developments enabled by the \upgradeone detector and Run~3 data-taking conditions. Following the trigger and preselection, candidate \Bmumu decays are identified using a boosted decision tree (BDT) classifier~\cite{Breiman,XGBoosting} trained to reduce combinatorial background. The signal yields of \Bsmumu and \Bdmumu decays are determined with an extended unbinned maximum-likelihood fit to the dimuon invariant mass, performed simultaneously in five subsamples split according to the response of the BDT classifier. The signal yields are converted into branching fractions using \BdToKpi and \BuToJpsiK with \Jpsimumu normalisation modes, which are chosen due to their large yields and final states similar to the signals. The efficiency ratios between the signal and normalisation modes are determined through a combination of simulation and data-driven techniques. A fit to the dimuon invariant-mass distribution is used to perform background subtraction via the \sPlot~\cite{Pivk:2004ty} technique, followed by a fit to the decay-time distribution. The decays \BdTopipi, \BsToKK, \BuToJpsiK and \BsToJpsiphi, with \Jpsimumu and \PhiKK, are used to validate the branching fraction and lifetime measurements and to correct simulation. To avoid experimenter's bias, the dimuon invariant-mass range \mbox{[5200, 5450]\mevcc}, where the \Bsmumu and \Bdmumu decays peak, was not studied until all methods had been finalised. 

In Run~3, the LHCb trigger system is divided into two stages, both implemented purely in software. In the first stage, data from the detector are read out at the LHC bunch-crossing frequency and events are passed to the second stage if they satisfy requirements on at least one of two multivariate classifiers. These are trained to identify single tracks with high transverse momenta (\pt) and impact parameter significances (\chisqip), or pairs of tracks forming a displaced vertex. In the second trigger stage, two-body $B$-meson decays are reconstructed by a single algorithm that combines two oppositely charged tracks with high $p$, \pt and \chisqip. The resulting $B$ meson candidates must have a good vertex-fit quality, a significant displacement from any primary vertex (PV) and have a momentum that is well aligned with the vector joining the decay vertex and its associated PV. Since this algorithm imposes no PID requirements on the tracks, it reconstructs both \Bmumu and \Bhh decays (where $h^{(')}=K, \pi$), reducing systematic uncertainties associated with their selection efficiencies. Similar, but distinct, trigger algorithms that impose additional requirements on the PID responses of the muon and kaon candidate tracks are used to reconstruct \BuToJpsiK and \BsToJpsiphi candidates.

The candidates reconstructed in the second stage of the trigger are stored, and passed through an offline selection in order to increase signal purity and to isolate different final states. Final-state tracks are required to fall within the geometrical acceptance of the RICH and muon detectors, and additional loose requirements are imposed on the kinematic properties of the tracks and $B$ candidates. A requirement on the output of a multivariate classifier is imposed to suppress combinatorial background while maintaining a high signal efficiency. This classifier is trained using simulated \Bsmm decays as signal proxy and a large simulated sample of inclusive $b\bar{b}\to X \mu^+ \mu^-$ processes as background. It combines information on the track impact parameter (IP) with respect to the associated PV, the distance of closest approach of the tracks, and the $B$ candidate IP, \chisqip, vertex fit \chisq and the angle between its momentum vector and the line joining the PV to its decay vertex. Finally, strict requirements on the PID information, provided predominantly by the RICH and muon systems, are used to separate two-body decays into $\mup\mun$, $K^+\pi^-$, $K^+K^-$ and $\pi^+\pi^-$ final states.

After these selection requirements, the \Bmumu signal sample remains dominated by combinatorial background, predominantly made of muons from semileptonic decays of two $b$ hadrons in the same event. This background is controlled using a BDT classifier, trained using simulated \Bsmumu decays as signal and inclusive $b\bar{b}\to X \mu^+ \mu^-$ events as background that have passed the trigger and offline selection. The most discriminating features used in the BDT are the responses of two dedicated isolation classifiers trained using tracks from the underlying event that do not form the signal candidate, exploiting the tendency of \Bmumu decays to be more isolated, \ie\ to have fewer additional nearby tracks, than background candidates. Additional training features for this BDT are: $\sqrt{\Delta\phi^2+\Delta\eta^2}$ where $\Delta\phi$ and $\Delta\eta$ are the differences between the azimuthal angle and pseudorapidity of the two tracks forming the $B$ candidate, the minimum \chisqip of the signal candidate tracks, the angle between the $B$ candidate momentum vector and the line between the PV and its decay vertex, the vertex fit \chisq of the $B$ candidate and the \chisqip of the $B$ candidate. The output of the BDT classifier is rescaled after training to give an approximately uniform distribution for signal \Bmumu decays in a range between zero and one. To enhance the signal sensitivity, the \Bsmumu and \Bdmumu yields are extracted through a simultaneous fit to the dimuon invariant mass performed in five bins of the BDT classifier response, with the bin edges 0.25, 0.40, 0.50, 0.60, 0.70 and 1.00. The background-dominated region below 0.25 is not included in the fit. The effective lifetime is measured using candidates passing the requirement $\text{BDT}>0.5$, which is optimised to give the maximum expected signal significance for \Bsmumu decays.

\section*{Branching fraction measurement}
\begin{figure}[t]
\centering
\includegraphics[width=0.49\textwidth]{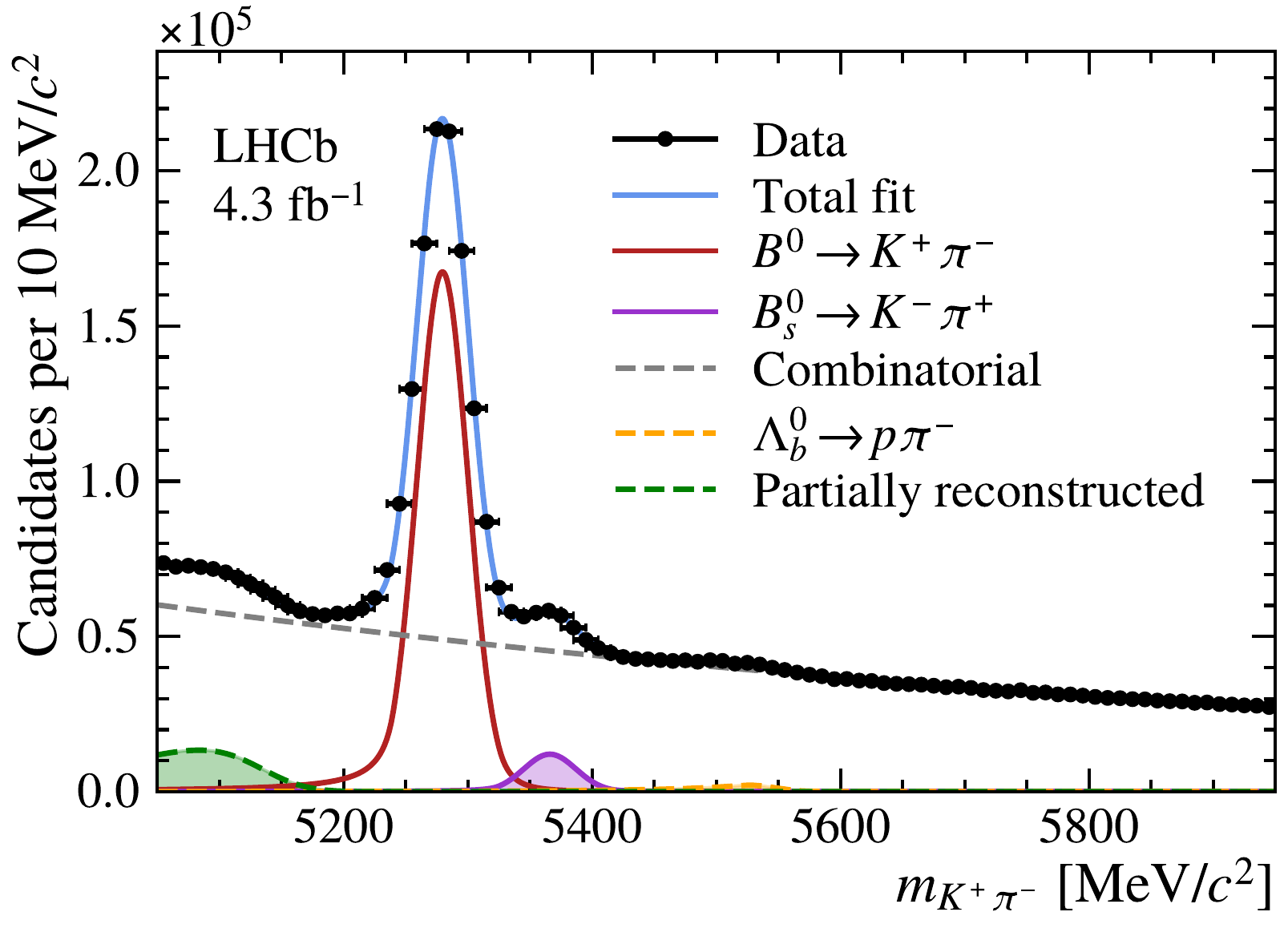} 
\includegraphics[width=0.49\textwidth]{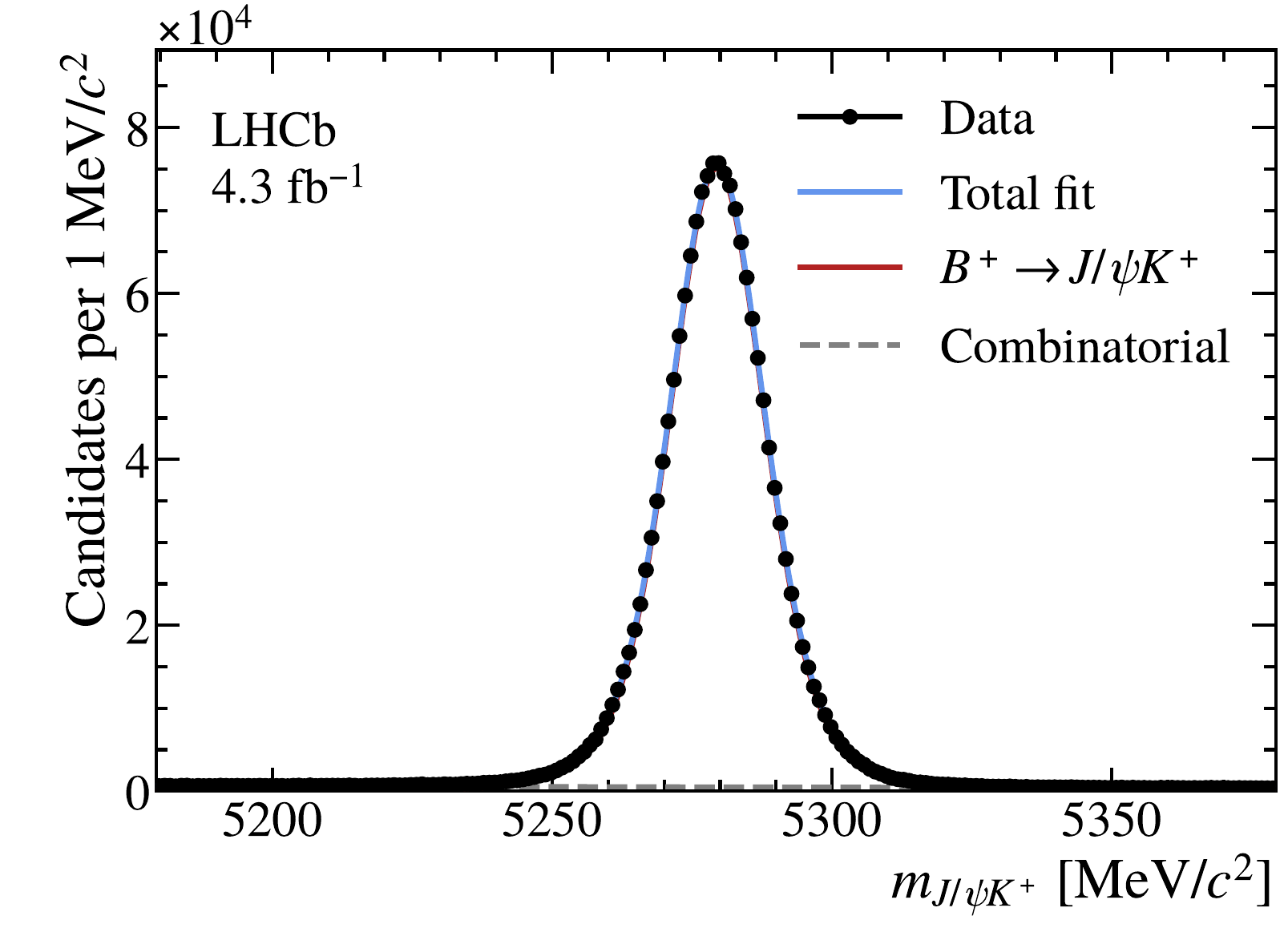}
\caption{Mass distribution of (left) $B^0 \to K^+\pi^-$ and (right) $B^+ \to \jpsi K^+$ events with the fit curve overlaid. The data sample corresponds to an integrated luminosity of $4.3~\rm{fb}^{-1}$ collected in 2025.}
\label{fig:norm}
\end{figure}

The signal yields $N_{\Bmumu}$ are converted into branching fractions via
\begin{equation}
    \mathcal{B}(\Bmumu) = N_{\Bmumu} \times
\frac{\mathcal{B}_{\text{norm}}}{N_{\text{norm}}}
\frac{\varepsilon_{\text{norm}}}{\varepsilon_{\text{sig}}}  
\times \frac{f_{\text{norm}}}{f_{\text{sig}}}
\equiv  N_{\Bmumu} \times \alpha^{\text{norm}}_{\Bmumu},
\end{equation}
where $\mathcal{B}_{\text{norm}}$ and $N_{\text{norm}}$ are the branching fraction~\cite{PDG2026} and yield of the normalisation mode (\BdToKpi or \BuToJpsiK), $\varepsilon_{\text{norm}}$ ($\varepsilon_{\text{sig}}$) is the efficiency for the normalisation (signal) mode, and $f_{\text{norm}}$ ($f_{\text{sig}}$) is the probability for a $b$ quark to fragment into the corresponding $B$ meson. Assuming $f_d=f_u$, the relevant ratio $f_s/f_d$ has been measured by LHCb in $pp$ collisions at $\sqrt{s}=13.6\tev$ to be $0.247\pm0.008$~\cite{LHCb-PAPER-2026-056}. The yields of \BdToKpi and \BuToJpsiK decays are evaluated from fits to data in the various data-taking periods into which Run~3 is divided, one of which is shown in Fig.~\ref{fig:norm}.

The values of $\alpha^{\text{norm}}_{\Bmumu}$ from the two normalisation channels are determined for 14 data-taking periods, and averaged according to the normalisation mode yields and ratios of signal and normalisation efficiencies. The values obtained of $\alpha^{\text{norm}}_{\Bmumu}$ for each normalisation mode are then combined, taking their correlations into account, to obtain the single-event sensitivities \mbox{$\alpha_{B^0_s\to \mu^+\mu^-}=(1.13\pm 0.04)\times 10^{-11}$} and \mbox{$\alpha_{B^0\to \mu^+\mu^-}=(2.76\pm 0.05)\times 10^{-12}$}. The expected signal yields, assuming SM branching fractions, in the BDT region used for the fit ($\rm{BDT}>0.25$) are $235\pm13$ \Bsmumu and $26\pm1$ \Bdmumu decays.

The efficiencies are evaluated using a combination of simulated signal and normalisation decays and calibration samples taken from data. Corrections are applied to the simulation to improve agreement with data. In particular, the kinematics of the \Bd (\Bs) candidate and the event multiplicity are corrected using \BuToJpsiK (\BsToJpsiphi) data, the track reconstruction and PID efficiencies are calibrated using data samples of hadrons and muons, and the BDT distribution of signal candidates is corrected with \BdToKpi data.

The branching fractions of \Bdmumu and \Bsmumu decays are extracted from an extended unbinned maximum-likelihood fit to the dimuon invariant mass ($m_{\mu^+\mu^-}$) in the range [4900, 5900]\mevcc. Three types of backgrounds are considered in the fit: \Bhh decays, semileptonic decays, and combinatorial background. The \Bhh background occurs when both hadrons are misidentified as muons and consists of $B^0 \to \pi^+\pi^-$, $B^0 \to K^+\pi^-$, $B^0_s \to K^-\pi^+$ and $B^0_s \to K^+K^-$ decays. The pion-to-muon and kaon-to-muon misidentification probabilities are estimated from a large data calibration sample of $\Lc \to pK^-\pi^+$ decays. These misidentification probabilities are verified using a \Bhh data sample in which one of the hadrons in the final state is misidentified as a muon. The expected yield is estimated using \BdToKpi as normalisation mode. Semileptonic decays of $b$ hadrons include $B^0 \to \pi^-\mu^+\nu_{\mu}$, $B^0_s \to K^-\mu^+\nu_{\mu}$, and $\Lb \to p\mu^-\neub_{\mu}$ decays where the hadron in the final state is misidentified as a muon, as well as $B^{0(+)} \to \pi^{0(+)}\mu^+\mu^-$ and $B^+_c \to \jpsi \mu^+ \nu_{\mu}$ decays with $\jpsi \to \mu^+ \mu^-$. The yield of each semileptonic mode is estimated with simulation, using \BuToJpsiK decays as a normalisation mode. The expected yields of each background source in the mass and BDT regions used for the fit are estimated to be $37\pm2$ \Bhh,
$250\pm34$ $B^0 \to \pi^-\mu^+\nu_{\mu}$,
$18\pm6$ $B^0_s \to K^-\mu^+\nu_{\mu}$,
$32\pm13$ $\Lb \to p\mu^-\neub_{\mu}$,
$123\pm12$ $B^{0(+)} \to \pi^{0(+)}\mu^+\mu^-$,
and $90\pm8$
$B^+_c \to \jpsi \mu^+ \nu_{\mu}$ decays.
The expected yields are Gaussian-constrained in the fit according to the values and uncertainties of their estimated efficiencies and branching fractions~\cite{PDG2026}. The contribution of $B^0_s \to \mu^+\mu^-\gamma$ decays is found to be negligible. The yield of the combinatorial background is free to vary in the fit in each BDT region.

The mass distributions of the \Bdmumu and \Bsmumu signals are described with double-sided Crystal Ball functions~\cite{Skwarnicki:1986xj} whose parameters are Gaussian-constrained in the fit. To account for possible biases in the track reconstruction, the mean values are obtained from data fits to $B^0 \to K^+\pi^-$ and $B^0_s \to K^+K^-$ decays. A resolution of $23\mevcc$ is obtained by interpolating the resolutions measured from mass fits to charmonium and bottomonium resonances decaying into two muons. The tail parameters, and the variation of the shape parameters across the BDT regions are obtained from corrected simulation. The mass distributions of the \Bhh, $B^0 \to \pi^-\mu^+\nu_{\mu}$ and $B^0_s \to K^-\mu^+\nu_{\mu}$ decays are obtained from simulation accounting for the change in momentum due to the decays of pions and kaons to muons. The resulting distribution is described using Gaussian kernel density estimates~\cite{CRANMER2001198}. The effect on the final results of using alternative models to parametrise the \Bhh mass distribution and the statistical uncertainty on its shape are found to be negligible. The remaining semileptonic backgrounds are also modelled with Gaussian kernel density estimates of simulated data in which the full event selection is applied. The combinatorial background is modelled with a single exponential function with its slope free to vary in the fit. The dimuon mass distribution with the projection of the fit overlaid is shown in Fig.~\ref{fig:bf_results} for $\text{BDT}>0.6$ and in Appendix~\ref{sec:fits} for the five individual BDT regions.

\begin{figure}
    \centering
    \includegraphics[width=0.95\linewidth]{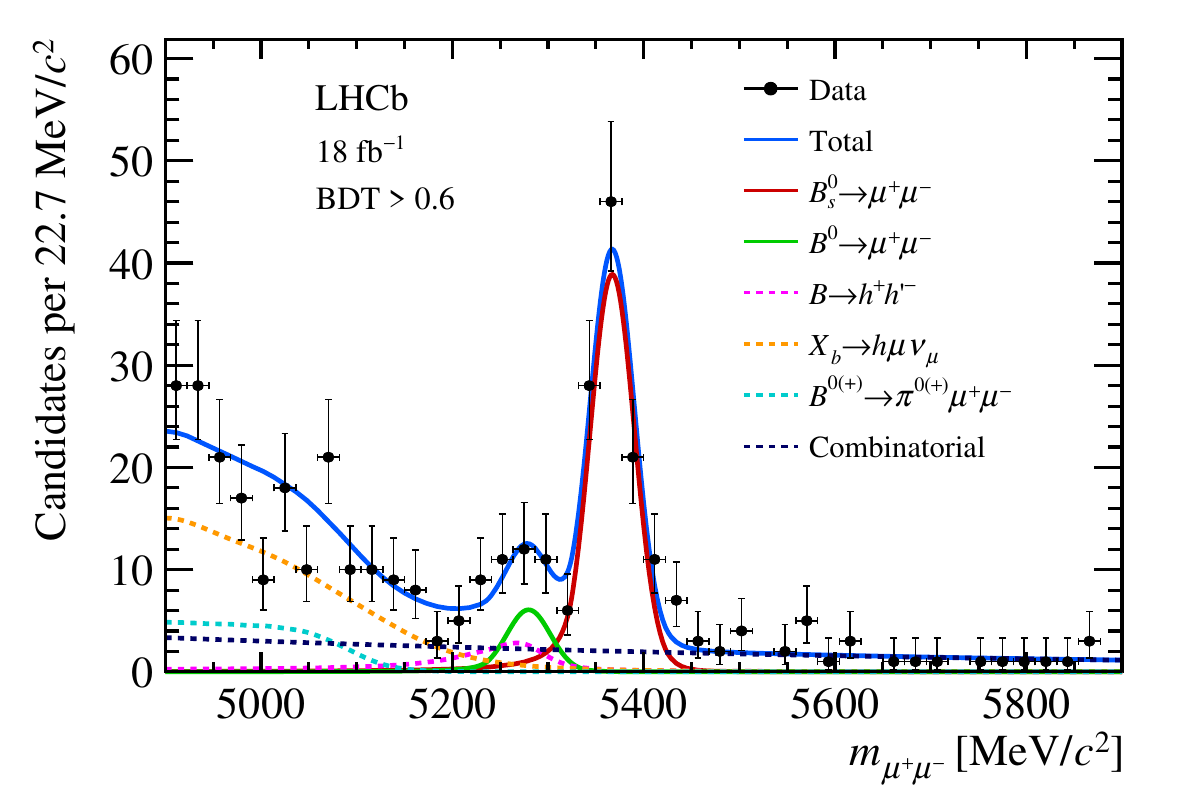}
    \caption{Dimuon invariant-mass distribution in the $\text{BDT}>0.6$ region. The data points are shown in black, the projection of the fit is overlaid in blue, and its components are indicated in the legend. Semileptonic backgrounds with a neutrino in the final state are grouped in the orange component.}
    \label{fig:bf_results}
\end{figure}

\section*{Effective lifetime measurement}
The effective lifetime is measured in a single bin of the BDT response. First, an unbinned maximum likelihood fit to the dimuon invariant mass in the range \mbox{[5200, 5900]\mevcc} is performed in order to extract the background-subtracted \Bsmm decay-time distribution using the \sPlot method~\cite{Pivk:2004ty}. Combinatorial background, \Bdmm decays and misidentified \Bhh decays are modelled in the fit, with the \Bhh yield Gaussian-constrained to the expected value, determined using the method described above, while all other yields are allowed to float freely. 

\begin{figure}[ht]
    \centering
    \includegraphics[width=0.75\linewidth]{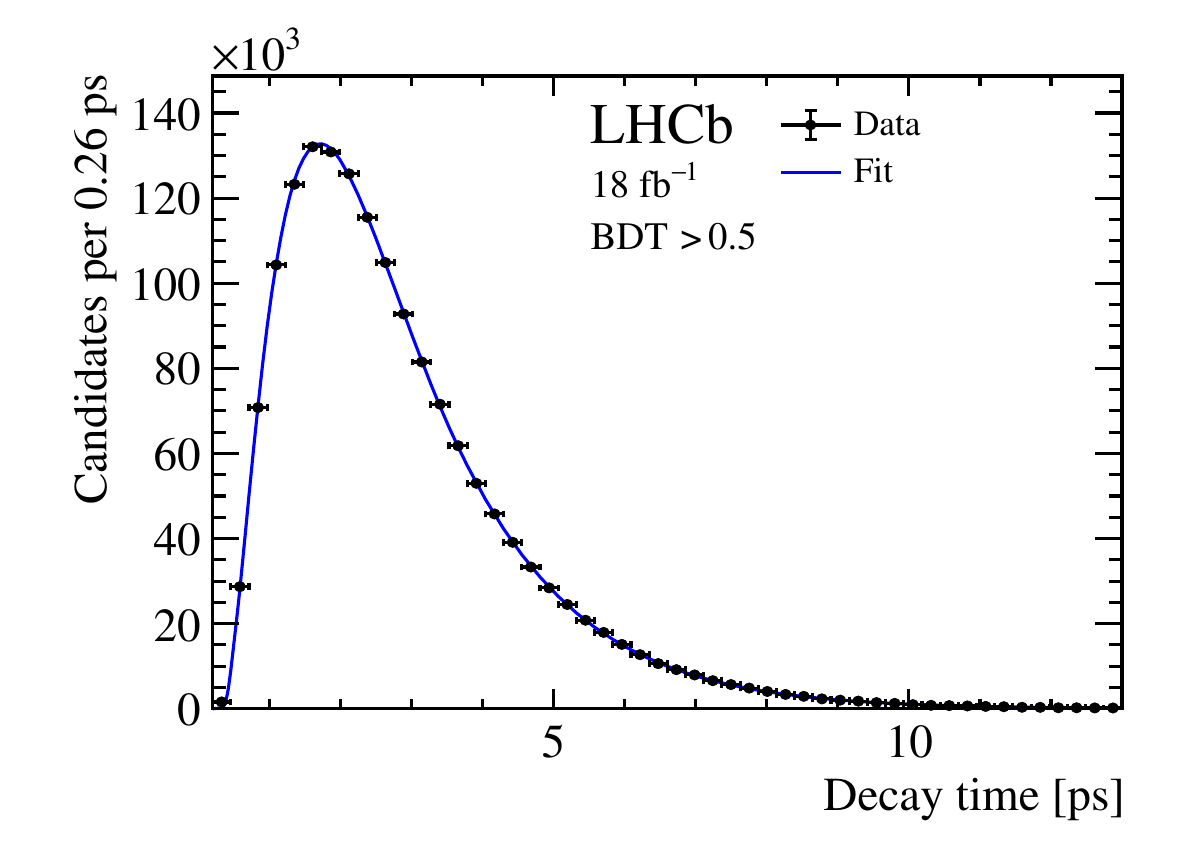}
    \caption{The $B^0 \to K^+\pi^-$ decay-time distribution extracted using the {\it{sPlot}} formalism. The data points are shown in black while the maximum-likelihood fit used to extract the parameters of the decay-time acceptance function is shown in blue.}
    \label{fig:BdToKpi_acceptance}
\end{figure}

A fit to the background-subtracted decay-time distribution is then performed in order to determine the effective lifetime. The efficiency to reconstruct and select \Bsmm decays as a function of decay time is parametrised as the exponential of a cubic spline, which is flexible enough to capture the features of the efficiency curve while being guaranteed to be positive. The parameters of this efficiency function are determined using a fit to the background-subtracted \BdToKpi decay-time distribution in data, shown in Fig.~\ref{fig:BdToKpi_acceptance}, where the \Bd lifetime is fixed to the known value~\cite{PDG2026} and the parameters of the acceptance function are allowed to float. To account for possible differences between \Bsmumu and \BdToKpi decays, this efficiency function is multiplied by a second function of the same type, fitted to the ratio of \Bsmumu over \BdToKpi decay-time efficiencies in simulation. The final fit to the background-subtracted \Bsmm decay-time data is performed using the resulting acceptance function multiplied by a single exponential $e^{-t/\tau_{\mu\mu}}$.

This method is validated using \BsToKK and \BdTopipi decays, where the effective lifetimes measured on the Run~3 data sample are found to be within 0.012\ps of the known lifetimes~\cite{PDG2026}. This difference is taken as a systematic uncertainty on the \Bsmm lifetime measurement. The fits to the validation channels are shown in Appendix~\ref{sec:lifetime}.

Other sources of systematic uncertainty included are the effects of residual background contamination from unmodelled semileptonic decays (0.006\ps), of a nonzero production asymmetry on the decay-time distribution (0.002\ps), of the decay-time acceptance on the relative admixture of light and heavy eigenstates as a function of \ADG (0.003\ps), and the residual bias of the lifetime fitting procedure evaluated with pseudoexperiments (0.008\ps). The effects of neglecting decay-time resolution in the fit model and of correlations between dimuon mass and decay time are found to be negligible.

\section*{Results}

The signal branching fractions are determined to be
\begin{align*}
   \BRof \Bsmumu &=
   \left(3.31^{\,+\,0.31\,+\,0.15}_{\,-\,0.30\,-\,0.13}\right)
   \times 10^{-9}, \\
   \BRof \Bdmumu &=
   \left(1.26^{\,+\,0.51\,+\,0.06}_{\,-\,0.47\,-\,0.05}\right)
   \times 10^{-10},
\end{align*}
where the first uncertainty is statistical and the second systematic. The statistical uncertainty is computed by repeating the fit after fixing all nuisance parameters to the post-fit values obtained in the default fit. The dominant systematic uncertainty on $\BRof \Bsmumu$ arises from $f_s/f_d$ and amounts to 3\% of the branching fraction, while that on $\BRof \Bdmumu$ arises from the knowledge of the \Bhh and semileptonic backgrounds and amounts to 4\%. The \Bsmumu signal is observed with high significance. An excess of \Bdmumu events is observed with respect to the background expectation with a significance of $3.0$ standard deviations, taking systematic uncertainties into account. The significance is computed from the difference in the likelihood value between fits with and without the signal component using Wilks' theorem~\cite{Wilks:1938dza}, and is validated with pseudoexperiments. The evidence for the \Bdmumu decay enables the ratio of branching fractions to be measured as
\begin{align*}
   \Rmumu &=
   0.038^{\,+\,0.016}_{\,-\,0.014}\pm0.002\ ,
\end{align*}
where the first uncertainty is statistical and the second systematic. 

The \Bsmumu selection efficiency depends on the lifetime~\cite{DeBruyn:2012wj}. In the fit, the value $\ADG=+1$ is assumed. The dependence on this assumption is evaluated by repeating the fit using $\ADG=0$ and $\ADG=-1$. An increase of $4.0\%$ and $9.2\%$ in $\BRof \Bsmumu$ is found, respectively.

\begin{figure}
    \centering
    \includegraphics[width=0.9\linewidth]{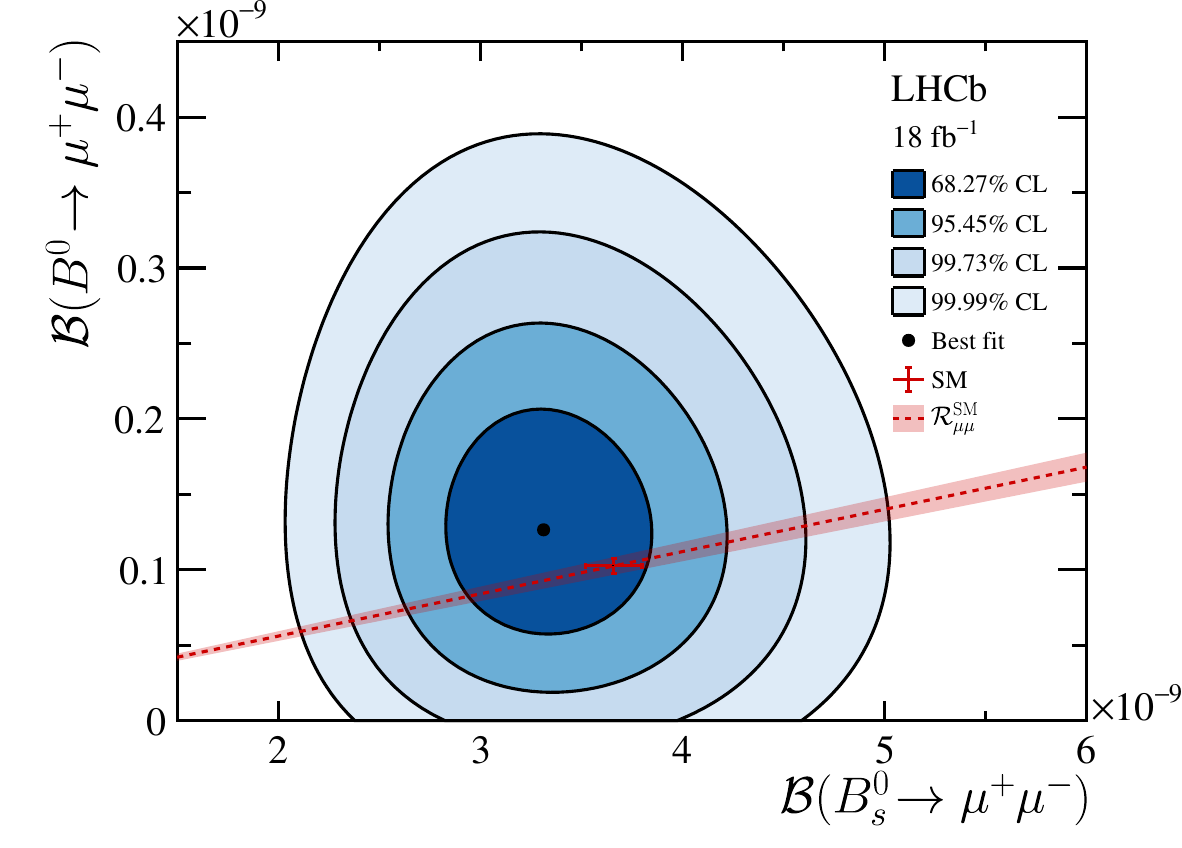}
    \caption{Two-dimensional profile likelihood of $\mathcal{B}(B_s^0 \to \mu^+\mu^-) \ \text{vs}\ \mathcal{B}(B^0 \to \mu^+\mu^-)$. The best fit value is indicated by the black dot, while the SM predictions~\cite{Beneke:2019slt} are indicated in red. The contours for $68.27\%$, $95.45\%$, $99.73\%$ and $99.99\%$ confidence level are overlaid.}
    \label{fig:2d_run3}
\end{figure}

The two-dimensional profile likelihood of $\BRof \Bsmumu$ and $\BRof \Bdmumu$ is shown in Fig.~\ref{fig:2d_run3}. It is obtained by taking the ratio between the likelihood obtained from a fit in which the two parameters of interest are fixed and the likelihood from the nominal fit. A linear correlation of $-3\%$ is observed between the \Bdmm and \Bsmumu branching fractions.

\begin{figure}[ht]
\centering
\includegraphics[width=0.49\textwidth]{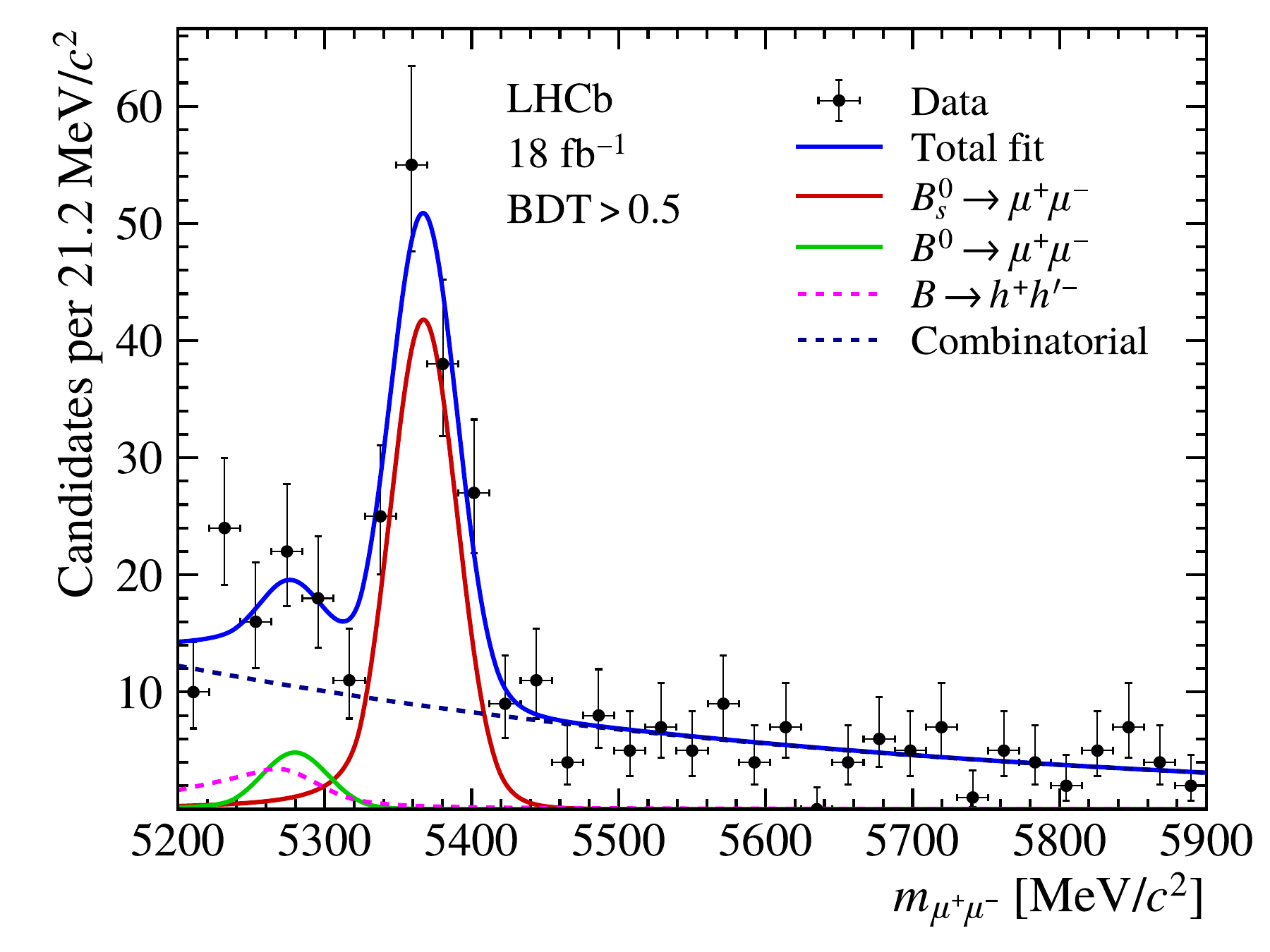} 
\includegraphics[width=0.49\textwidth]{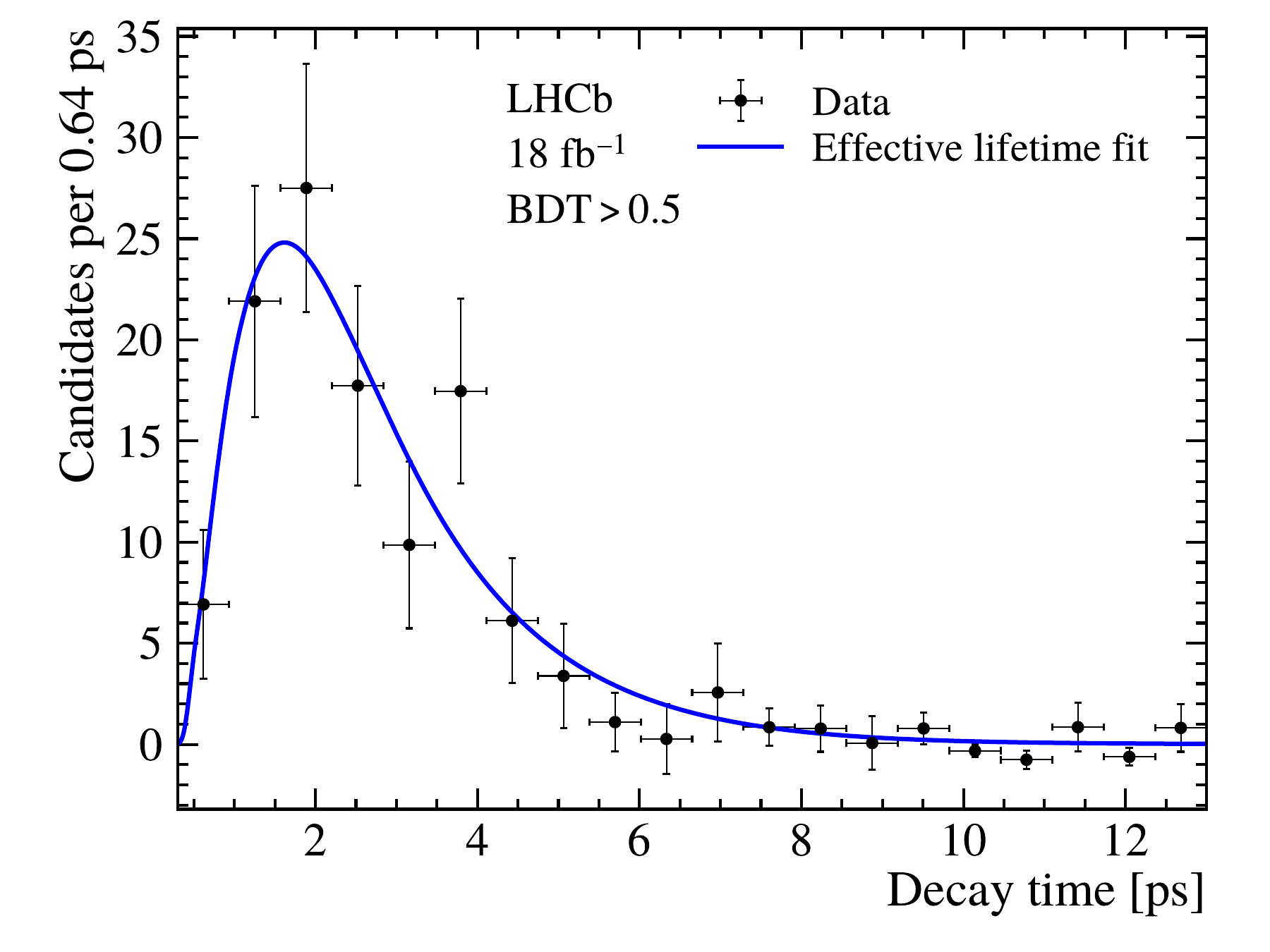}
\caption{(Left) Dimuon invariant-mass distribution and total fit curve used to determine sWeights for the effective $B^0_s \to \mu^+\mu^-$ lifetime measurement. (Right) Background-subtracted $B^0_s \to \mu^+\mu^-$ decay-time distribution with the fit curve overlaid.}
\label{fig:lifetime}
\end{figure}

The mass fit used in the lifetime measurement and the resulting background-subtracted decay-time distribution are shown in Fig.~\ref{fig:lifetime}. The best-fit value for the effective lifetime is 
\begin{align*}
    \tau_{\mu\mu} = 1.42 \pm 0.18 \pm 0.02\ps,
\end{align*}
where the first uncertainty is statistical and the second systematic. Using the known values of $\Gamma_s$ and $\Delta\Gamma_s$~\cite{HFLAV:GsDGs}, this result can be recast as
\begin{equation*}
    \mathcal{A}_{\Delta\Gamma_s}^{\mu\mu}=-1.14\pm1.78\pm0.15\pm0.06,
\end{equation*}
where the first uncertainty is statistical, the second systematic and the third due to the uncertainties on the external inputs where correlations are taken into account. This estimate is not constrained to the physical range $[-1,+1]$. The \Bsmumu signal yields obtained in the effective-lifetime and branching-fraction mass fits are compatible within 1.7 standard deviations, as evaluated with pseudoexperiments accounting for the correlations between the input parameters and for the different mass and BDT ranges used in each fit.

The branching fraction results are combined with those presented in Refs.~\cite{LHCb-PAPER-2021-007,LHCb-PAPER-2021-008} using a simultaneous fit to the Run~1, 2 and 3 data, whose mass distributions are shown in Appendix~\ref{sec:fits}. The \Bmumu branching fractions and nuisance parameters common to the three samples are shared in the fit, while parameters specific to Run~1 and 2 data are the same as those used in the previous measurements~\cite{LHCb-PAPER-2021-008}. The results of the combined fit are
\begin{align*}
   \BRof \Bsmumu &=
   \left(3.25^{\,+\,0.26\,+\,0.13}_{\,-\,0.25\,-\,0.12}\right)
   \times 10^{-9}, \\
   \BRof \Bdmumu &=
   \left(1.14^{\,+\,0.42\,+\,0.06}_{\,-\,0.39\,-\,0.05}\right)
   \times 10^{-10}, \\
   \Rmumu &=
   0.035^{\,+\,0.014\,+\,0.002}_{\,-\,0.012\,-\,0.001}\ , 
\end{align*}
where the first uncertainty is statistical and the second systematic. The significance of the \Bdmumu signal in the combined fit, computed using Wilks' theorem, is $3.2$ standard deviations, and is validated using pseudoexperiments.

The measurement of the effective lifetime is statistically dominated and the systematic uncertainty is uncorrelated with that of the result presented in Refs.~\cite{LHCb-PAPER-2021-007,LHCb-PAPER-2021-008}. Therefore, the Run~1--2 and Run~3 results are combined using an inverse-variance weighted average, yielding
\begin{align*}
    \tau_{\mu\mu}=1.60 \pm 0.16 \pm 0.01\ps,
\end{align*}
where the first uncertainty is statistical and the second systematic. This corresponds to
\begin{equation*}
    \mathcal{A}_{\Delta\Gamma_s}^{\mu\mu}=0.91 \pm 1.92 \pm 0.18  \pm0.10.
\end{equation*}

\section*{Conclusions}

Enabled by the unique capabilities of the LHCb \upgradeone detector, evidence for the \Bdmm decay is reported with a significance of over three standard deviations, along with the world's most precise determinations of the \Bsmm and \Bdmm branching fractions and the \Bsmm effective lifetime. This work marks the first analysis of \Bds meson decays into two muons performed with the new detector, demonstrating its capabilities to perform precision flavour physics at high luminosity. The results are combined with those from the previous generation LHCb detector. All results are consistent with SM expectations and constrain new physics contributions at energy scales beyond those accessible through direct production at the LHC.

\clearpage

\appendix
\section{Validation of the effective lifetime measurement}
\label{sec:lifetime}

\begin{figure}[ht]
    \centering
    \includegraphics[width=0.49\linewidth]{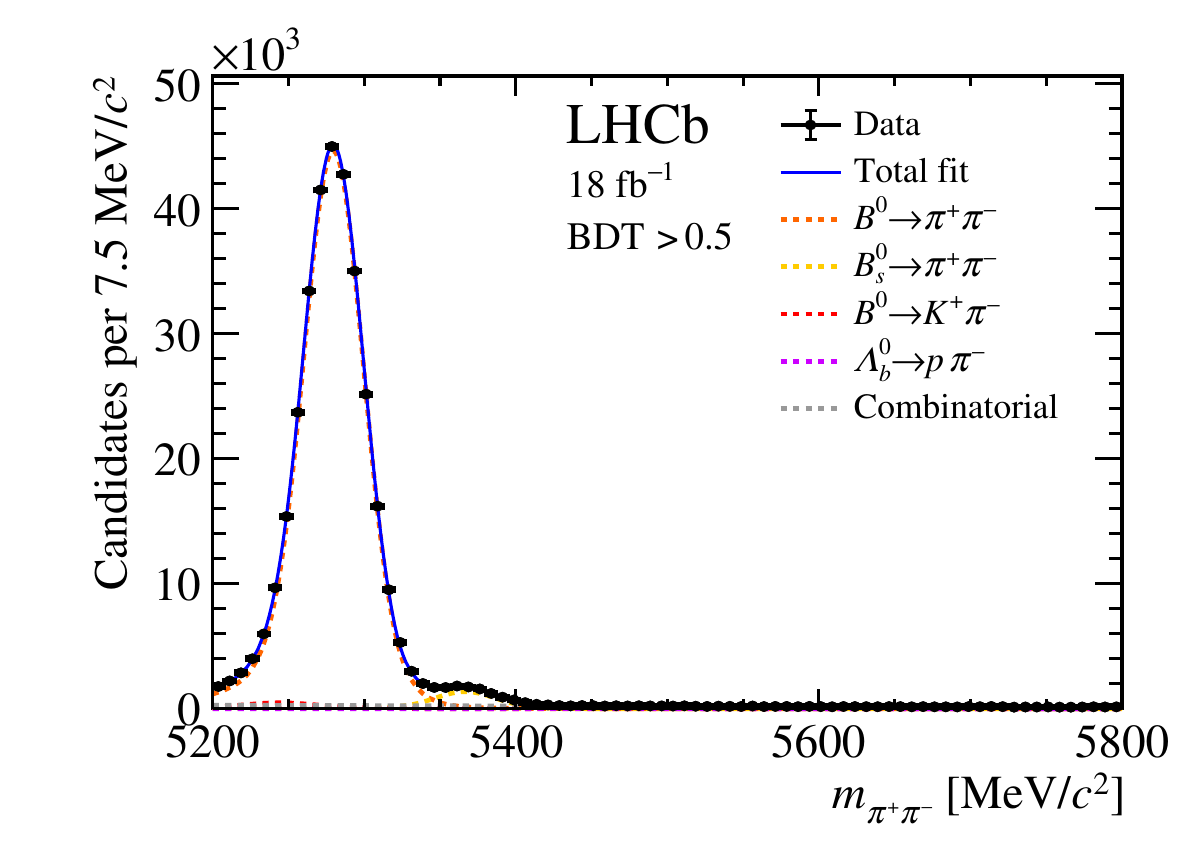}
    \includegraphics[width=0.49\linewidth]{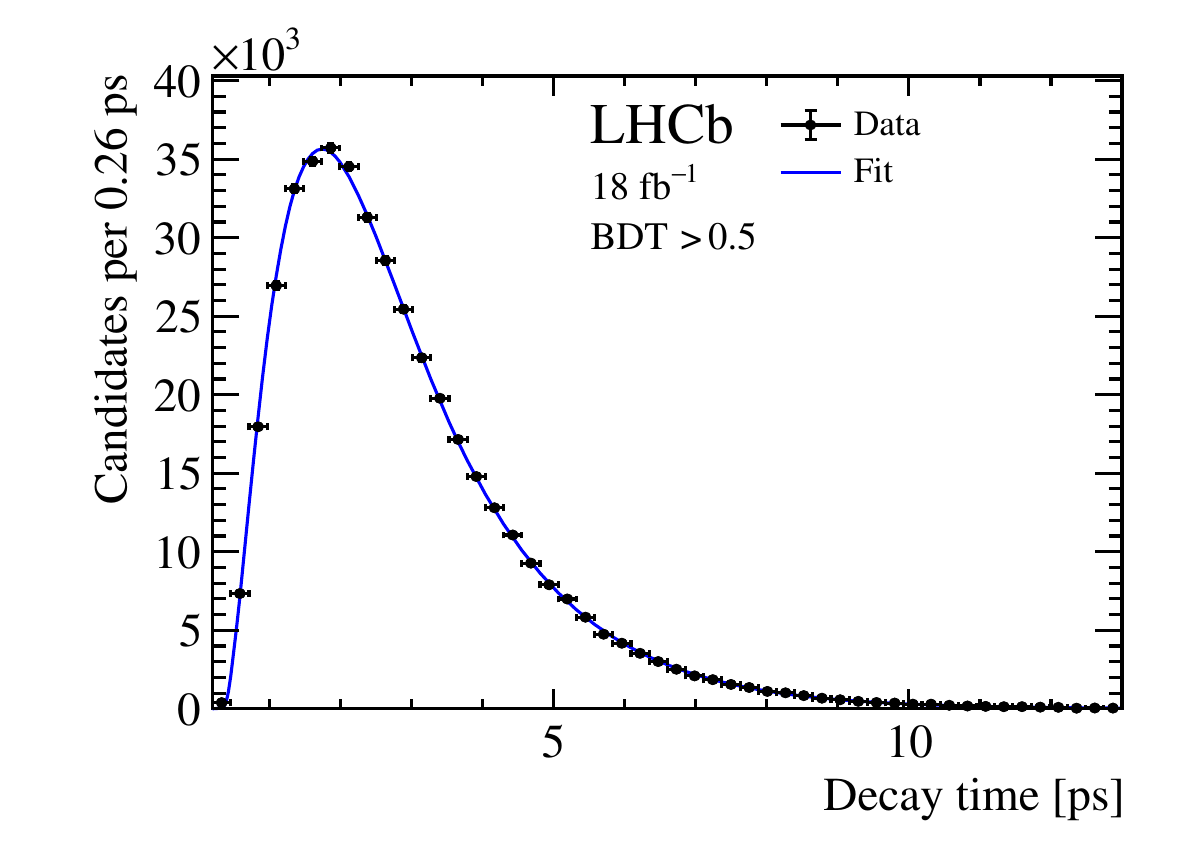}
    \caption{(Left) Invariant-mass distribution with fit used to extract weights for $B^0 \to \pi^+ \pi^-$ decays. (Right) Background-subtracted $B^0 \to \pi^+ \pi^-$ decay-time distribution with fit used to determine the mean $B^0$ lifetime.}
    \label{fig:LT_BdTopipi}
\end{figure}

\begin{figure}[ht]
    \centering
    \includegraphics[width=0.49\linewidth]{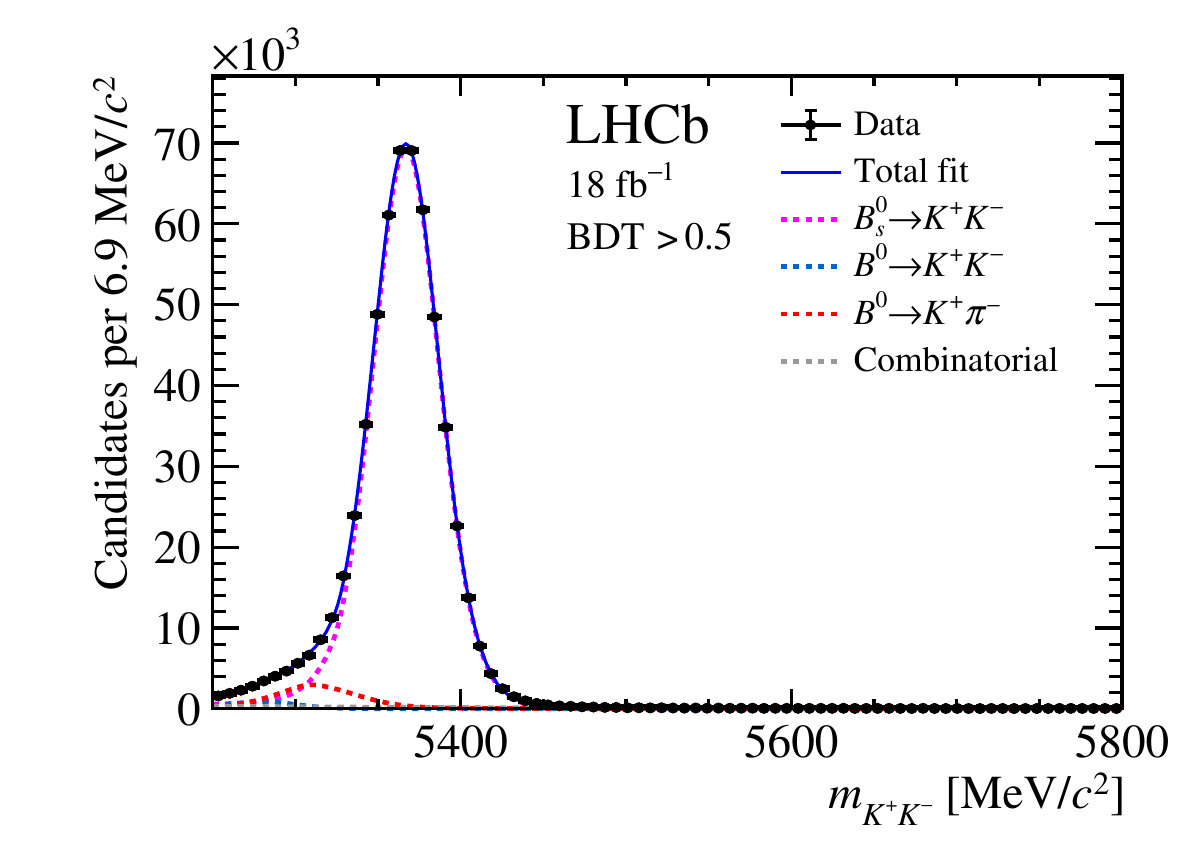}
    \includegraphics[width=0.49\linewidth]{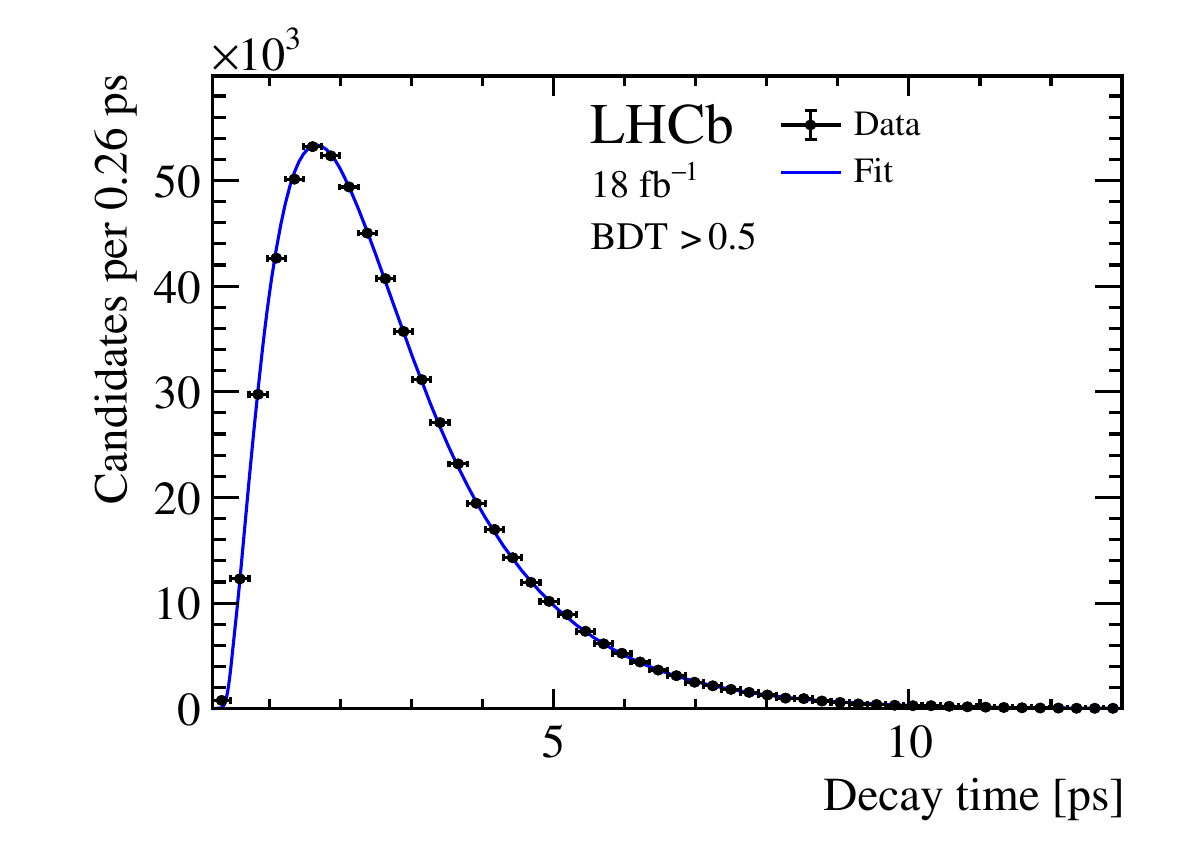}
    \caption{(Left) Invariant-mass distribution with fit used to extract weights for $B_s^0 \to K^+ K^-$ decays. (Right) Background-subtracted $B_s^0 \to K^+ K^-$ decay-time distribution with fit used to determine the effective $B_s^0 \to K^+ K^-$ lifetime.}
    \label{fig:LT_BsToKK}
\end{figure}

Figures~\ref{fig:LT_BdTopipi} and \ref{fig:LT_BsToKK} show the invariant-mass fits used to obtain the background-subtraction weights for $B^0\to\pi^+\pi^-$ and $B_s^0\to K^+K^-$ decays, respectively, and the subsequent decay-time fits used to validate the acceptance determination.

\section{Invariant-mass fits for the branching fractions}
\label{sec:fits}

\begin{figure}[ht]
    \centering
    \includegraphics[width=0.49\linewidth]{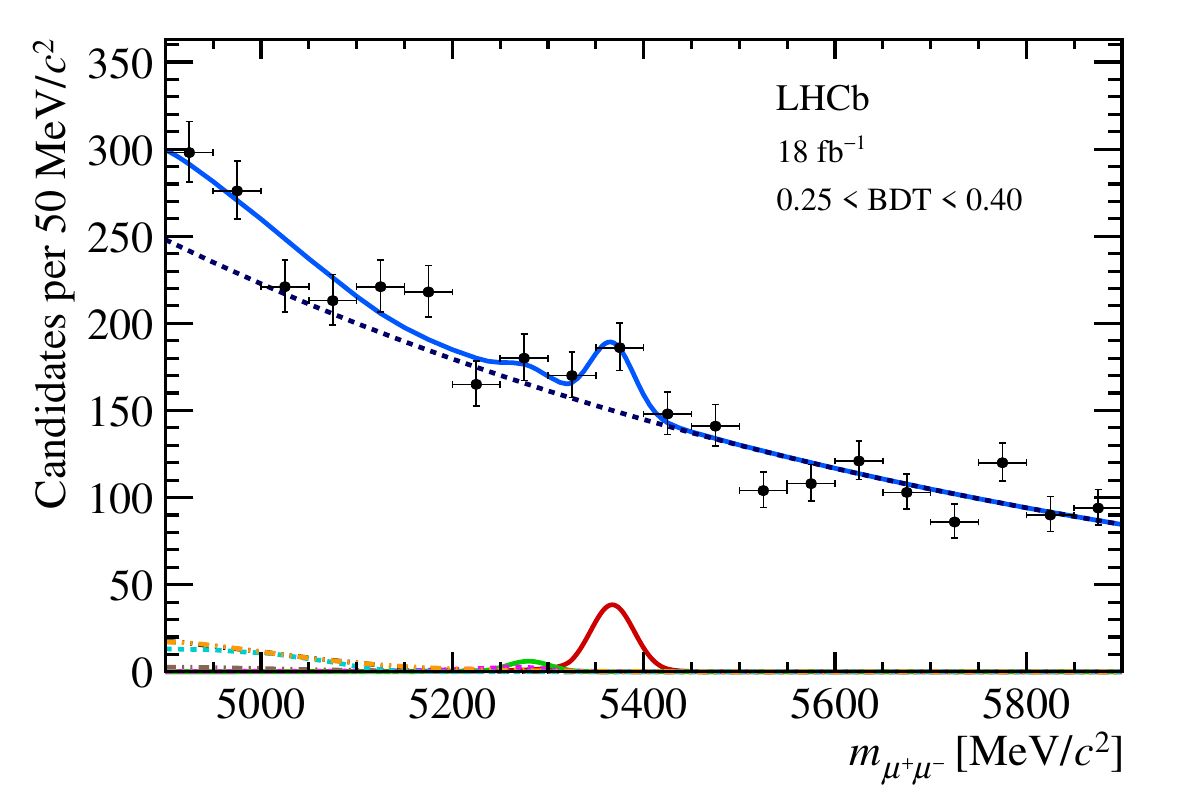}
    \includegraphics[width=0.49\linewidth]{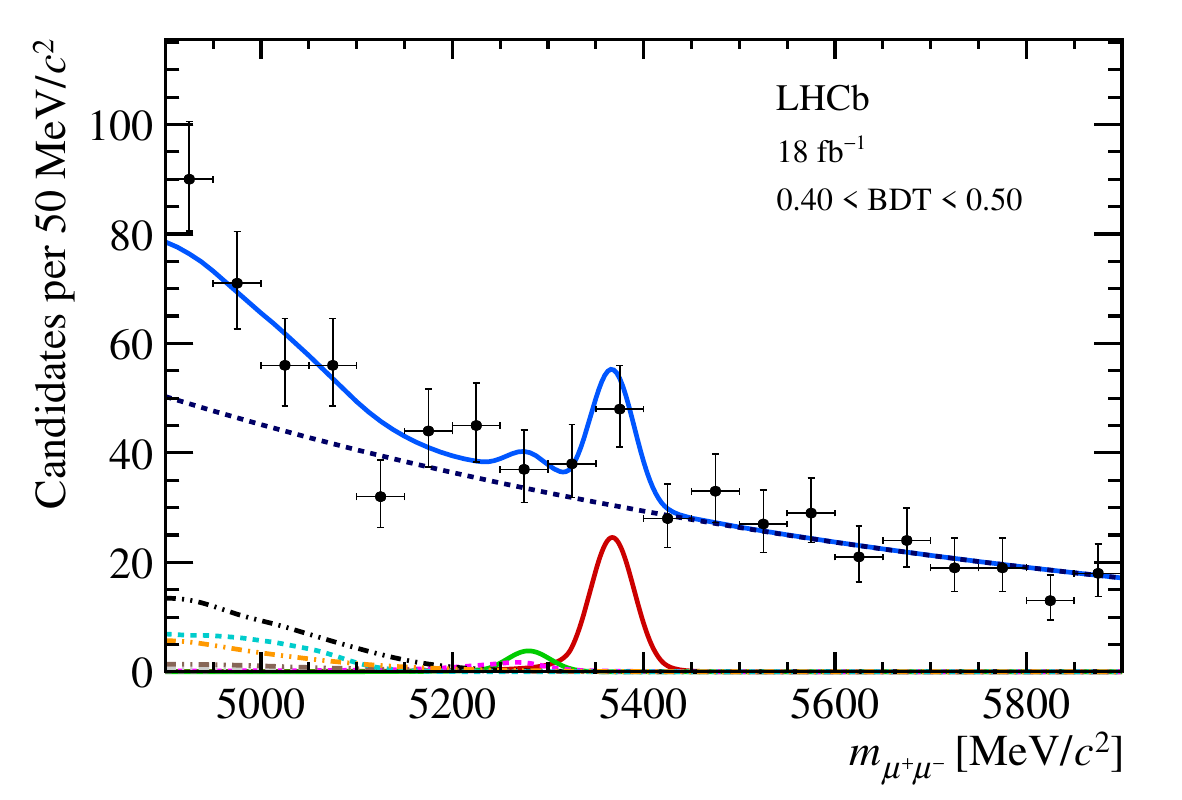}
    \includegraphics[width=0.49\linewidth]{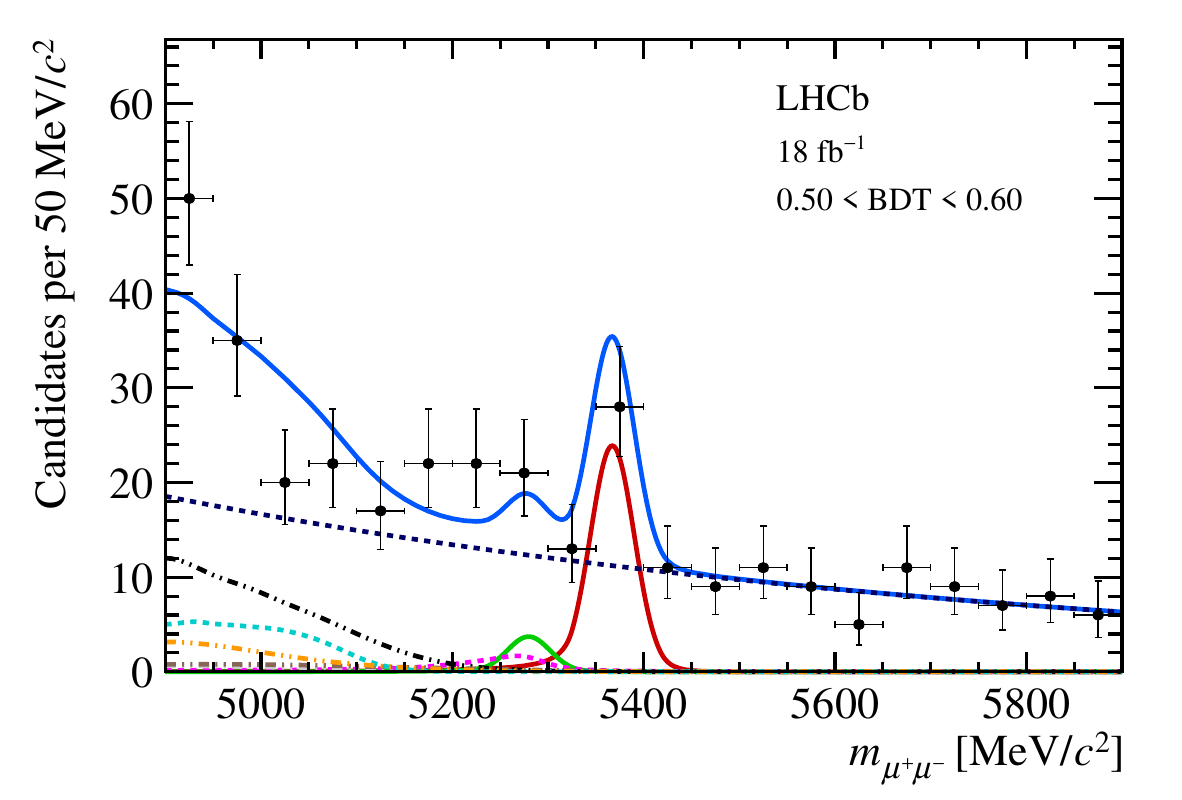}
    \includegraphics[width=0.49\linewidth]{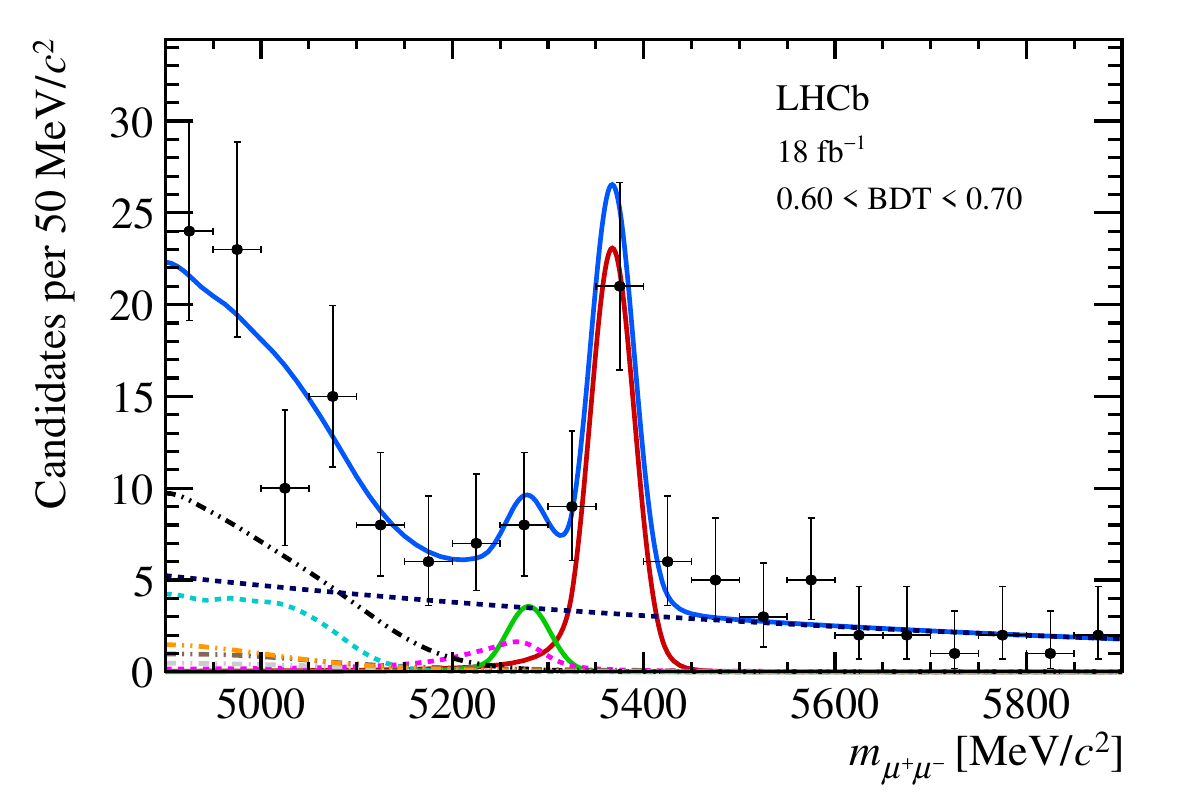}
    \includegraphics[width=0.49\linewidth]{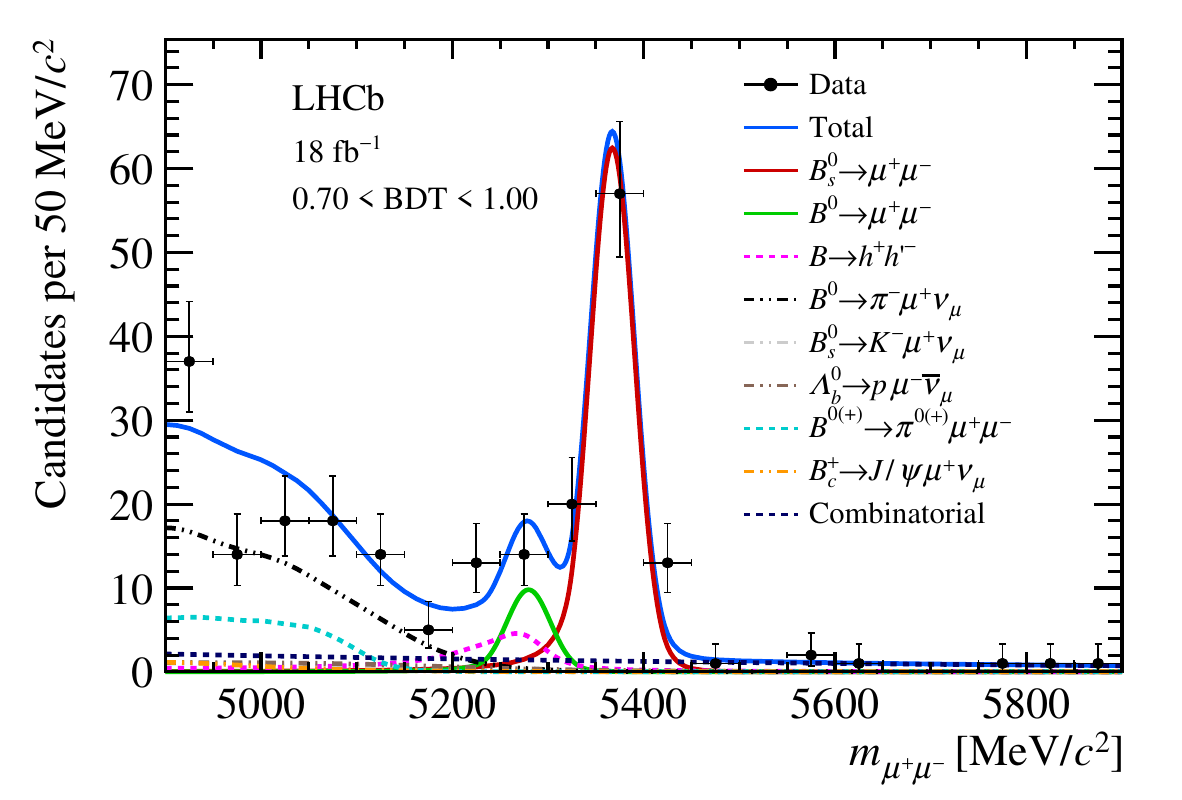}
    \caption{Dimuon invariant-mass distributions from the Run 3 fit in all BDT regions. The data points are shown in black, the total fit curve is overlaid in blue, and its components are indicated in the legend.}
    \label{fig:bf_results_low_bins}
\end{figure}

\begin{figure}[ht]
    \centering
\includegraphics[width=0.32\linewidth]{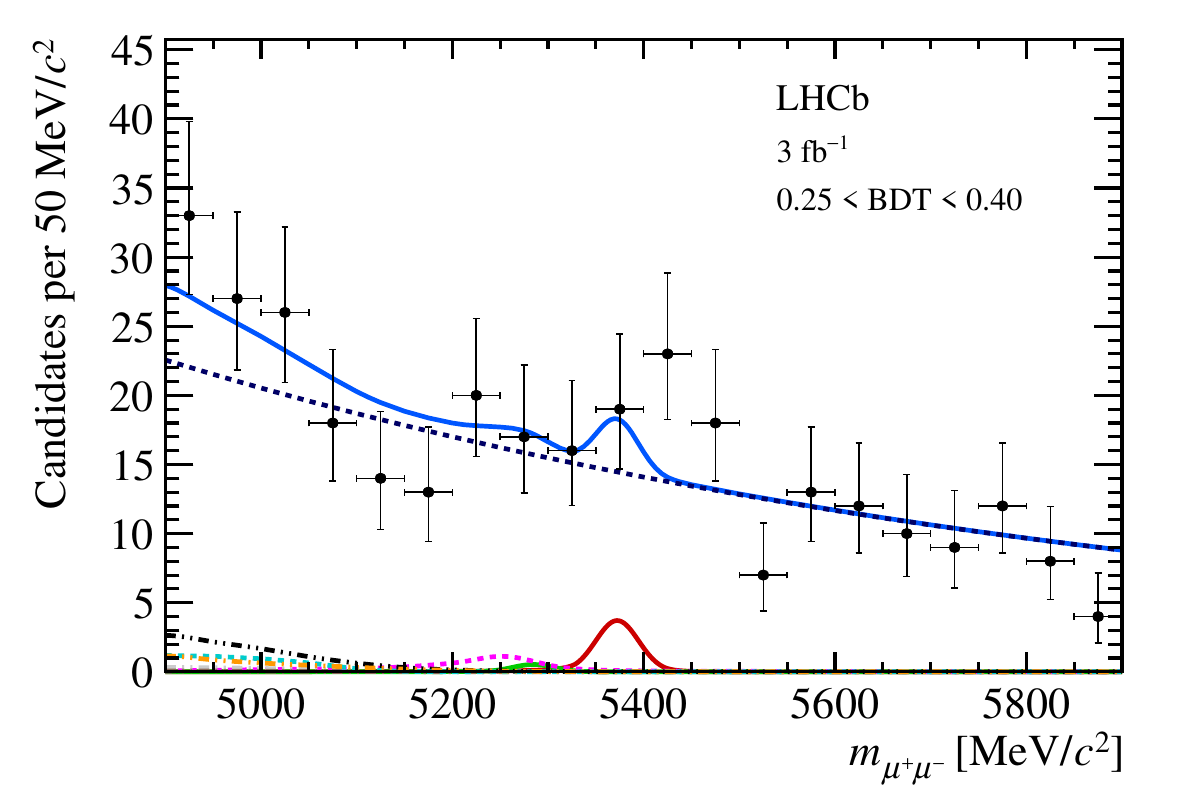}
\includegraphics[width=0.32\linewidth]{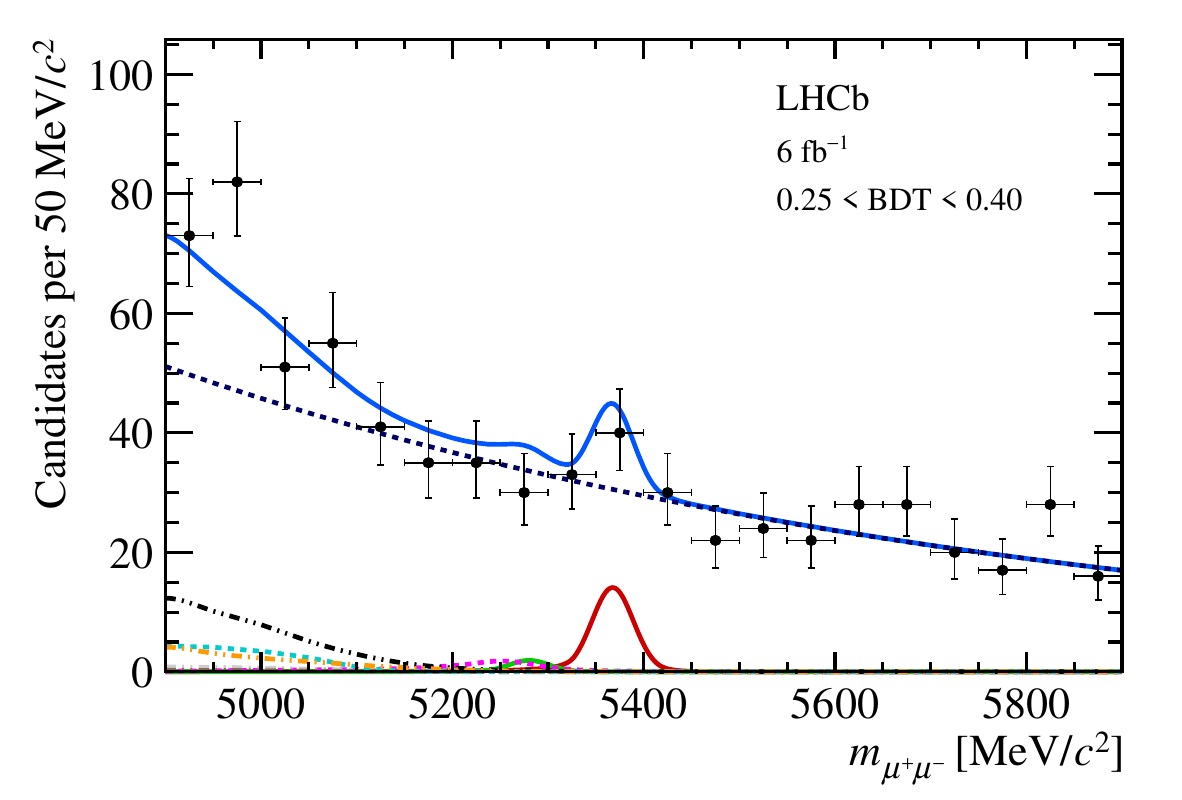}
\includegraphics[width=0.32\linewidth]{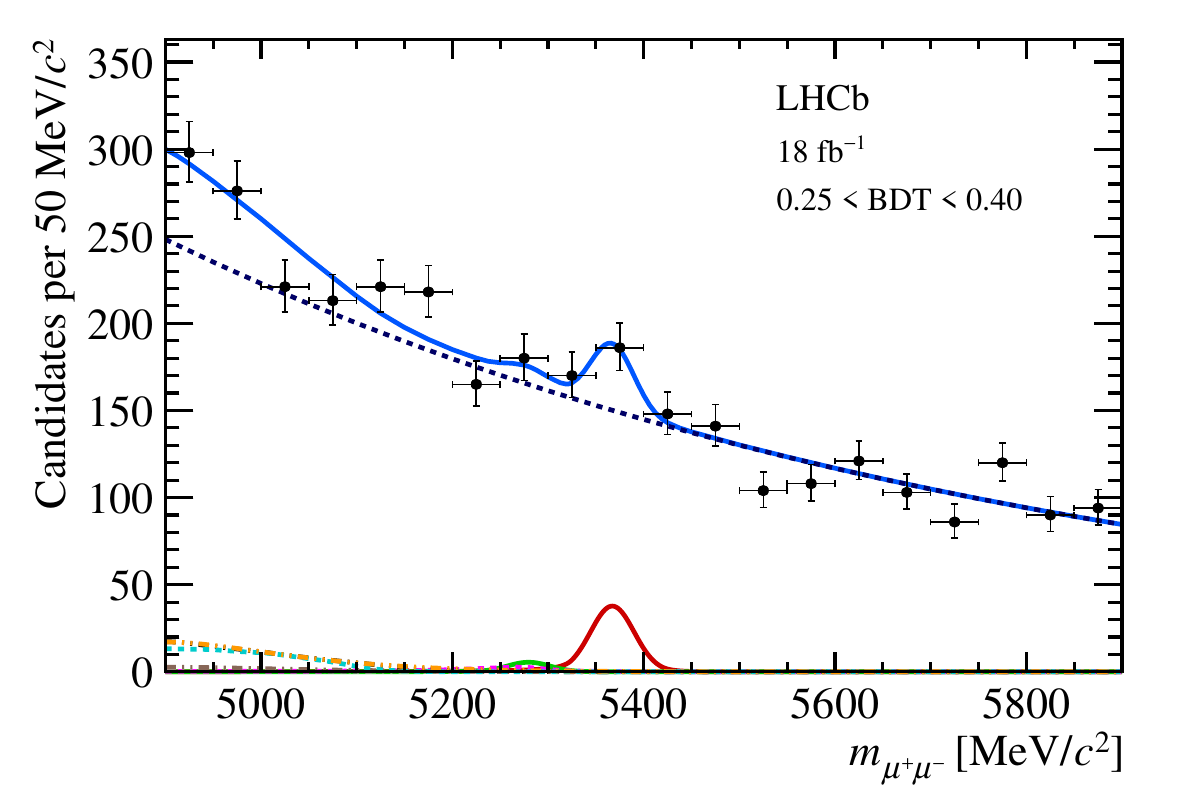}

\includegraphics[width=0.32\linewidth]{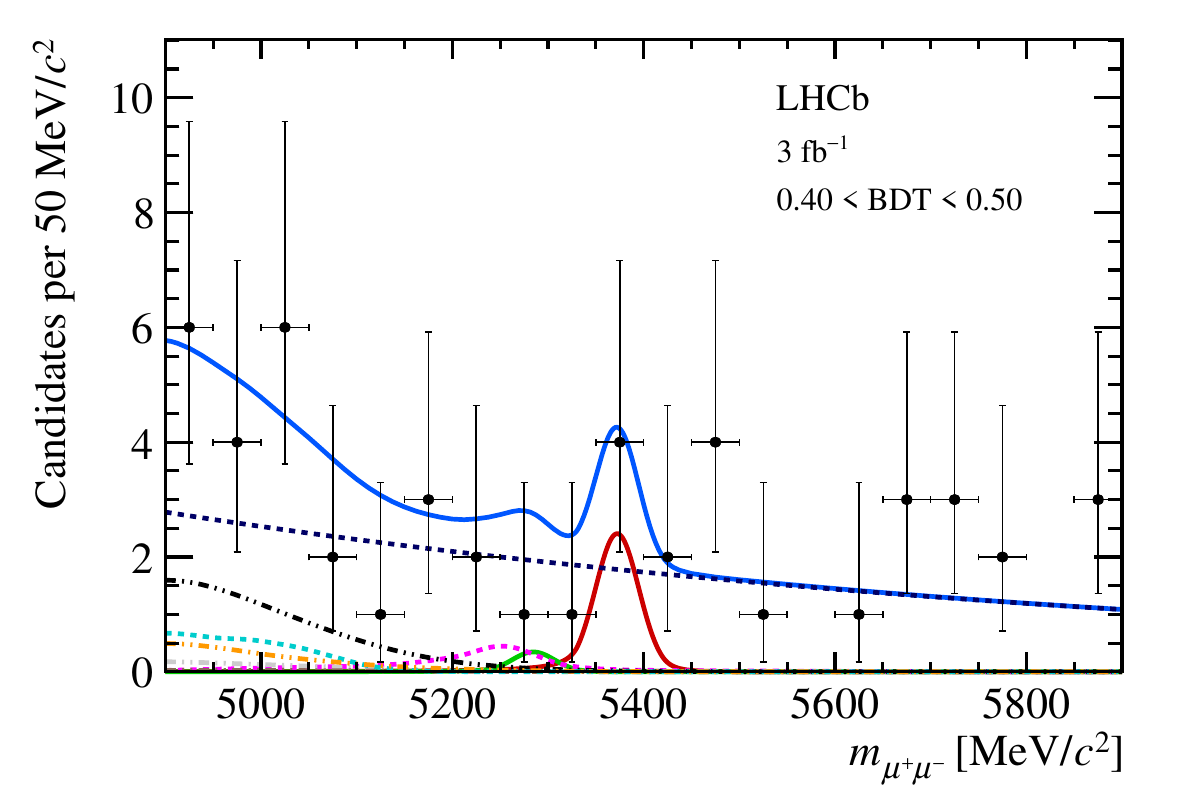}
\includegraphics[width=0.32\linewidth]{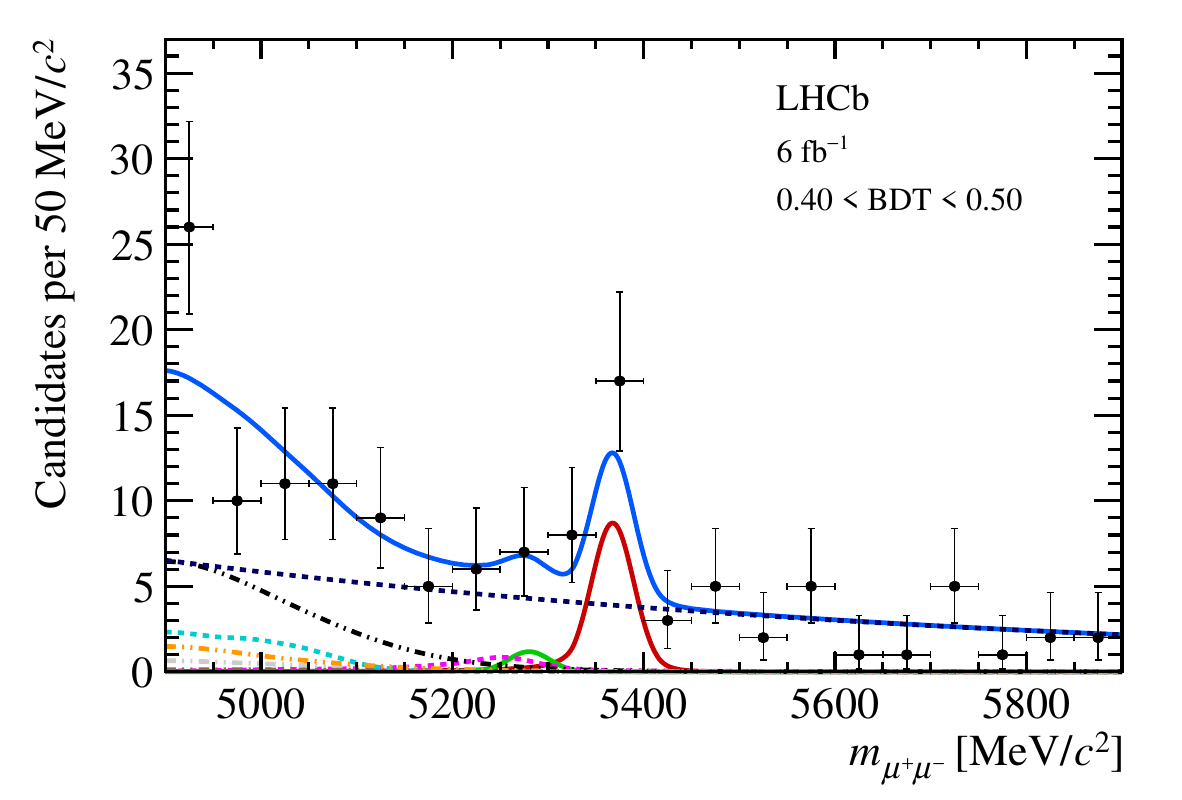}
\includegraphics[width=0.32\linewidth]{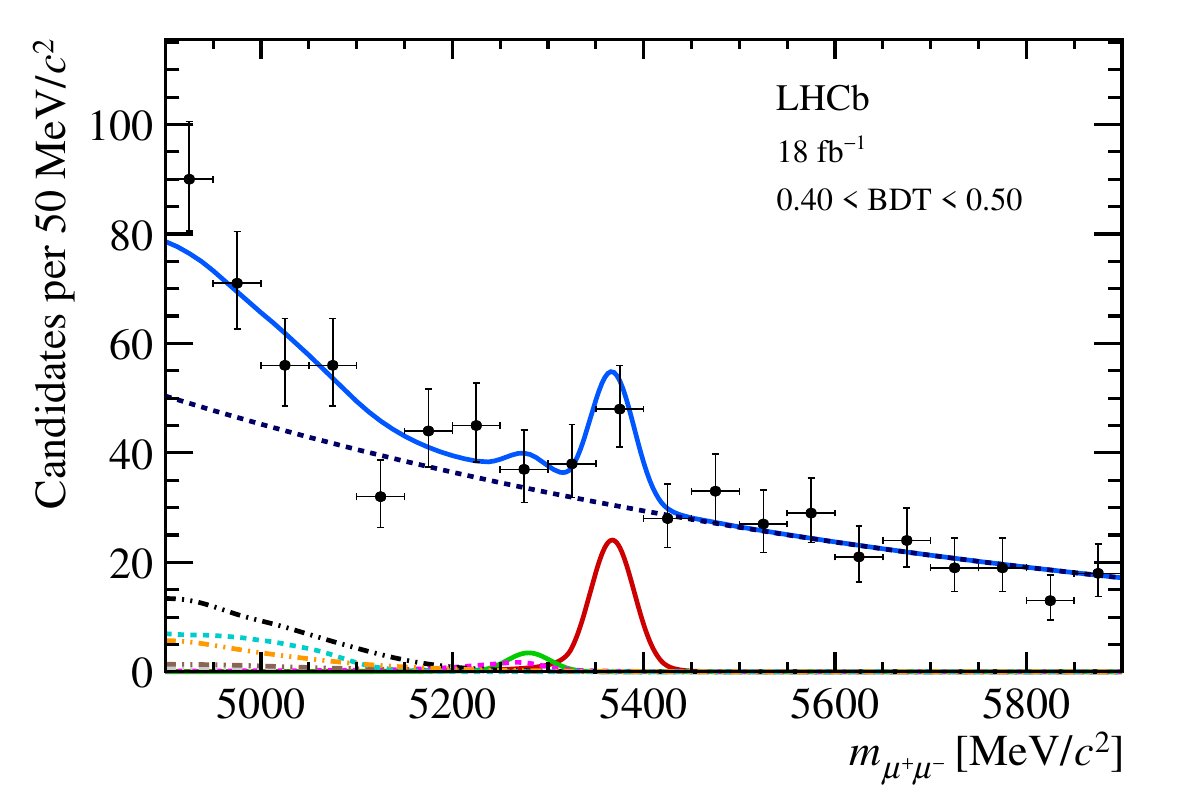}

\includegraphics[width=0.32\linewidth]{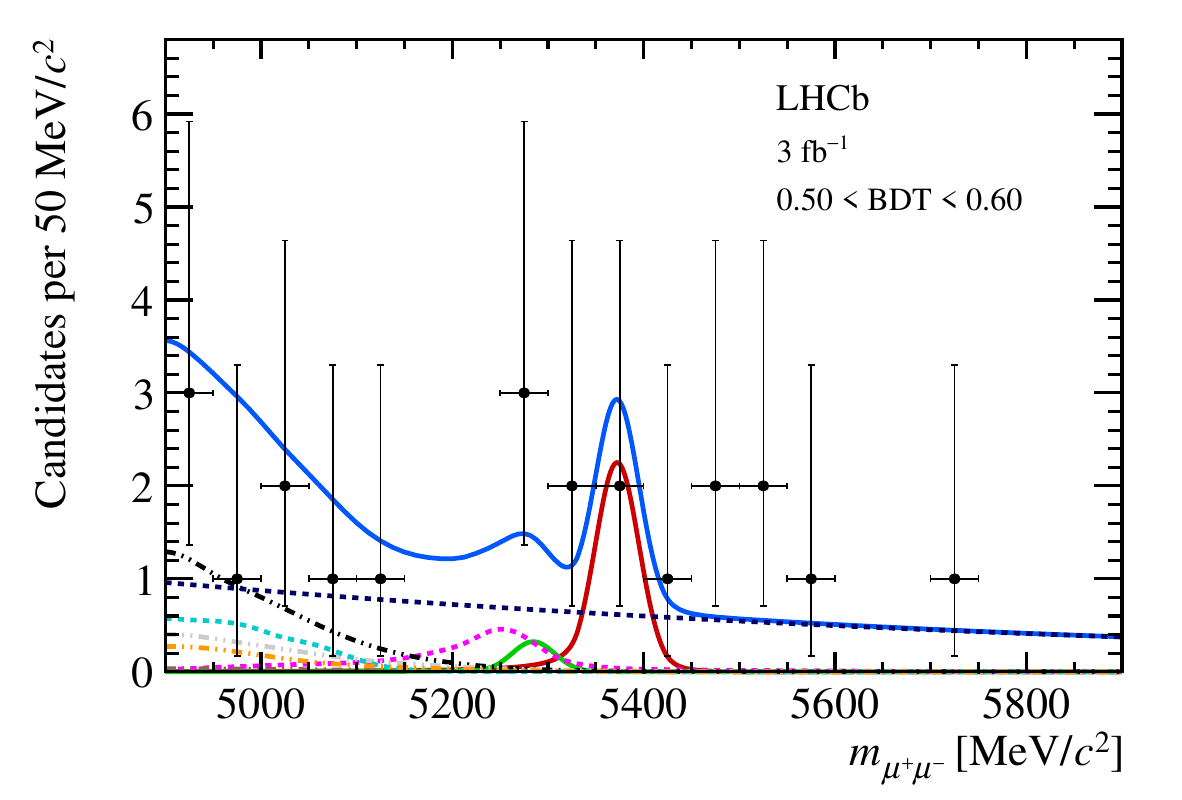}
\includegraphics[width=0.32\linewidth]{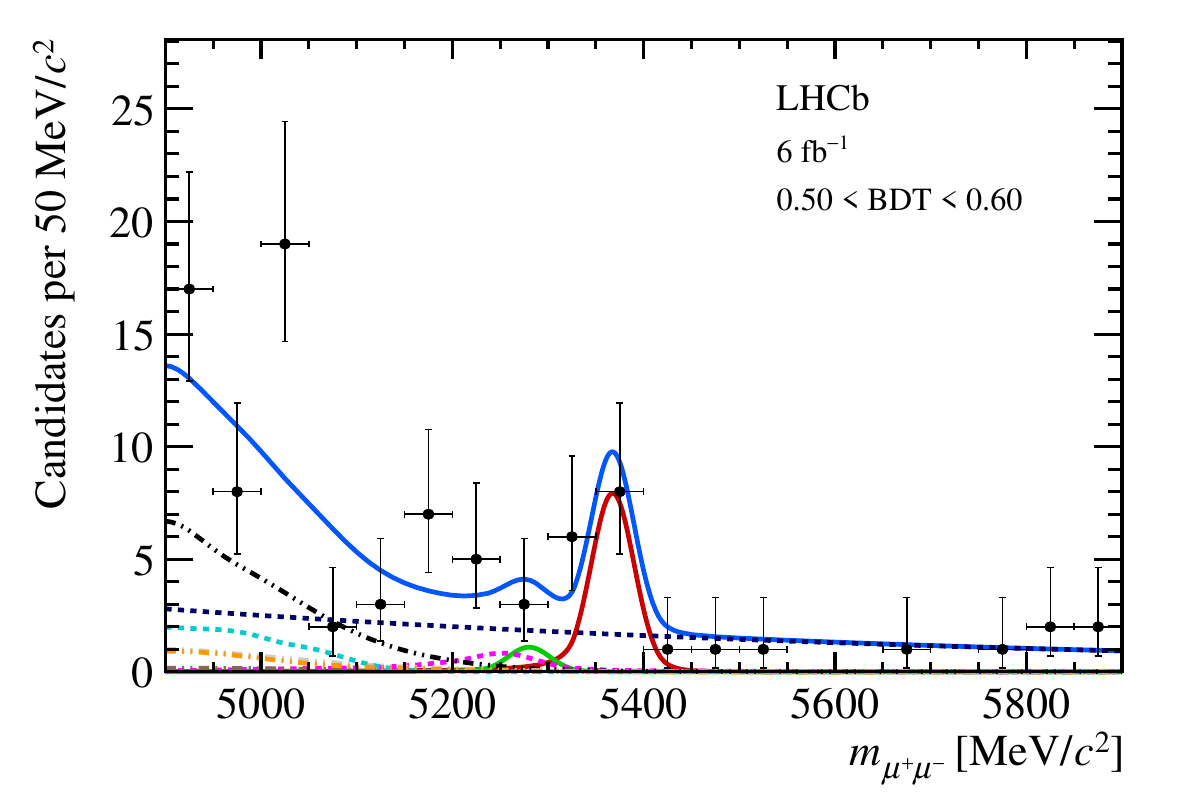}
\includegraphics[width=0.32\linewidth]{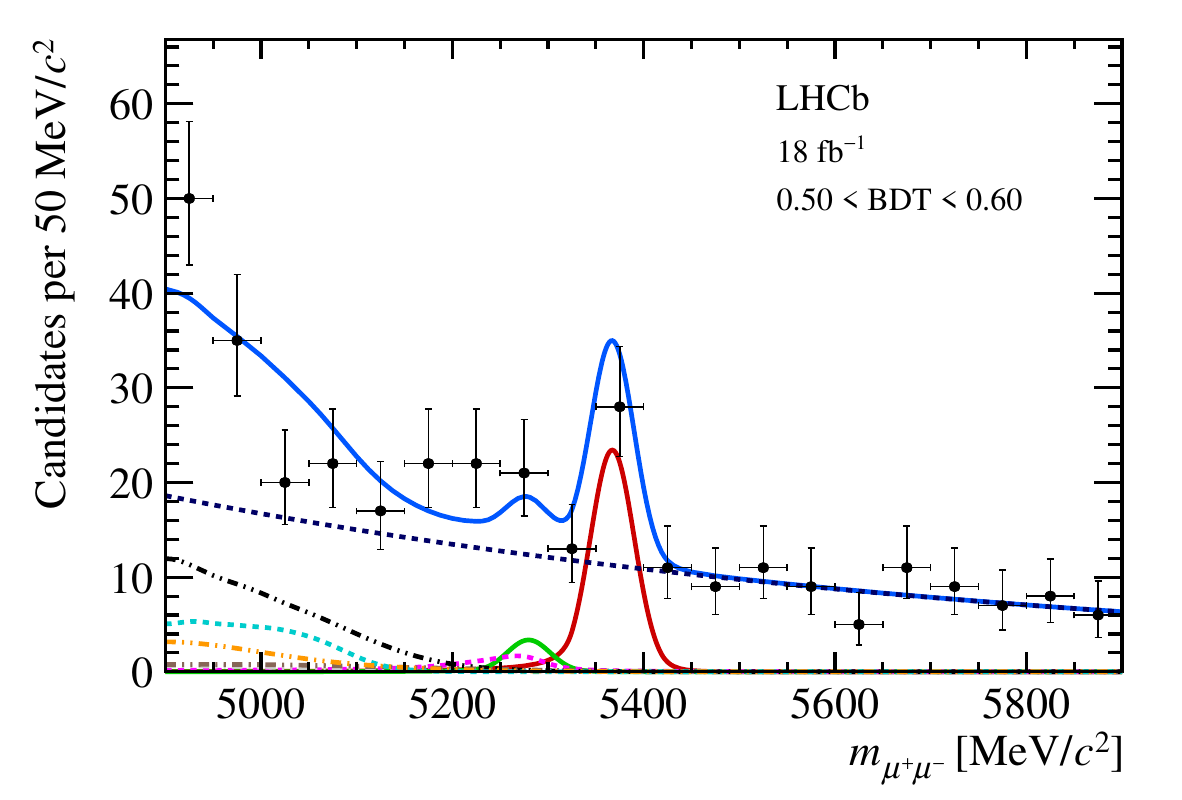}

\includegraphics[width=0.32\linewidth]{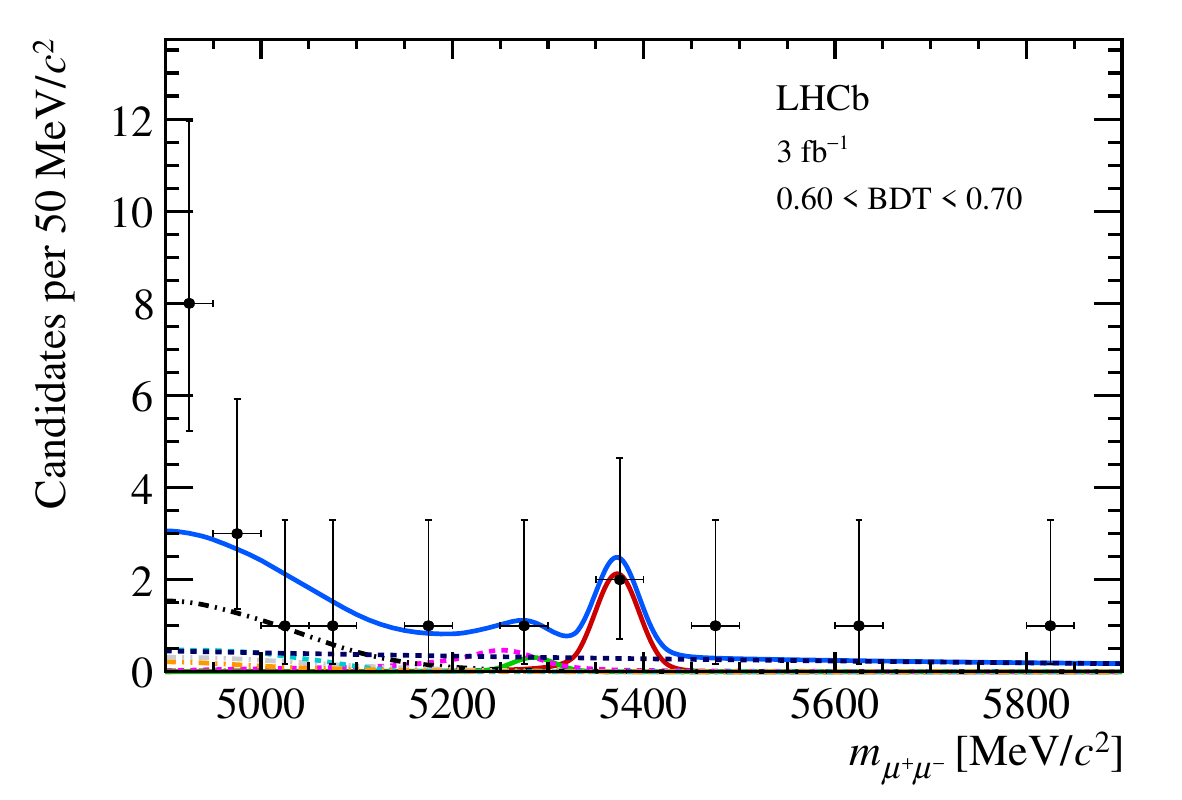}
\includegraphics[width=0.32\linewidth]{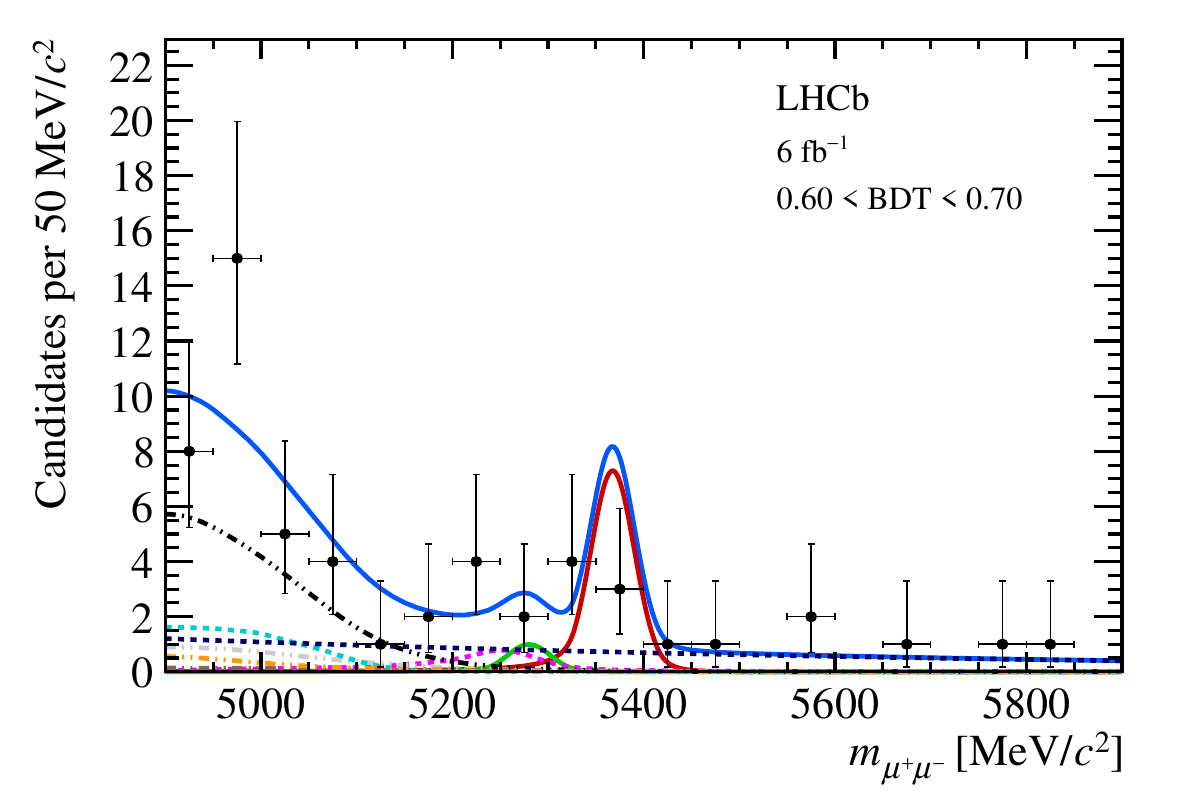}
\includegraphics[width=0.32\linewidth]{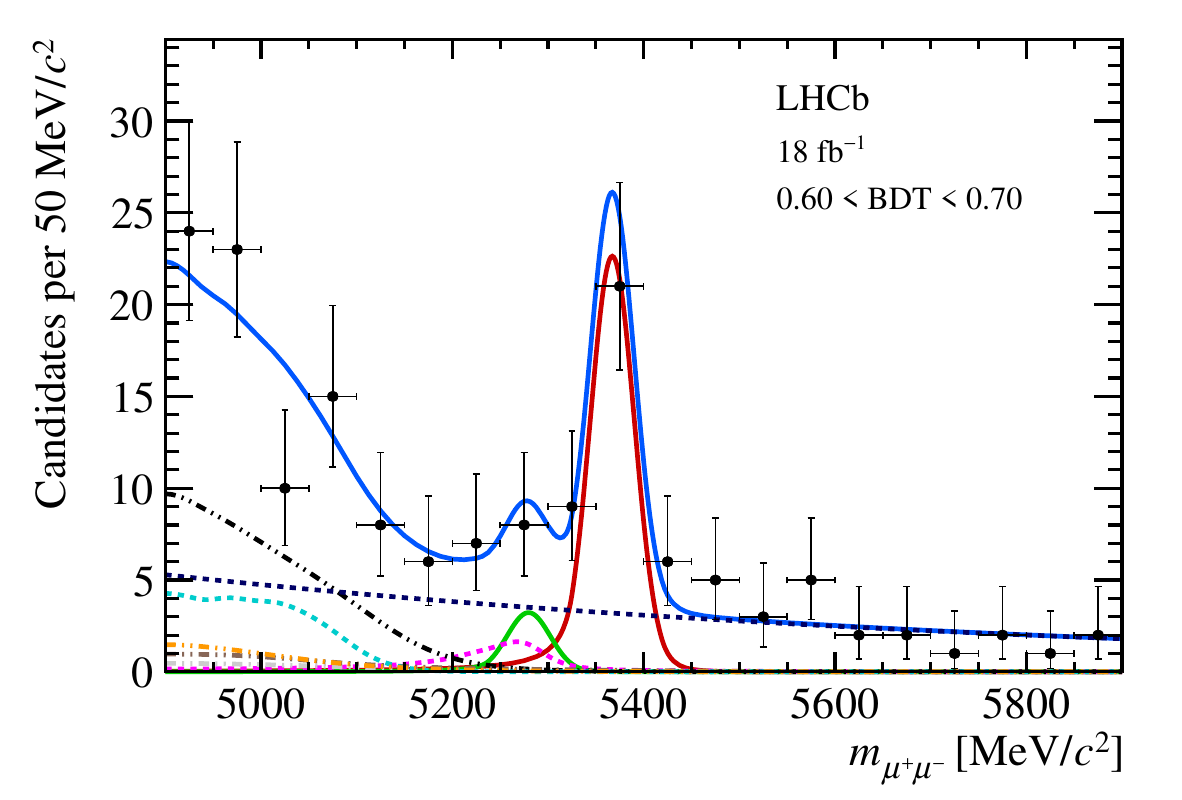}

\includegraphics[width=0.32\linewidth]{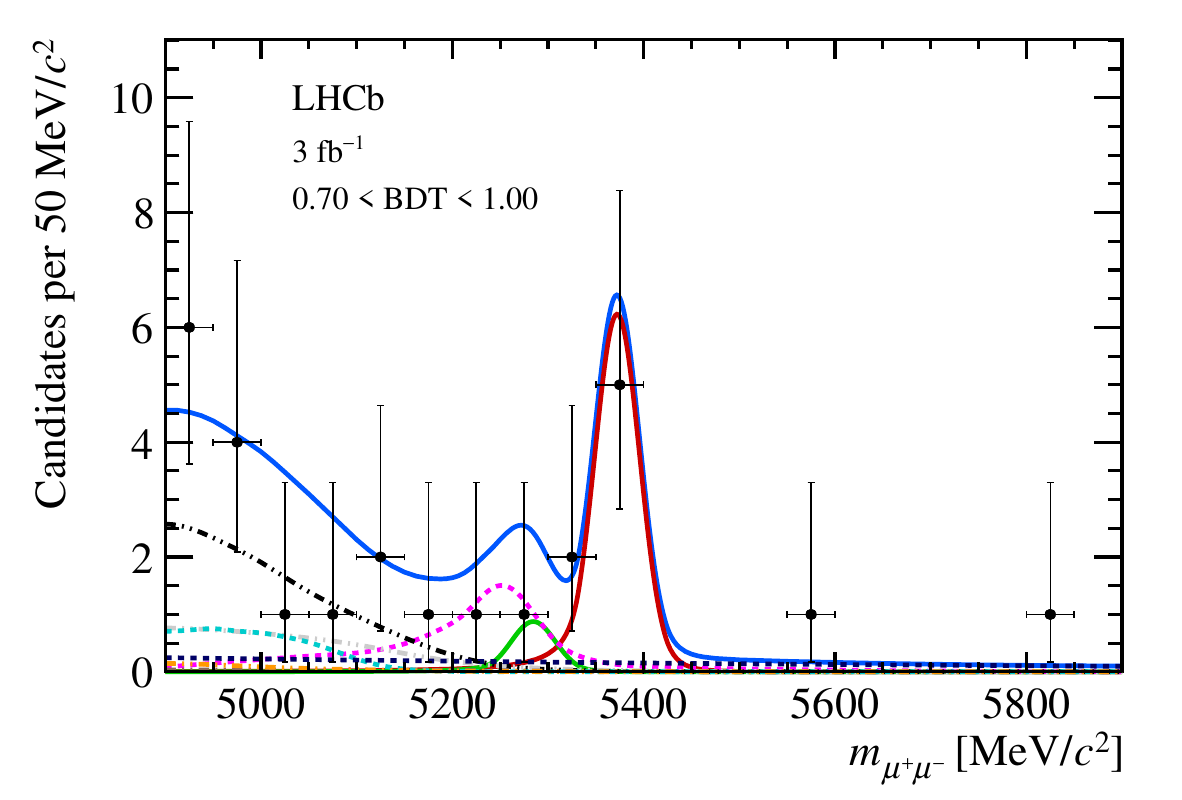}
\includegraphics[width=0.32\linewidth]{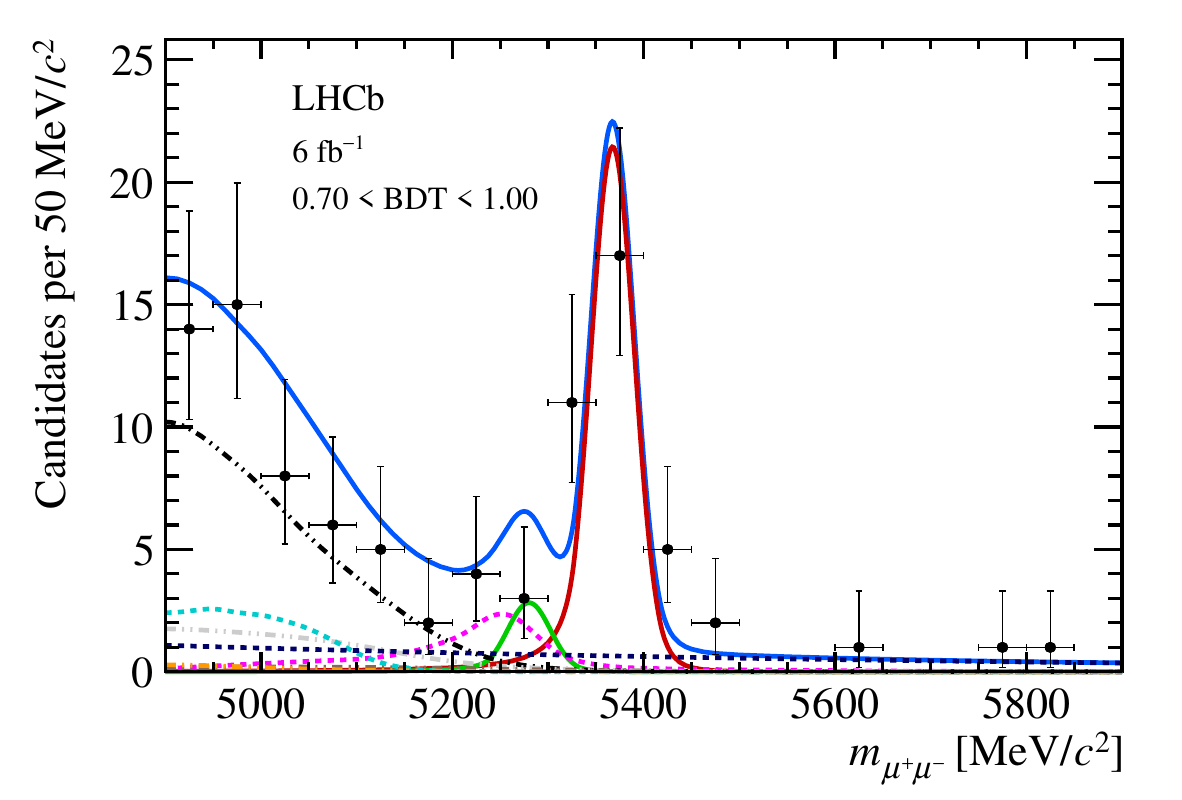}
\includegraphics[width=0.32\linewidth]{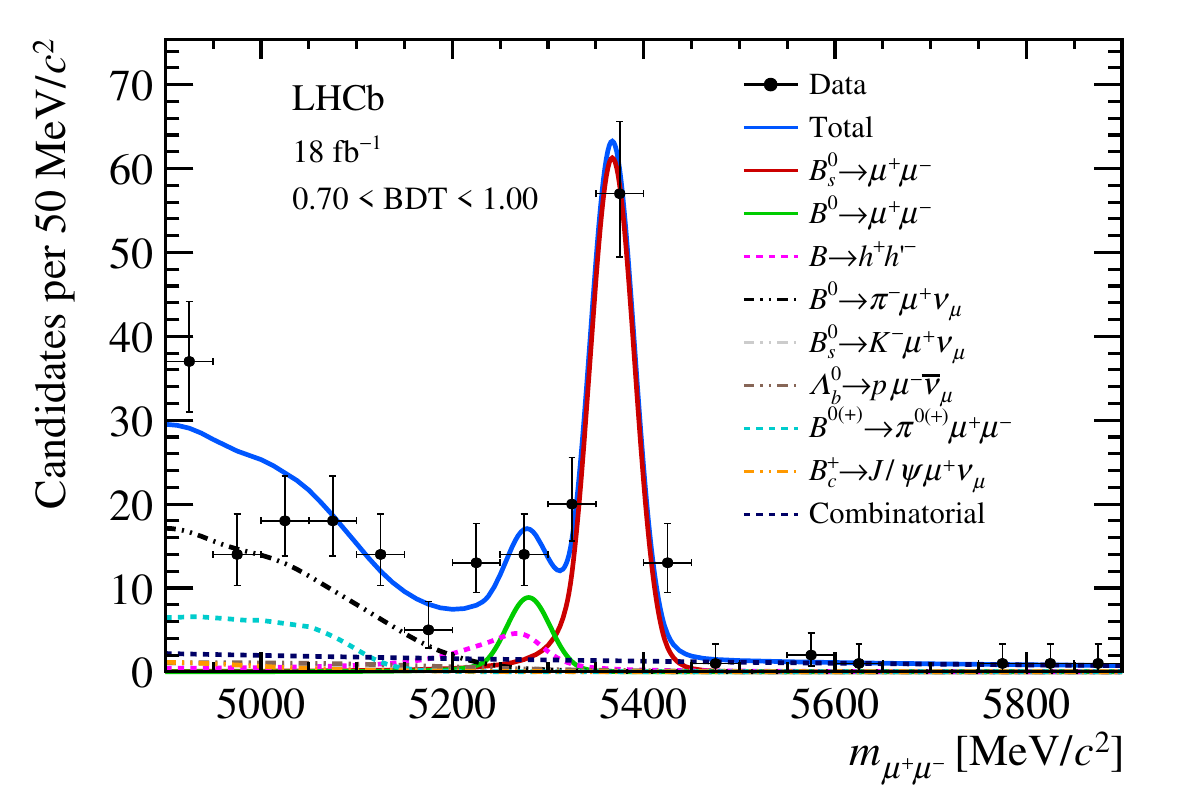}
    \caption{Dimuon invariant-mass distributions from the combined fit in all BDT regions in (left) Run 1, (centre) Run 2 and (right) Run 3. The data points are shown in black, the total fit curve is overlaid in blue, and its components are indicated in the legend.}
    \label{fig:bf_results_Run123_bins}
\end{figure}

The dimuon invariant-mass fits in all BDT categories are shown in Fig.~\ref{fig:bf_results_low_bins} and Fig.~\ref{fig:bf_results_Run123_bins} for the Run~3 and combined Run~1, 2 and 3 data, respectively.

\clearpage

\section*{Acknowledgements}
%
%
\noindent We express our gratitude to our colleagues in the CERN
accelerator departments for the excellent performance of the LHC. We
thank the technical and administrative staff at the LHCb
institutes.
We acknowledge support from CERN and from the national agencies:
ARC (Australia);
CAPES, CNPq, FAPERJ and FINEP (Brazil); 
MOST and NSFC (China); 
CNRS/IN2P3 and CEA (France);  
BMFTR, DFG and MPG (Germany);
NKFIH (Hungary);              
INFN (Italy); 
NWO (Netherlands); 
MNiSW and NCN (Poland); 
MEC/IFA (Romania); 
MICIU and AEI (Spain);
SNSF and SER (Switzerland); 
NASU (Ukraine); 
STFC (United Kingdom); 
DOE NP and NSF (USA).
We acknowledge the computing resources that are provided by ARDC (Australia), 
CBPF (Brazil),
CERN, 
IHEP and LZU (China),
IN2P3 (France), 
KIT and DESY (Germany), 
INFN (Italy), 
SURF (Netherlands),
Polish WLCG (Poland),
IFIN-HH (Romania), 
PIC (Spain), CSCS (Switzerland), 
GridPP (United Kingdom),
and NSF (USA).  
We are indebted to the communities behind the multiple open-source
software packages on which we depend.
Individual groups or members have received support from
RTP (Australia), 
FWO Odysseus grant G0ASD25N (Belgium), 
Key Research Program of Frontier Sciences of CAS, CAS PIFI, CAS CCEPP (China); 
Minciencias (Colombia);
EPLANET, Marie Sk\l{}odowska-Curie Actions, ERC and NextGenerationEU (European Union);
A*MIDEX, ANR, IPhU and Labex P2IO, and R\'{e}gion Auvergne-Rh\^{o}ne-Alpes (France);
Alexander-von-Humboldt Foundation (Germany);
ICSC (Italy); 
Severo Ochoa and Mar\'ia de Maeztu Units of Excellence, GVA, XuntaGal, GENCAT, InTalent-Inditex and Prog.~Atracci\'on Talento CM (Spain);
the Leverhulme Trust, the Royal Society and UKRI (United Kingdom).

\addcontentsline{toc}{section}{References}
\bibliographystyle{LHCb/LHCb}
 
\bibliography{main,LHCb/standard,LHCb/LHCb-PAPER,LHCb/LHCb-CONF,LHCb/LHCb-DP,LHCb/LHCb-TDR,LHCb/LHCb-PUB}

\ifx\mcitethebibliography\mciteundefinedmacro
\PackageError{LHCb.bst}{mciteplus.sty has not been loaded}
{This bibstyle requires the use of the mciteplus package.}\fi
\providecommand{\href}[2]{#2}
\begin{mcitethebibliography}{10}
\mciteSetBstSublistMode{n}
\mciteSetBstMaxWidthForm{subitem}{\alph{mcitesubitemcount})}
\mciteSetBstSublistLabelBeginEnd{\mcitemaxwidthsubitemform\space}
{\relax}{\relax}

\bibitem{ATLAS:2012yve}
ATLAS collaboration, G.~Aad {\em et~al.}, \ifthenelse{\boolean{articletitles}}{\emph{{Observation of a new particle in the search for the Standard Model Higgs boson with the ATLAS detector at the LHC}}, }{}\href{https://doi.org/10.1016/j.physletb.2012.08.020}{Phys.\ Lett.\  \textbf{B716} (2012) 1}, \href{http://arxiv.org/abs/1207.7214}{{\normalfont\ttfamily arXiv:1207.7214}}\relax
\mciteBstWouldAddEndPuncttrue
\mciteSetBstMidEndSepPunct{\mcitedefaultmidpunct}
{\mcitedefaultendpunct}{\mcitedefaultseppunct}\relax
\EndOfBibitem
\bibitem{CMS:2012qbp}
CMS collaboration, S.~Chatrchyan {\em et~al.}, \ifthenelse{\boolean{articletitles}}{\emph{{Observation of a new boson at a mass of 125 GeV with the CMS experiment at the LHC}}, }{}\href{https://doi.org/10.1016/j.physletb.2012.08.021}{Phys.\ Lett.\  \textbf{B716} (2012) 30}, \href{http://arxiv.org/abs/1207.7235}{{\normalfont\ttfamily arXiv:1207.7235}}\relax
\mciteBstWouldAddEndPuncttrue
\mciteSetBstMidEndSepPunct{\mcitedefaultmidpunct}
{\mcitedefaultendpunct}{\mcitedefaultseppunct}\relax
\EndOfBibitem
\bibitem{Planck:2018vyg}
Planck collaboration, N.~Aghanim {\em et~al.}, \ifthenelse{\boolean{articletitles}}{\emph{{Planck 2018 results. VI. Cosmological parameters}}, }{}\href{https://doi.org/10.1051/0004-6361/201833910}{Astron.\ Astrophys.\  \textbf{641} (2020) A6}, Erratum \href{https://doi.org/10.1051/0004-6361/201833910e}{ibid.\   \textbf{652} (2021) C4}, \href{http://arxiv.org/abs/1807.06209}{{\normalfont\ttfamily arXiv:1807.06209}}\relax
\mciteBstWouldAddEndPuncttrue
\mciteSetBstMidEndSepPunct{\mcitedefaultmidpunct}
{\mcitedefaultendpunct}{\mcitedefaultseppunct}\relax
\EndOfBibitem
\bibitem{Sakharov:1967dj}
A.~D. Sakharov, \ifthenelse{\boolean{articletitles}}{\emph{{Violation of \CP invariance, $C$ asymmetry, and baryon asymmetry of the universe}}, }{}\href{https://doi.org/10.1070/PU1991v034n05ABEH002497}{Pis'ma v ZhETF \textbf{5} (1967) 32}\relax
\mciteBstWouldAddEndPuncttrue
\mciteSetBstMidEndSepPunct{\mcitedefaultmidpunct}
{\mcitedefaultendpunct}{\mcitedefaultseppunct}\relax
\EndOfBibitem
\bibitem{Glashow:1970gm}
S.~L. Glashow, J.~Iliopoulos, and L.~Maiani, \ifthenelse{\boolean{articletitles}}{\emph{{Weak interactions with lepton-hadron symmetry}}, }{}\href{https://doi.org/10.1103/PhysRevD.2.1285}{Phys.\ Rev.\  \textbf{D2} (1970) 1285}\relax
\mciteBstWouldAddEndPuncttrue
\mciteSetBstMidEndSepPunct{\mcitedefaultmidpunct}
{\mcitedefaultendpunct}{\mcitedefaultseppunct}\relax
\EndOfBibitem
\bibitem{E598:1974sol}
E598 collaboration, J.~J. Aubert {\em et~al.}, \ifthenelse{\boolean{articletitles}}{\emph{{Experimental observation of a heavy particle $J$}}, }{}\href{https://doi.org/10.1103/PhysRevLett.33.1404}{Phys.\ Rev.\ Lett.\  \textbf{33} (1974) 1404}\relax
\mciteBstWouldAddEndPuncttrue
\mciteSetBstMidEndSepPunct{\mcitedefaultmidpunct}
{\mcitedefaultendpunct}{\mcitedefaultseppunct}\relax
\EndOfBibitem
\bibitem{SLAC-SP-017:1974ind}
SLAC-SP-017 collaboration, J.-E. Augustin {\em et~al.}, \ifthenelse{\boolean{articletitles}}{\emph{{Discovery of a narrow resonance in $e^+ e^-$ annihilation}}, }{}\href{https://doi.org/10.1103/PhysRevLett.33.1406}{Phys.\ Rev.\ Lett.\  \textbf{33} (1974) 1406}\relax
\mciteBstWouldAddEndPuncttrue
\mciteSetBstMidEndSepPunct{\mcitedefaultmidpunct}
{\mcitedefaultendpunct}{\mcitedefaultseppunct}\relax
\EndOfBibitem
\bibitem{ARGUS:1987xtv}
ARGUS collaboration, H.~Albrecht {\em et~al.}, \ifthenelse{\boolean{articletitles}}{\emph{{Observation of $B^0$--$\bar{B}^0$ mixing}}, }{}\href{https://doi.org/10.1016/0370-2693(87)91177-4}{Phys.\ Lett.\  \textbf{B192} (1987) 245}\relax
\mciteBstWouldAddEndPuncttrue
\mciteSetBstMidEndSepPunct{\mcitedefaultmidpunct}
{\mcitedefaultendpunct}{\mcitedefaultseppunct}\relax
\EndOfBibitem
\bibitem{Langacker:1991an}
P.~Langacker and M.-x. Luo, \ifthenelse{\boolean{articletitles}}{\emph{{Implications of precision electroweak experiments for $M_t$, $\rho_{0}$, $\sin^2\theta_W$ and grand unification}}, }{}\href{https://doi.org/10.1103/PhysRevD.44.817}{Phys.\ Rev.\  \textbf{D44} (1991) 817}\relax
\mciteBstWouldAddEndPuncttrue
\mciteSetBstMidEndSepPunct{\mcitedefaultmidpunct}
{\mcitedefaultendpunct}{\mcitedefaultseppunct}\relax
\EndOfBibitem
\bibitem{CDF:1995wbb}
CDF collaboration, F.~Abe {\em et~al.}, \ifthenelse{\boolean{articletitles}}{\emph{{Observation of top quark production in $\bar{p}p$ collisions}}, }{}\href{https://doi.org/10.1103/PhysRevLett.74.2626}{Phys.\ Rev.\ Lett.\  \textbf{74} (1995) 2626}, \href{http://arxiv.org/abs/hep-ex/9503002}{{\normalfont\ttfamily arXiv:hep-ex/9503002}}\relax
\mciteBstWouldAddEndPuncttrue
\mciteSetBstMidEndSepPunct{\mcitedefaultmidpunct}
{\mcitedefaultendpunct}{\mcitedefaultseppunct}\relax
\EndOfBibitem
\bibitem{D0:1995jca}
D0 collaboration, S.~Abachi {\em et~al.}, \ifthenelse{\boolean{articletitles}}{\emph{{Observation of the top quark}}, }{}\href{https://doi.org/10.1103/PhysRevLett.74.2632}{Phys.\ Rev.\ Lett.\  \textbf{74} (1995) 2632}, \href{http://arxiv.org/abs/hep-ex/9503003}{{\normalfont\ttfamily arXiv:hep-ex/9503003}}\relax
\mciteBstWouldAddEndPuncttrue
\mciteSetBstMidEndSepPunct{\mcitedefaultmidpunct}
{\mcitedefaultendpunct}{\mcitedefaultseppunct}\relax
\EndOfBibitem
\bibitem{Altmannshofer:2012az}
W.~Altmannshofer and D.~M. Straub, \ifthenelse{\boolean{articletitles}}{\emph{{Cornering new physics in $b \to s$ transitions}}, }{}\href{https://doi.org/10.1007/JHEP08(2012)121}{JHEP \textbf{08} (2012) 121}, \href{http://arxiv.org/abs/1206.0273}{{\normalfont\ttfamily arXiv:1206.0273}}\relax
\mciteBstWouldAddEndPuncttrue
\mciteSetBstMidEndSepPunct{\mcitedefaultmidpunct}
{\mcitedefaultendpunct}{\mcitedefaultseppunct}\relax
\EndOfBibitem
\bibitem{LHCb-PAPER-2014-006}
LHCb collaboration, R.~Aaij {\em et~al.}, \ifthenelse{\boolean{articletitles}}{\emph{{Differential branching fractions and isospin asymmetries of \mbox{\decay{\B}{K^{(*)}\mumu}} decays}}, }{}\href{https://doi.org/10.1007/JHEP06(2014)133}{JHEP \textbf{06} (2014) 133}, \href{http://arxiv.org/abs/1403.8044}{{\normalfont\ttfamily arXiv:1403.8044}}\relax
\mciteBstWouldAddEndPuncttrue
\mciteSetBstMidEndSepPunct{\mcitedefaultmidpunct}
{\mcitedefaultendpunct}{\mcitedefaultseppunct}\relax
\EndOfBibitem
\bibitem{LHCb-PAPER-2021-014}
LHCb collaboration, R.~Aaij {\em et~al.}, \ifthenelse{\boolean{articletitles}}{\emph{{Branching fraction measurements of the rare $\Bs \to \phiz \mup\mun$ and $\Bs \to f_2^\prime(1525) \mup\mun$ decays}}, }{}\href{https://doi.org/10.1103/PhysRevLett.127.151801}{Phys.\ ~Rev.\ ~Lett.\  \textbf{127} (2021) 151801}, \href{http://arxiv.org/abs/2105.14007}{{\normalfont\ttfamily arXiv:2105.14007}}\relax
\mciteBstWouldAddEndPuncttrue
\mciteSetBstMidEndSepPunct{\mcitedefaultmidpunct}
{\mcitedefaultendpunct}{\mcitedefaultseppunct}\relax
\EndOfBibitem
\bibitem{LHCb-PAPER-2025-041}
LHCb collaboration, R.~Aaij {\em et~al.}, \ifthenelse{\boolean{articletitles}}{\emph{{Comprehensive analysis of the $\Bz\to\Kstarz\mup\mun$ decay}}, }{}\href{https://doi.org/10.1103/24g9-yn9d}{{Phys.\ ~Rev.\ ~Lett.\ } \textbf{137} (2026) 021802}, \href{http://arxiv.org/abs/2512.18053}{{\normalfont\ttfamily arXiv:2512.18053}}\relax
\mciteBstWouldAddEndPuncttrue
\mciteSetBstMidEndSepPunct{\mcitedefaultmidpunct}
{\mcitedefaultendpunct}{\mcitedefaultseppunct}\relax
\EndOfBibitem
\bibitem{CMS:2024syx}
CMS collaboration, A.~Hayrapetyan {\em et~al.}, \ifthenelse{\boolean{articletitles}}{\emph{{Test of lepton flavor universality in B$^{\pm}$$\to$ K$^{\pm}\mu^+\mu^-$ and B$^{\pm}$$\to$ K$^{\pm}$e$^+$e$^-$ decays in proton-proton collisions at $\sqrt{s}$ = 13\tev}}, }{}\href{https://doi.org/10.1088/1361-6633/ad4e65}{Rept.\ Prog.\ Phys.\  \textbf{87} (2024) 077802}, \href{http://arxiv.org/abs/2401.07090}{{\normalfont\ttfamily arXiv:2401.07090}}\relax
\mciteBstWouldAddEndPuncttrue
\mciteSetBstMidEndSepPunct{\mcitedefaultmidpunct}
{\mcitedefaultendpunct}{\mcitedefaultseppunct}\relax
\EndOfBibitem
\bibitem{LHCb-PAPER-2020-041}
LHCb collaboration, R.~Aaij {\em et~al.}, \ifthenelse{\boolean{articletitles}}{\emph{{Angular analysis of the $\Bp\to\Kstarp\mup\mun$ decay}}, }{}\href{https://doi.org/10.1103/PhysRevLett.126.161802}{Phys.\ ~Rev.\ ~Lett.\  \textbf{126} (2021) 161802}, \href{http://arxiv.org/abs/2012.13241}{{\normalfont\ttfamily arXiv:2012.13241}}\relax
\mciteBstWouldAddEndPuncttrue
\mciteSetBstMidEndSepPunct{\mcitedefaultmidpunct}
{\mcitedefaultendpunct}{\mcitedefaultseppunct}\relax
\EndOfBibitem
\bibitem{LHCb-PAPER-2021-022}
LHCb collaboration, R.~Aaij {\em et~al.}, \ifthenelse{\boolean{articletitles}}{\emph{{Angular analysis of the rare decay $\Bs \to \Pphi \mup \mun$}}, }{}\href{https://doi.org/10.1007/JHEP11(2021)043}{JHEP \textbf{11} (2021) 043}, \href{http://arxiv.org/abs/2107.13428}{{\normalfont\ttfamily arXiv:2107.13428}}\relax
\mciteBstWouldAddEndPuncttrue
\mciteSetBstMidEndSepPunct{\mcitedefaultmidpunct}
{\mcitedefaultendpunct}{\mcitedefaultseppunct}\relax
\EndOfBibitem
\bibitem{CMS:2024atz}
CMS collaboration, A.~Hayrapetyan {\em et~al.}, \ifthenelse{\boolean{articletitles}}{\emph{{Angular analysis of the $B^0\to K^*(892)^0\mu^+\mu^-$ decay in proton-proton collisions at $\sqrt{s}$ = 13\tev}}, }{}\href{https://doi.org/10.1016/j.physletb.2025.139406}{Phys.\ Lett.\  \textbf{B864} (2025) 139406}, \href{http://arxiv.org/abs/2411.11820}{{\normalfont\ttfamily arXiv:2411.11820}}\relax
\mciteBstWouldAddEndPuncttrue
\mciteSetBstMidEndSepPunct{\mcitedefaultmidpunct}
{\mcitedefaultendpunct}{\mcitedefaultseppunct}\relax
\EndOfBibitem
\bibitem{ATLAS:2018gqc}
ATLAS collaboration, M.~Aaboud {\em et~al.}, \ifthenelse{\boolean{articletitles}}{\emph{{Angular analysis of $B^0_d \rightarrow K^{*}\mu^+\mu^-$ decays in $pp$ collisions at $\sqrt{s}= 8$\tev with the ATLAS detector}}, }{}\href{https://doi.org/10.1007/JHEP10(2018)047}{JHEP \textbf{10} (2018) 047}, \href{http://arxiv.org/abs/1805.04000}{{\normalfont\ttfamily arXiv:1805.04000}}\relax
\mciteBstWouldAddEndPuncttrue
\mciteSetBstMidEndSepPunct{\mcitedefaultmidpunct}
{\mcitedefaultendpunct}{\mcitedefaultseppunct}\relax
\EndOfBibitem
\bibitem{Belle:2016fev}
Belle collaboration, S.~Wehle {\em et~al.}, \ifthenelse{\boolean{articletitles}}{\emph{{Lepton-flavor-dependent angular analysis of $B\to K^\ast \ell^+\ell^-$}}, }{}\href{https://doi.org/10.1103/PhysRevLett.118.111801}{Phys.\ Rev.\ Lett.\  \textbf{118} (2017) 111801}, \href{http://arxiv.org/abs/1612.05014}{{\normalfont\ttfamily arXiv:1612.05014}}\relax
\mciteBstWouldAddEndPuncttrue
\mciteSetBstMidEndSepPunct{\mcitedefaultmidpunct}
{\mcitedefaultendpunct}{\mcitedefaultseppunct}\relax
\EndOfBibitem
\bibitem{Horgan:2013hoa}
R.~R. Horgan, Z.~Liu, S.~Meinel, and M.~Wingate, \ifthenelse{\boolean{articletitles}}{\emph{{Lattice QCD calculation of form factors describing the rare decays $B \to K^* \ell^+ \ell^-$ and $B_s \to \phi \ell^+ \ell^-$}}, }{}\href{https://doi.org/10.1103/PhysRevD.89.094501}{Phys.\ Rev.\  \textbf{D89} (2014) 094501}, \href{http://arxiv.org/abs/1310.3722}{{\normalfont\ttfamily arXiv:1310.3722}}\relax
\mciteBstWouldAddEndPuncttrue
\mciteSetBstMidEndSepPunct{\mcitedefaultmidpunct}
{\mcitedefaultendpunct}{\mcitedefaultseppunct}\relax
\EndOfBibitem
\bibitem{Bharucha:2015bzk}
A.~Bharucha, D.~M. Straub, and R.~Zwicky, \ifthenelse{\boolean{articletitles}}{\emph{{$B\to V\ell^+\ell^-$ in the Standard Model from light-cone sum rules}}, }{}\href{https://doi.org/10.1007/JHEP08(2016)098}{JHEP \textbf{08} (2016) 098}, \href{http://arxiv.org/abs/1503.05534}{{\normalfont\ttfamily arXiv:1503.05534}}\relax
\mciteBstWouldAddEndPuncttrue
\mciteSetBstMidEndSepPunct{\mcitedefaultmidpunct}
{\mcitedefaultendpunct}{\mcitedefaultseppunct}\relax
\EndOfBibitem
\bibitem{Gubernari:2018wyi}
N.~Gubernari, A.~Kokulu, and D.~van Dyk, \ifthenelse{\boolean{articletitles}}{\emph{{$B\to P$ and $B\to V$ form factors from $B$-meson light-cone sum rules beyond leading twist}}, }{}\href{https://doi.org/10.1007/JHEP01(2019)150}{JHEP \textbf{01} (2019) 150}, \href{http://arxiv.org/abs/1811.00983}{{\normalfont\ttfamily arXiv:1811.00983}}\relax
\mciteBstWouldAddEndPuncttrue
\mciteSetBstMidEndSepPunct{\mcitedefaultmidpunct}
{\mcitedefaultendpunct}{\mcitedefaultseppunct}\relax
\EndOfBibitem
\bibitem{Parrott:2022zte}
HPQCD collaboration, W.~G. Parrott, C.~Bouchard, and C.~T.~H. Davies, \ifthenelse{\boolean{articletitles}}{\emph{{Standard Model predictions for $B\to K\ell^+\ell^-$, $B\to K\ell_1^-\ell_2^+$ and $B\to K\nu\bar\nu$ using form factors from $N_f=2+1+1$ lattice QCD}}, }{}\href{https://doi.org/10.1103/PhysRevD.107.014511}{Phys.\ Rev.\  \textbf{D107} (2023) 014511}, Erratum \href{https://doi.org/10.1103/PhysRevD.107.119903}{ibid.\   \textbf{D107} (2023) 119903}, \href{http://arxiv.org/abs/2207.13371}{{\normalfont\ttfamily arXiv:2207.13371}}\relax
\mciteBstWouldAddEndPuncttrue
\mciteSetBstMidEndSepPunct{\mcitedefaultmidpunct}
{\mcitedefaultendpunct}{\mcitedefaultseppunct}\relax
\EndOfBibitem
\bibitem{Khodjamirian:2010vf}
A.~Khodjamirian, T.~Mannel, A.~A. Pivovarov, and Y.-M. Wang, \ifthenelse{\boolean{articletitles}}{\emph{{Charm-loop effect in $B \to K^{(*)} \ell^{+} \ell^{-}$ and $B\to K^*\gamma$}}, }{}\href{https://doi.org/10.1007/JHEP09(2010)089}{JHEP \textbf{09} (2010) 089}, \href{http://arxiv.org/abs/1006.4945}{{\normalfont\ttfamily arXiv:1006.4945}}\relax
\mciteBstWouldAddEndPuncttrue
\mciteSetBstMidEndSepPunct{\mcitedefaultmidpunct}
{\mcitedefaultendpunct}{\mcitedefaultseppunct}\relax
\EndOfBibitem
\bibitem{Lyon:2014hpa}
J.~Lyon and R.~Zwicky, \ifthenelse{\boolean{articletitles}}{\emph{{Resonances gone topsy turvy - the charm of QCD or new physics in $b \to s \ell^+ \ell^-$?}}, }{}\href{http://arxiv.org/abs/1406.0566}{{\normalfont\ttfamily arXiv:1406.0566}}\relax
\mciteBstWouldAddEndPuncttrue
\mciteSetBstMidEndSepPunct{\mcitedefaultmidpunct}
{\mcitedefaultendpunct}{\mcitedefaultseppunct}\relax
\EndOfBibitem
\bibitem{Ciuchini:2015qxb}
M.~Ciuchini {\em et~al.}, \ifthenelse{\boolean{articletitles}}{\emph{{$B\to K^* \ell^+ \ell^-$ decays at large recoil in the Standard Model: a theoretical reappraisal}}, }{}\href{https://doi.org/10.1007/JHEP06(2016)116}{JHEP \textbf{06} (2016) 116}, \href{http://arxiv.org/abs/1512.07157}{{\normalfont\ttfamily arXiv:1512.07157}}\relax
\mciteBstWouldAddEndPuncttrue
\mciteSetBstMidEndSepPunct{\mcitedefaultmidpunct}
{\mcitedefaultendpunct}{\mcitedefaultseppunct}\relax
\EndOfBibitem
\bibitem{Gubernari:2020eft}
N.~Gubernari, D.~van Dyk, and J.~Virto, \ifthenelse{\boolean{articletitles}}{\emph{{Non-local matrix elements in $B_{(s)}\to \{K^{(*)},\phi\}\ell^+\ell^-$}}, }{}\href{https://doi.org/10.1007/JHEP02(2021)088}{JHEP \textbf{02} (2021) 088}, \href{http://arxiv.org/abs/2011.09813}{{\normalfont\ttfamily arXiv:2011.09813}}\relax
\mciteBstWouldAddEndPuncttrue
\mciteSetBstMidEndSepPunct{\mcitedefaultmidpunct}
{\mcitedefaultendpunct}{\mcitedefaultseppunct}\relax
\EndOfBibitem
\bibitem{Gubernari:2022hxn}
N.~Gubernari, M.~Reboud, D.~van Dyk, and J.~Virto, \ifthenelse{\boolean{articletitles}}{\emph{{Improved theory predictions and global analysis of exclusive $b \to s\mu^+\mu^-$ processes}}, }{}\href{https://doi.org/10.1007/JHEP09(2022)133}{JHEP \textbf{09} (2022) 133}, \href{http://arxiv.org/abs/2206.03797}{{\normalfont\ttfamily arXiv:2206.03797}}\relax
\mciteBstWouldAddEndPuncttrue
\mciteSetBstMidEndSepPunct{\mcitedefaultmidpunct}
{\mcitedefaultendpunct}{\mcitedefaultseppunct}\relax
\EndOfBibitem
\bibitem{LHCb-PAPER-2024-011}
LHCb collaboration, R.~Aaij {\em et~al.}, \ifthenelse{\boolean{articletitles}}{\emph{{Comprehensive analysis of local and nonlocal amplitudes in the $\Bd\to\Kstarz\mumu$ decay}}, }{}\href{https://doi.org/10.1007/JHEP09(2024)026}{{JHEP} \textbf{09} (2024) 026}, Erratum \href{https://doi.org/10.1007/JHEP05(2025)208}{ibid.\   \textbf{05} (2025) 208}, \href{http://arxiv.org/abs/2405.17347}{{\normalfont\ttfamily arXiv:2405.17347}}\relax
\mciteBstWouldAddEndPuncttrue
\mciteSetBstMidEndSepPunct{\mcitedefaultmidpunct}
{\mcitedefaultendpunct}{\mcitedefaultseppunct}\relax
\EndOfBibitem
\bibitem{LHCb-PAPER-2023-033}
LHCb collaboration, R.~Aaij {\em et~al.}, \ifthenelse{\boolean{articletitles}}{\emph{{Amplitude analysis of the $\Bz \to \Kstarz\mup\mun$ decay}}, }{}\href{https://doi.org/10.1103/PhysRevLett.132.131801}{Phys.\ ~Rev.\ ~Lett.\  \textbf{132} (2024) 131801}, \href{http://arxiv.org/abs/2312.09115}{{\normalfont\ttfamily arXiv:2312.09115}}\relax
\mciteBstWouldAddEndPuncttrue
\mciteSetBstMidEndSepPunct{\mcitedefaultmidpunct}
{\mcitedefaultendpunct}{\mcitedefaultseppunct}\relax
\EndOfBibitem
\bibitem{Beneke:2019slt}
M.~Beneke, C.~Bobeth, and R.~Szafron, \ifthenelse{\boolean{articletitles}}{\emph{{Power-enhanced leading-logarithmic QED corrections to $B_q\to\mu^+\mu^-$}}, }{}\href{https://doi.org/10.1007/JHEP10(2019)232}{JHEP \textbf{10} (2019) 232}, Erratum \href{https://doi.org/10.1007/JHEP11(2022)099}{ibid.\   \textbf{11} (2022) 099}, \href{http://arxiv.org/abs/1908.07011}{{\normalfont\ttfamily arXiv:1908.07011}}\relax
\mciteBstWouldAddEndPuncttrue
\mciteSetBstMidEndSepPunct{\mcitedefaultmidpunct}
{\mcitedefaultendpunct}{\mcitedefaultseppunct}\relax
\EndOfBibitem
\bibitem{FLAG2024}
Flavour Lattice Averaging Group, Y.~Aoki {\em et~al.}, \ifthenelse{\boolean{articletitles}}{\emph{{FLAG Review 2024}}, }{}\href{https://doi.org/10.1103/nfzp-p5dn}{Phys.\ Rev.\  \textbf{D113} (2026) 014508}, \href{http://arxiv.org/abs/2411.04268}{{\normalfont\ttfamily arXiv:2411.04268}}\relax
\mciteBstWouldAddEndPuncttrue
\mciteSetBstMidEndSepPunct{\mcitedefaultmidpunct}
{\mcitedefaultendpunct}{\mcitedefaultseppunct}\relax
\EndOfBibitem
\bibitem{Buras:2024per}
A.~J. Buras, \ifthenelse{\boolean{articletitles}}{\emph{{Hunting new animalcula with rare K and B decays}}, }{}\href{https://doi.org/10.1051/epjconf/202431400002}{EPJ Web Conf.\  \textbf{314} (2024) 00002}, \href{http://arxiv.org/abs/2411.03440}{{\normalfont\ttfamily arXiv:2411.03440}}\relax
\mciteBstWouldAddEndPuncttrue
\mciteSetBstMidEndSepPunct{\mcitedefaultmidpunct}
{\mcitedefaultendpunct}{\mcitedefaultseppunct}\relax
\EndOfBibitem
\bibitem{DAmbrosio:2002vsn}
G.~D'Ambrosio, G.~F. Giudice, G.~Isidori, and A.~Strumia, \ifthenelse{\boolean{articletitles}}{\emph{{Minimal flavor violation: an effective field theory approach}}, }{}\href{https://doi.org/10.1016/S0550-3213(02)00836-2}{Nucl.\ Phys.\  \textbf{B645} (2002) 155}, \href{http://arxiv.org/abs/hep-ph/0207036}{{\normalfont\ttfamily arXiv:hep-ph/0207036}}\relax
\mciteBstWouldAddEndPuncttrue
\mciteSetBstMidEndSepPunct{\mcitedefaultmidpunct}
{\mcitedefaultendpunct}{\mcitedefaultseppunct}\relax
\EndOfBibitem
\bibitem{DeBruyn:2012wk}
K.~De~Bruyn {\em et~al.}, \ifthenelse{\boolean{articletitles}}{\emph{{Probing New Physics via the $B_s^0\to\mu^+\mu^-$ effective lifetime}}, }{}\href{https://doi.org/10.1103/PhysRevLett.109.041801}{Phys.\ Rev.\ Lett.\  \textbf{109} (2012) 041801}, \href{http://arxiv.org/abs/1204.1737}{{\normalfont\ttfamily arXiv:1204.1737}}\relax
\mciteBstWouldAddEndPuncttrue
\mciteSetBstMidEndSepPunct{\mcitedefaultmidpunct}
{\mcitedefaultendpunct}{\mcitedefaultseppunct}\relax
\EndOfBibitem
\bibitem{CLEO:1984nby}
CLEO collaboration, R.~Giles {\em et~al.}, \ifthenelse{\boolean{articletitles}}{\emph{{Two-body decays of B mesons}}, }{}\href{https://doi.org/10.1103/PhysRevD.30.2279}{Phys.\ Rev.\  \textbf{D30} (1984) 2279}\relax
\mciteBstWouldAddEndPuncttrue
\mciteSetBstMidEndSepPunct{\mcitedefaultmidpunct}
{\mcitedefaultendpunct}{\mcitedefaultseppunct}\relax
\EndOfBibitem
\bibitem{LHCb-PAPER-2014-049}
CMS and LHCb collaborations, V.~Khachatryan {\em et~al.}, \ifthenelse{\boolean{articletitles}}{\emph{{Observation of the rare \mbox{\decay{\Bs}{\mumu}} decay from the combined analysis of CMS and LHCb data}}, }{}\href{https://doi.org/10.1038/nature14474}{Nature \textbf{522} (2015) 68}, \href{http://arxiv.org/abs/1411.4413}{{\normalfont\ttfamily arXiv:1411.4413}}\relax
\mciteBstWouldAddEndPuncttrue
\mciteSetBstMidEndSepPunct{\mcitedefaultmidpunct}
{\mcitedefaultendpunct}{\mcitedefaultseppunct}\relax
\EndOfBibitem
\bibitem{LHCb-PAPER-2017-001}
LHCb collaboration, R.~Aaij {\em et~al.}, \ifthenelse{\boolean{articletitles}}{\emph{{Measurement of the \mbox{\decay{\Bs}{\mumu}} branching fraction and effective lifetime and search for \mbox{\decay{\Bz}{\mumu}} decays}}, }{}\href{https://doi.org/10.1103/PhysRevLett.118.191801}{Phys.\ ~Rev.\ ~Lett.\  \textbf{118} (2017) 191801}, \href{http://arxiv.org/abs/1703.05747}{{\normalfont\ttfamily arXiv:1703.05747}}\relax
\mciteBstWouldAddEndPuncttrue
\mciteSetBstMidEndSepPunct{\mcitedefaultmidpunct}
{\mcitedefaultendpunct}{\mcitedefaultseppunct}\relax
\EndOfBibitem
\bibitem{ATLAS:2018cur}
ATLAS collaboration, M.~Aaboud {\em et~al.}, \ifthenelse{\boolean{articletitles}}{\emph{{Study of the rare decays of $B_s^0$ and $B^0$ mesons into muon pairs using data collected during 2015 and 2016 with the ATLAS detector}}, }{}\href{https://doi.org/10.1007/JHEP04(2019)098}{JHEP \textbf{04} (2019) 098}, \href{http://arxiv.org/abs/1812.03017}{{\normalfont\ttfamily arXiv:1812.03017}}\relax
\mciteBstWouldAddEndPuncttrue
\mciteSetBstMidEndSepPunct{\mcitedefaultmidpunct}
{\mcitedefaultendpunct}{\mcitedefaultseppunct}\relax
\EndOfBibitem
\bibitem{CMS:2022mgd}
CMS collaboration, A.~Tumasyan {\em et~al.}, \ifthenelse{\boolean{articletitles}}{\emph{{Measurement of the $B_s^0\to\mu^+\mu^-$ decay properties and search for the $B^0\to\mu^+\mu^-$ decay in proton-proton collisions at $\sqrt{s}=13$\tev}}, }{}\href{https://doi.org/10.1016/j.physletb.2023.137955}{Phys.\ Lett.\  \textbf{B842} (2023) 137955}, \href{http://arxiv.org/abs/2212.10311}{{\normalfont\ttfamily arXiv:2212.10311}}\relax
\mciteBstWouldAddEndPuncttrue
\mciteSetBstMidEndSepPunct{\mcitedefaultmidpunct}
{\mcitedefaultendpunct}{\mcitedefaultseppunct}\relax
\EndOfBibitem
\bibitem{LHCb-PAPER-2021-007}
LHCb collaboration, R.~Aaij {\em et~al.}, \ifthenelse{\boolean{articletitles}}{\emph{{Analysis of neutral $B$-meson decays into two muons}}, }{}\href{https://doi.org/10.1103/PhysRevLett.128.041801}{Phys.\ ~Rev.\ ~Lett.\  \textbf{128} (2022) 041801}, \href{http://arxiv.org/abs/2108.09284}{{\normalfont\ttfamily arXiv:2108.09284}}\relax
\mciteBstWouldAddEndPuncttrue
\mciteSetBstMidEndSepPunct{\mcitedefaultmidpunct}
{\mcitedefaultendpunct}{\mcitedefaultseppunct}\relax
\EndOfBibitem
\bibitem{LHCb-PAPER-2021-008}
LHCb collaboration, R.~Aaij {\em et~al.}, \ifthenelse{\boolean{articletitles}}{\emph{{Measurement of the $\Bs \to \mup \mun$ decay properties and search for the $\Bd \to \mup \mun$ and $\Bs \to \mup \mun \gamma$ decays}}, }{}\href{https://doi.org/10.1103/PhysRevD.105.012010}{Phys.\ ~Rev.\  \textbf{D105} (2022) 012010}, \href{http://arxiv.org/abs/2108.09283}{{\normalfont\ttfamily arXiv:2108.09283}}\relax
\mciteBstWouldAddEndPuncttrue
\mciteSetBstMidEndSepPunct{\mcitedefaultmidpunct}
{\mcitedefaultendpunct}{\mcitedefaultseppunct}\relax
\EndOfBibitem
\bibitem{LHCb-CONF-2020-002}
{ATLAS, CMS and LHCb collaborations}, \ifthenelse{\boolean{articletitles}}{\emph{{Combination of the ATLAS, CMS and LHCb results on the $B_{(s)}^0 \to \mu^+\mu^-$ decays}}}{}
\newblock \href{https://cds.cern.ch/record/2727207}{ATLAS-CONF-2020-049, CMS PAS BPH-20-003, LHCb-CONF-2020-002}, 2020\relax
\mciteBstWouldAddEndPuncttrue
\mciteSetBstMidEndSepPunct{\mcitedefaultmidpunct}
{\mcitedefaultendpunct}{\mcitedefaultseppunct}\relax
\EndOfBibitem
\bibitem{ATLAS:2023trk}
ATLAS collaboration, G.~Aad {\em et~al.}, \ifthenelse{\boolean{articletitles}}{\emph{{Measurement of the $B_s^0\to\mu^+\mu^-$ effective lifetime with the ATLAS detector}}, }{}\href{https://doi.org/10.1007/JHEP09(2023)199}{JHEP \textbf{09} (2023) 199}, \href{http://arxiv.org/abs/2308.01171}{{\normalfont\ttfamily arXiv:2308.01171}}\relax
\mciteBstWouldAddEndPuncttrue
\mciteSetBstMidEndSepPunct{\mcitedefaultmidpunct}
{\mcitedefaultendpunct}{\mcitedefaultseppunct}\relax
\EndOfBibitem
\bibitem{LHCb-DP-2022-002}
LHCb collaboration, R.~Aaij {\em et~al.}, \ifthenelse{\boolean{articletitles}}{\emph{{The LHCb Upgrade I}}, }{}\href{https://doi.org/10.1088/1748-0221/19/05/P05065}{{JINST} \textbf{19} (2024) P05065}, \href{http://arxiv.org/abs/2305.10515}{{\normalfont\ttfamily arXiv:2305.10515}}\relax
\mciteBstWouldAddEndPuncttrue
\mciteSetBstMidEndSepPunct{\mcitedefaultmidpunct}
{\mcitedefaultendpunct}{\mcitedefaultseppunct}\relax
\EndOfBibitem
\bibitem{LHCb-TDR-015}
LHCb collaboration, \ifthenelse{\boolean{articletitles}}{\emph{{LHCb Tracker Upgrade Technical Design Report}}, }{} \href{https://cds.cern.ch/search?p=CERN-LHCC-2014-001&f=reportnumber&action_search=Search&c=LHCb} {CERN-LHCC-2014-001}, 2014\relax
\mciteBstWouldAddEndPuncttrue
\mciteSetBstMidEndSepPunct{\mcitedefaultmidpunct}
{\mcitedefaultendpunct}{\mcitedefaultseppunct}\relax
\EndOfBibitem
\bibitem{LHCb-TDR-014}
LHCb collaboration, \ifthenelse{\boolean{articletitles}}{\emph{{LHCb PID Upgrade Technical Design Report}}, }{} \href{https://cds.cern.ch/search?p=CERN-LHCC-2013-022&f=reportnumber&action_search=Search&c=LHCb} {CERN-LHCC-2013-022}, 2013\relax
\mciteBstWouldAddEndPuncttrue
\mciteSetBstMidEndSepPunct{\mcitedefaultmidpunct}
{\mcitedefaultendpunct}{\mcitedefaultseppunct}\relax
\EndOfBibitem
\bibitem{LHCb-DP-2026-006}
LHCb collaboration, R.~Aaij {\em et~al.}, \ifthenelse{\boolean{articletitles}}{\emph{{Performance of the LHCb Run 3 detector}}, }{} LHCb-DP-2026-006, {in preparation}\relax
\mciteBstWouldAddEndPuncttrue
\mciteSetBstMidEndSepPunct{\mcitedefaultmidpunct}
{\mcitedefaultendpunct}{\mcitedefaultseppunct}\relax
\EndOfBibitem
\bibitem{LHCb-TDR-016}
LHCb collaboration, \ifthenelse{\boolean{articletitles}}{\emph{{LHCb Trigger and Online Upgrade Technical Design Report}}, }{} \href{https://cds.cern.ch/search?p=CERN-LHCC-2014-016&f=reportnumber&action_search=Search&c=LHCb} {CERN-LHCC-2014-016}, 2014\relax
\mciteBstWouldAddEndPuncttrue
\mciteSetBstMidEndSepPunct{\mcitedefaultmidpunct}
{\mcitedefaultendpunct}{\mcitedefaultseppunct}\relax
\EndOfBibitem
\bibitem{LHCb-DP-2020-002}
L.~Anderlini {\em et~al.}, \ifthenelse{\boolean{articletitles}}{\emph{{Muon identification for LHCb Run 3}}, }{}\href{https://doi.org/10.1088/1748-0221/15/12/T12005}{JINST \textbf{15} (2020) T12005}, \href{http://arxiv.org/abs/2008.01579}{{\normalfont\ttfamily arXiv:2008.01579}}\relax
\mciteBstWouldAddEndPuncttrue
\mciteSetBstMidEndSepPunct{\mcitedefaultmidpunct}
{\mcitedefaultendpunct}{\mcitedefaultseppunct}\relax
\EndOfBibitem
\bibitem{LHCb-DP-2025-007}
P.~Albicocco {\em et~al.}, \ifthenelse{\boolean{articletitles}}{\emph{{Performance of the LHCb muon detector in Run 3}}, }{}\href{https://doi.org/10.1016/j.nima.2026.171588}{Nucl.\ Instrum.\ Meth.\  \textbf{A1089} (2026) 171588}, \href{http://arxiv.org/abs/2604.21075}{{\normalfont\ttfamily arXiv:2604.21075}}\relax
\mciteBstWouldAddEndPuncttrue
\mciteSetBstMidEndSepPunct{\mcitedefaultmidpunct}
{\mcitedefaultendpunct}{\mcitedefaultseppunct}\relax
\EndOfBibitem
\bibitem{Sjostrand:2007gs}
T.~Sj\"{o}strand, S.~Mrenna, and P.~Skands, \ifthenelse{\boolean{articletitles}}{\emph{{A brief introduction to PYTHIA 8.1}}, }{}\href{https://doi.org/10.1016/j.cpc.2008.01.036}{Comput.\ Phys.\ Commun.\  \textbf{178} (2008) 852}, \href{http://arxiv.org/abs/0710.3820}{{\normalfont\ttfamily arXiv:0710.3820}}\relax
\mciteBstWouldAddEndPuncttrue
\mciteSetBstMidEndSepPunct{\mcitedefaultmidpunct}
{\mcitedefaultendpunct}{\mcitedefaultseppunct}\relax
\EndOfBibitem
\bibitem{Lange:2001uf}
D.~J. Lange, \ifthenelse{\boolean{articletitles}}{\emph{{The EvtGen particle decay simulation package}}, }{}\href{https://doi.org/10.1016/S0168-9002(01)00089-4}{Nucl.\ Instrum.\ Meth.\  \textbf{A462} (2001) 152}\relax
\mciteBstWouldAddEndPuncttrue
\mciteSetBstMidEndSepPunct{\mcitedefaultmidpunct}
{\mcitedefaultendpunct}{\mcitedefaultseppunct}\relax
\EndOfBibitem
\bibitem{davidson2015photos}
N.~Davidson, T.~Przedzinski, and Z.~Was, \ifthenelse{\boolean{articletitles}}{\emph{{PHOTOS interface in C++: Technical and physics documentation}}, }{}\href{https://doi.org/https://doi.org/10.1016/j.cpc.2015.09.013}{Comput.\ Phys.\ Commun.\  \textbf{199} (2016) 86}, \href{http://arxiv.org/abs/1011.0937}{{\normalfont\ttfamily arXiv:1011.0937}}\relax
\mciteBstWouldAddEndPuncttrue
\mciteSetBstMidEndSepPunct{\mcitedefaultmidpunct}
{\mcitedefaultendpunct}{\mcitedefaultseppunct}\relax
\EndOfBibitem
\bibitem{Agostinelli:2002hh}
Geant4 collaboration, S.~Agostinelli {\em et~al.}, \ifthenelse{\boolean{articletitles}}{\emph{{Geant4: A simulation toolkit}}, }{}\href{https://doi.org/10.1016/S0168-9002(03)01368-8}{Nucl.\ Instrum.\ Meth.\  \textbf{A506} (2003) 250}\relax
\mciteBstWouldAddEndPuncttrue
\mciteSetBstMidEndSepPunct{\mcitedefaultmidpunct}
{\mcitedefaultendpunct}{\mcitedefaultseppunct}\relax
\EndOfBibitem
\bibitem{Allison:2006ve}
Geant4 collaboration, J.~Allison {\em et~al.}, \ifthenelse{\boolean{articletitles}}{\emph{{Geant4 developments and applications}}, }{}\href{https://doi.org/10.1109/TNS.2006.869826}{IEEE Trans.\ Nucl.\ Sci.\  \textbf{53} (2006) 270}\relax
\mciteBstWouldAddEndPuncttrue
\mciteSetBstMidEndSepPunct{\mcitedefaultmidpunct}
{\mcitedefaultendpunct}{\mcitedefaultseppunct}\relax
\EndOfBibitem
\bibitem{LHCb-PROC-2011-006}
M.~Clemencic {\em et~al.}, \ifthenelse{\boolean{articletitles}}{\emph{{The \lhcb simulation application, Gauss: Design, evolution and experience}}, }{}\href{https://doi.org/10.1088/1742-6596/331/3/032023}{J.\ Phys.\ Conf.\ Ser.\  \textbf{331} (2011) 032023}\relax
\mciteBstWouldAddEndPuncttrue
\mciteSetBstMidEndSepPunct{\mcitedefaultmidpunct}
{\mcitedefaultendpunct}{\mcitedefaultseppunct}\relax
\EndOfBibitem
\bibitem{Breiman}
L.~Breiman, J.~H. Friedman, R.~A. Olshen, and C.~J. Stone, {\em Classification and regression trees}, Wadsworth international group, Belmont, California, USA, 1984\relax
\mciteBstWouldAddEndPuncttrue
\mciteSetBstMidEndSepPunct{\mcitedefaultmidpunct}
{\mcitedefaultendpunct}{\mcitedefaultseppunct}\relax
\EndOfBibitem
\bibitem{XGBoosting}
T.~Chen and C.~Guestrin, \ifthenelse{\boolean{articletitles}}{\emph{{XGBoost}: A scalable tree boosting system}, }{} in {\em Proceedings of the 22nd ACM SIGKDD international conference on knowledge Discovery and data mining}, \href{https://doi.org/10.1145/2939672.2939785}{ KDD '16, (New York, NY, USA), 785--794, ACM, 2016}, \href{http://arxiv.org/abs/1603.02754}{{\normalfont\ttfamily arXiv:1603.02754}}\relax
\mciteBstWouldAddEndPuncttrue
\mciteSetBstMidEndSepPunct{\mcitedefaultmidpunct}
{\mcitedefaultendpunct}{\mcitedefaultseppunct}\relax
\EndOfBibitem
\bibitem{Pivk:2004ty}
M.~Pivk and F.~R. Le~Diberder, \ifthenelse{\boolean{articletitles}}{\emph{{sPlot: A statistical tool to unfold data distributions}}, }{}\href{https://doi.org/10.1016/j.nima.2005.08.106}{Nucl.\ Instrum.\ Meth.\  \textbf{A555} (2005) 356}, \href{http://arxiv.org/abs/physics/0402083}{{\normalfont\ttfamily arXiv:physics/0402083}}\relax
\mciteBstWouldAddEndPuncttrue
\mciteSetBstMidEndSepPunct{\mcitedefaultmidpunct}
{\mcitedefaultendpunct}{\mcitedefaultseppunct}\relax
\EndOfBibitem
\bibitem{PDG2026}
Particle Data Group, F.~Takahashi {\em et~al.}, \ifthenelse{\boolean{articletitles}}{\emph{{\href{http://pdg.lbl.gov/}{Review of particle physics}}}, }{}\href{https://doi.org/10.1142/S0217751X26300115}{Mod.\ Phys.\  \textbf{A41} (2026) 2630011}\relax
\mciteBstWouldAddEndPuncttrue
\mciteSetBstMidEndSepPunct{\mcitedefaultmidpunct}
{\mcitedefaultendpunct}{\mcitedefaultseppunct}\relax
\EndOfBibitem
\bibitem{LHCb-PAPER-2026-056}
LHCb collaboration, R.~Aaij {\em et~al.}, \ifthenelse{\boolean{articletitles}}{\emph{{First measurement of the \Bs versus \Bu meson production ratio in $pp$ collisions at $\sqs = 13.6\tev$}}, }{} LHCb-PAPER-2026-056, {in preparation, to be submitted to JHEP}\relax
\mciteBstWouldAddEndPuncttrue
\mciteSetBstMidEndSepPunct{\mcitedefaultmidpunct}
{\mcitedefaultendpunct}{\mcitedefaultseppunct}\relax
\EndOfBibitem
\bibitem{Skwarnicki:1986xj}
T.~Skwarnicki, {\em {A study of the radiative cascade transitions between the Upsilon-prime and Upsilon resonances}}, PhD thesis, Institute of Nuclear Physics, Krakow, 1986, {\href{http://inspirehep.net/record/230779/}{DESY-F31-86-02}}\relax
\mciteBstWouldAddEndPuncttrue
\mciteSetBstMidEndSepPunct{\mcitedefaultmidpunct}
{\mcitedefaultendpunct}{\mcitedefaultseppunct}\relax
\EndOfBibitem
\bibitem{CRANMER2001198}
K.~Cranmer, \ifthenelse{\boolean{articletitles}}{\emph{Kernel estimation in high-energy physics}, }{}\href{https://doi.org/https://doi.org/10.1016/S0010-4655(00)00243-5}{Comput.\ Phys.\ Commun.\  \textbf{136} (2001) 198}\relax
\mciteBstWouldAddEndPuncttrue
\mciteSetBstMidEndSepPunct{\mcitedefaultmidpunct}
{\mcitedefaultendpunct}{\mcitedefaultseppunct}\relax
\EndOfBibitem
\bibitem{Wilks:1938dza}
S.~S. Wilks, \ifthenelse{\boolean{articletitles}}{\emph{{The large-sample distribution of the likelihood ratio for testing composite hypotheses}}, }{}\href{https://doi.org/10.1214/aoms/1177732360}{Ann.\ Math.\ Stat.\  \textbf{9} (1938) 60}\relax
\mciteBstWouldAddEndPuncttrue
\mciteSetBstMidEndSepPunct{\mcitedefaultmidpunct}
{\mcitedefaultendpunct}{\mcitedefaultseppunct}\relax
\EndOfBibitem
\bibitem{DeBruyn:2012wj}
K.~De~Bruyn {\em et~al.}, \ifthenelse{\boolean{articletitles}}{\emph{{Branching Ratio Measurements of $B_s$ Decays}}, }{}\href{https://doi.org/10.1103/PhysRevD.86.014027}{Phys.\ Rev.\  \textbf{D86} (2012) 014027}, \href{http://arxiv.org/abs/1204.1735}{{\normalfont\ttfamily arXiv:1204.1735}}\relax
\mciteBstWouldAddEndPuncttrue
\mciteSetBstMidEndSepPunct{\mcitedefaultmidpunct}
{\mcitedefaultendpunct}{\mcitedefaultseppunct}\relax
\EndOfBibitem
\bibitem{HFLAV:GsDGs}
Heavy Flavor Averaging Group (HFLAV), S.~Banerjee {\em et~al.}, \ifthenelse{\boolean{articletitles}}{\emph{{Averages of $b$-hadron, $c$-hadron, and $\tau$-lepton properties as of 2023}}, }{}\href{https://doi.org/10.1103/x87q-tld5}{Phys.\ Rev.\  \textbf{D113} (2026) 012008}, \href{http://arxiv.org/abs/2411.18639}{{\normalfont\ttfamily arXiv:2411.18639}}, with specific result from \href{https://doi.org/10.5281/zenodo.19446771}{\texttt{doi:10.5281/zenodo.19446771}}\relax
\mciteBstWouldAddEndPuncttrue
\mciteSetBstMidEndSepPunct{\mcitedefaultmidpunct}
{\mcitedefaultendpunct}{\mcitedefaultseppunct}\relax
\EndOfBibitem
\end{mcitethebibliography}

\newpage

\FloatBarrier
\newpage
\centerline
{\large\bf LHCb collaboration}
\begin
{flushleft}
\small
R.~Aaij$^{41}$\lhcborcid{0000-0003-0533-1952},
M.~Abdelfatah$^{74}$,
A.S.W.~Abdelmotteleb$^{62}$\lhcborcid{0000-0001-7905-0542},
C.~Abellan~Beteta$^{56}$\lhcborcid{0009-0009-0869-6798},
F.~Abudin\'en$^{64}$\lhcborcid{0000-0002-6737-3528},
T.~Ackernley$^{66}$\lhcborcid{0000-0002-5951-3498},
A.A.~Adefisoye$^{74}$\lhcborcid{0000-0003-2448-1550},
B.~Adeva$^{52}$\lhcborcid{0000-0001-9756-3712},
M.~Adinolfi$^{60}$\lhcborcid{0000-0002-1326-1264},
P.~Adlarson$^{91,46}$\lhcborcid{0000-0001-6280-3851}
% ...

{\footnotesize \it

% ...
$^{41}$Nikhef National Institute for Subatomic Physics, Amsterdam, Netherlands\\
% ...
$^{46}$National Center for Nuclear Research (NCBJ), Warsaw, Poland\\
% ...
$^{52}$Instituto Galego de F{\'\i}sica de Altas Enerx{\'\i}as (IGFAE), Universidade de Santiago de Compostela, Santiago de Compostela, Spain\\
% ...
$^{56}$Physik-Institut, Universit{\"a}t Z{\"u}rich, Z{\"u}rich, Switzerland\\
% ...
$^{60}$H.H. Wills Physics Laboratory, University of Bristol, Bristol, United Kingdom\\
% ...
$^{62}$Department of Physics, University of Warwick, Coventry, United Kingdom\\
% ...
$^{64}$School of Physics and Astronomy, University of Edinburgh, Edinburgh, United Kingdom\\
$^{65}$School of Physics and Astronomy, University of Glasgow, Glasgow, United Kingdom\\
$^{66}$Oliver Lodge Laboratory, University of Liverpool, Liverpool, United Kingdom\\
% ...
$^{74}$Syracuse University, Syracuse, NY, United States\\
% ...
$^{91}$Department of Physics and Astronomy, Uppsala University, Uppsala, Sweden, associated to $^{65}$\\
% ...
}
\end{flushleft}

\end{document}